\documentclass[aps,prx,showpacs,twocolumn,superscriptaddress,nobibnotes,floatfix,preprintnumbers]{revtex4-2}
\UseRawInputEncoding
\RequirePackage{subfigure}
\usepackage{lipsum}
\usepackage{graphicx}
\usepackage{color}
\usepackage{enumerate}
\usepackage{amsmath}
\usepackage{bm}
\usepackage{diagbox}
\usepackage{amssymb}
\usepackage{stmaryrd}
\usepackage{multirow}
\usepackage{bbm}
\usepackage[normalem]{ulem}
\usepackage{float}
\usepackage{longtable,booktabs}
\usepackage[dvipsnames,table,xcdraw]{xcolor}
\usepackage[normalem]{ulem}
\usepackage{physics}
\usepackage{url}
\usepackage{soul} 
\usepackage[colorinlistoftodos]{todonotes}
\usepackage{verbatim}
\usepackage{graphicx}
\usepackage{subfigure}
\graphicspath{{figures/}}
\usepackage{appendix}
\usepackage{siunitx}
\definecolor{darkblue}{rgb}{0.1,0.2,0.6} \definecolor{darkred}{rgb}{0.8,0.1,0.2}
\usepackage[colorlinks,citecolor=darkblue,linkcolor=darkred,urlcolor=darkblue]{hyperref}
\usepackage[all]{hypcap} 
\usepackage[draft]{pdfcomment}

\begin{document}

\title{Simulating 2+1D Lattice Quantum Electrodynamics at Finite Density \\
with Neural Flow Wavefunctions}

\author{Zhuo Chen}
\thanks{Co-first author.}
\affiliation{Center for Theoretical Physics, Massachusetts Institute of Technology, Cambridge, MA 02139, USA}
\affiliation{The NSF AI Institute for Artificial Intelligence and Fundamental Interactions}
\author{Di Luo}
\thanks{Co-first and corresponding author: diluo@mit.edu}
\affiliation{Center for Theoretical Physics, Massachusetts Institute of Technology, Cambridge, MA 02139, USA}
\affiliation{The NSF AI Institute for Artificial Intelligence and Fundamental Interactions}
\affiliation{Department of Physics, Harvard University, Cambridge, MA 02138, USA}
\author{Kaiwen Hu}
\affiliation{Department of Physics, University of Michigan}
\author{Bryan K. Clark}
\affiliation{Institute for Condensed Matter Theory and IQUIST and NCSA Center for Artificial Intelligence Innovation and Department of Physics, University of Illinois, Urbana-Champaign}

\begin{abstract}

We present a neural flow wavefunction, Gauge-Fermion FlowNet, and use it to simulate 2+1D lattice compact quantum electrodynamics with finite density dynamical fermions. 
The gauge field is represented by a neural network which parameterizes a discretized flow-based transformation of the amplitude while the fermionic sign structure is represented by a neural net backflow. This approach directly represents the $U(1)$ degree of freedom without any truncation, obeys Guass's law by construction, samples autoregressively avoiding any equilibration time, and variationally simulates Gauge-Fermion systems with sign problems accurately. In this model, we investigate confinement and string breaking phenomena in different fermion density and hopping regimes. We study the phase transition from the charge crystal phase to the vacuum phase at zero density, and observe the phase seperation and the net charge penetration blocking effect under magnetic interaction at finite density. In addition, we investigate a magnetic phase transition due to the competition effect between the kinetic energy of fermions and the magnetic energy of the gauge field. With our method, we further note potential differences on the order of the phase transitions between a continuous $U(1)$ system and  one with finite truncation. Our state-of-the-art neural network approach opens up new possibilities to study different gauge theories coupled to dynamical matter in higher dimensions. 

\end{abstract}

\preprint{MIT-CTP/5497}
\maketitle

\section{Introduction}

Gauge theory coupled to dynamical matter plays a fundamental role in physics. For example, quantum electrodynamics (QED) describes the light-matter interaction while  quantum chromodynamics (QCD) describes the quark-gluon interaction, which are the important components of the Standard Model. Meanwhile, exotic gauge theories with matter also arise in theories of condensed matter and AMO systems~\cite{Kitaev_2003,Hamma_2005,Vijay_2016}. When a dynamical gauge theory is discretized and placed on a lattice in the Hamiltonian formulation, Gauss's law needs to be explicitly imposed. There are various open questions on understanding both the phase diagram and the real-time evolution of dynamical gauge theory, which are challenging to address in simulations particularly in high spatial dimensions and in the finite charge density regimes.

Early attempts on simulating lattice gauge theory start with Monte Carlo simulations which achieve great success for scenarios without sign problems~\cite{rebbi_1983,Xu_2019}. In this case, the path integral could be expressed as a high dimensional distribution and observables can be computed through importance sampling. However, the Monte Carlo approach becomes challenging with high sample complexity when there exist sign problems in the complex-valued action, such as the real-time dynamics and the fermionic case. Besides Monte Carlo simulations, tensor networks provide another powerful tool for simulating lattice gauge theory. The tensor network approach is variational and hence has no sign problem but is mainly  efficient for problems in 1+1D~\cite{Ba_uls_2020}. While there are recent attempts on extending the tensor network approach to 2+1D and 3+1D simulations \cite{emonts2020variational,qlm_2dphase,emonts2022finding,zohar2018combining,PhysRevX.10.041040, Magnifico_2021}, the tensor network methods are usually more constrained in higher dimensions as well as real time dynamics where the entanglement grows with system size. It is also an open question on how to utilize the tensor network approach for simulating gauge theories with continuous or infinite degrees of freedom without imposing a cutoff.  With the recent development of quantum technologies, quantum computation provides another paradigm for lattice gauge theory simulations~\cite{mazzola2021gauge,haase2021resource,Ba_uls_2020,PRXQuantum.2.030334,PRXQuantum.3.020324,luo2020framework,zhou2022thermalization,yang2020observation,osborne2022large,kan2021investigating,PhysRevD.101.074512,PhysRevD.103.094501,PhysRevD.104.074505,zohar2013simulating,PhysRevLett.127.276402, PhysRevX.10.021041}. Despite the nice proposals and progress of the field, the performance of the near-term quantum algorithms are still limited due to the noisy nature of the current quantum devices and the large depth and qubit numbers required.

Meanwhile, with the advancement of machine learning, new attempts have been proposed to simulate lattice field theories with symmetries. One direction is to utilize the Lagrangian approach and learn the probability distribution or observables with neural networks \cite{favoni2022lattice,Boyda_2021, PhysRevLett.125.121601,albergo2019flow, PhysRevResearch.2.023369, https://doi.org/10.48550/arxiv.2207.08945}. This approach significantly speeds up the sampling of independent configurations but is still not applicable for models where the sign problem exists. Another direction is to work with the Hamiltonian formulation of lattice gauge theories and develop neural network quantum states that fulfill the gauge symmetries \cite{PhysRevLett.127.276402, https://doi.org/10.48550/arxiv.2101.07243,luo2022gauge}. In the past few years, the neural network quantum states have been used to solve quantum many-body physics in various contexts and demonstrate successes, including ground state properties~\cite{doi:10.1126/science.aag2302, Hibat_Allah_2020, PhysRevLett.124.020503, Irikura_2020, PhysRevResearch.3.023095, Han_2020,ferminet,Choo_2019,rnn_wavefunction,paulinet,Glasser_2018,Stokes_2020,Nomura_2017,martyn2022variational,Luo_2019}, finite temperature and real time dynamics \cite{xie2021ab,wang2021spacetime,py2021, gutierrez2020real, Schmitt_2020,Vicentini_2019,PhysRevB.99.214306,PhysRevLett.122.250502,PhysRevLett.122.250501,luo_gauge_inv,luo_povm}.   Although the variational approach is more flexible for handling the sign problem, previous research on variational quantum states with gauge symmetries has been limited to gauge theories with discrete gauge freedom or no formulation currently exists for coupling gauge fields to dynamical fermions beyond 1+1D \cite{PhysRevResearch.2.043145}. In this work, we develop a novel neural network architecture, Gauge-Fermion FlowNet (GFFN),  which consists of two parts.  The first part involves a flow-based generative model, which maps a simple normalized probability distribution (i.e. a gaussian) into the probability distribution over the gauge and fermionic degrees of freedom.  This mapping has an autoregressive construction which imposes the gauge symmetries of Gauss's law in a system with fermions, and provide efficient and exact sampling, which avoids autocorrelation time compared to the traditional Markov chain Monte Carlo method.  Using this approach, we are able to directly encode the $U(1)$ degree of freedom without any finite truncation, as is typically needed in other approaches such as tensor networks.  Because probability distributions are non-negative, an additional piece is needed to represent the sign-structure induced by the Fermions. To accomplish this, we use the neural-network backflow to represent the phase of the wave-function.   This approach doesn't spoil the autoregressive nature of the entire wave-function but supplements the probability distribution with an accurate sign-structure, allowing simulations of gauge fields coupled to finite-density dynamical fermions with sign problems that go beyond the sign-free Lagrangian approach.

The paper is structured as follows: In Sec.~\ref{sec:ham}, we review the Kogut-Susskind Hamiltonian formulation of 2+1D lattice compact QED. In Sec.~\ref{sec:nn} we introduce our novel gauge invariant flow-based neural network which simultaneously encodes the continuous gauge degrees of freedom and the fermion degrees of freedom while exactly satisfying Gauss's law.  In Sec.~\ref{sec:vmc}, we provide an overview of the variational Monte Carlo approach for solving the ground state with neural network wave function. Sec.~\ref{sec:string} studies the string breaking and confinement phenomena at different fermion densities and hopping amplitudes.  Sec.~\ref{sec:phase} investigates the phase transition between the charge crystal phase and the vacuum phase in zero density, as well as the effect of magnetic interactions on the net charge penetration blocking in finite dnesity. Sec.~\ref{sec:topo_phase} studies the magnetic phase transition resulting from the magnetic energy and the fermionic kinetic energy competition. Finally we draw conclusions and outlook in Sec.~\ref{sec:con}.

\section{Hamiltonian Formulation of 2+1D Lattice Compact QED}\label{sec:ham}

\begin{figure}[h]
    \centering
    \includegraphics[width=0.4\linewidth]{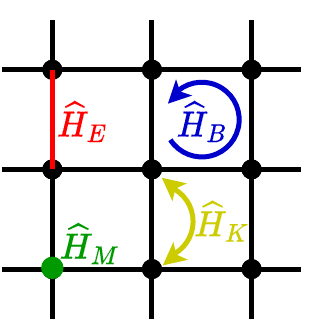}
    \caption{Illustration of the four terms in the Kogut-Susskind Hamiltonian in Eq.~\ref{eq:ham_terms}.
    }
    \label{fig:ham}
\end{figure}

We start with the Kogut-Susskind formulation of the 2+1D lattice compact QED, where the gauge field resides on the links and fermions reside on the vertices. 
The theory is described by the following Hamiltonian (Fig.~\ref{fig:ham}),
\begin{equation}\label{eq:ham}
    \hat{H} = \hat H_E + \hat H_B + \hat H_M + \hat H_K,
\end{equation}
where
\begin{align} \label{eq:ham_terms}
\begin{split}
    \hat H_E =& \frac{g_E^2}{2}\sum_{i, j} \qty[\qty(\hat {E}_{i,j}^{(x)})^2 + \qty(\hat {E}_{i,j}^{(y)})^2], \\
    \hat H_B =& -\frac{g_B^2}{2}\sum_{i, j} \qty(\hat P_{i,j} + \hat P_{i,j}^\dagger - 2), \\
    \hat H_M =& m \sum_{i, j} \bigg\{\frac{1}{2}\qty[(-1)^{i+j} + 1] \hat \psi_{i,j}^\dagger \hat \psi_{i,j} \\ &\ \ \ \ \ \ \ -  \frac{1}{2}\qty[(-1)^{i+j} - 1] \hat\psi_{i,j} \hat\psi_{i,j}^\dagger\bigg\}, \\
    \hat H_K =& -\kappa \sum_{i, j} \bigg[\hat \psi_{i,j}^\dagger \qty(\hat U_{i,j}^{(x)})^\dagger \hat \psi_{i+1, j} \\& \ \ \ \ \ \ \  \ \ + \hat \psi_{i,j}^\dagger \qty(\hat U_{i,j}^{(y)})^\dagger \hat \psi_{i, j+1} + \text{H.C.}\bigg].
\end{split}
\end{align}
Here, $\hat E_{i,j}^{(x)}$ (or $\hat E_{i,j}^{(y)}$) refers to the electric operator acting on the link between $(i, j)$ and $(i+1, j)$ (or $(i, j+1)$). Similarly for $\hat U_{i,j}^{(x)}$ (or $\hat U_{i,j}^{(y)}$). The plaquette operator acting on plaquette $(i, j)$ is
\begin{equation}
    \hat P_{i, j} = \hat U_{i,j}^{(x)} \hat U_{i+1,j}^{(y)} \qty(\hat U_{i,j+1}^{(x)})^\dagger \qty(\hat U_{i,j}^{(y)})^\dagger.
\end{equation}
In addition, $\hat E_{i,j}^{(w)}$ and  $\hat U_{i,j}^{(w)}$ obeys the commutation relation that
\begin{equation}
    [\hat E_{i,j}^{(w)}, \hat U_{i',j'}^{(w')}] = -\delta_{i, i'}\delta_{j, j'}\delta_{w, w'} \hat U_{i',j'}^{(w')}.
\end{equation}
Moreover, there is a Gauss' law on each vertex as
\begin{equation}
     \hat E_{i,j}^{(x)} + \hat E_{i,j}^{(y)} - \hat E_{i-1,j}^{(x)} - \hat E_{i,j-1}^{(y)} = \hat q_{i, j} \quad \forall (i, j),
\end{equation}
where
\begin{equation} \label{eq:stager_charge}
    \hat q_{i, j} = \hat\psi_{i, j}^\dagger \hat\psi_{i, j} + \frac{1}{2}\qty[(-1)^{i+j} - 1].
\end{equation}
Intuitively, a fermion creates a positive charge on an even site while no charges on an odd site, whereas a hole (no fermion) creates a negative charge on an odd site while no charges on an even site.

For the physical scenario of the standard QED, $g_E^2 g_B^2 = 8\kappa^2$, but one can also relax the constraint to study the effects of different coupling constants. In this paper, our variational wavefunction works in the eigenbasis $\ket{E}$ of the $\hat E$ operator. According to the commutation relation, we have $\hat U \ket{E} = \ket{E-1}$ and $\hat U^\dagger \ket{E} = \ket{E+1}$. We note that our definition of $H_B$ and $H_M$ might be different from other definitions by a shift of identity operator. For the Hamiltonian described in Eq.~\ref{eq:ham}, the sign problem exists at both zero density and finite density, though the sign problem can be avoided if there are an even number of fermion species or the gauge field is $\mathbb{Z}_2$\cite{Xu_2019}.

\section{Gauge Invariant Autoregressive Flow Architecture}\label{sec:nn}

In this work, we design a gauge invariant autoregressive flow neural network, Gauge-Fermion FlowNet that incorporates both the gauge fields' infinite degrees of freedom ($E\in \mathbb{Z}$) as well as the fermionic degrees of freedom ($f \in \{0, 1\}$, the occupation number). The neural network can be thought as a variantional ansatz with parameters $\bm\theta$ such that, when a particular configuration of the fermions and gauge fields $(\bm f, \bm E)$ is provided (throughout bold-face indicates vectors), the neural network returns the corresponding (normalized) wave function coefficient $\psi_{\bm\theta} (\bm f, \bm E)$. We define the overall wave function as
\begin{equation}
    \psi_{\bm\theta}(\bm f, \bm E) = \sqrt{p_{\bm\theta_A}(\bm f, \bm E)} e^{i \phi_{\bm\theta_B}(\bm f, \bm E)}
\end{equation}

where the amplitude part $p_{\bm{\theta}_A}$ and the phase part $\phi_{\bm{\theta}_B}$ are parameterized using different neural networks.

\begin{figure}[h]
    \centering
    \includegraphics[width=\linewidth]{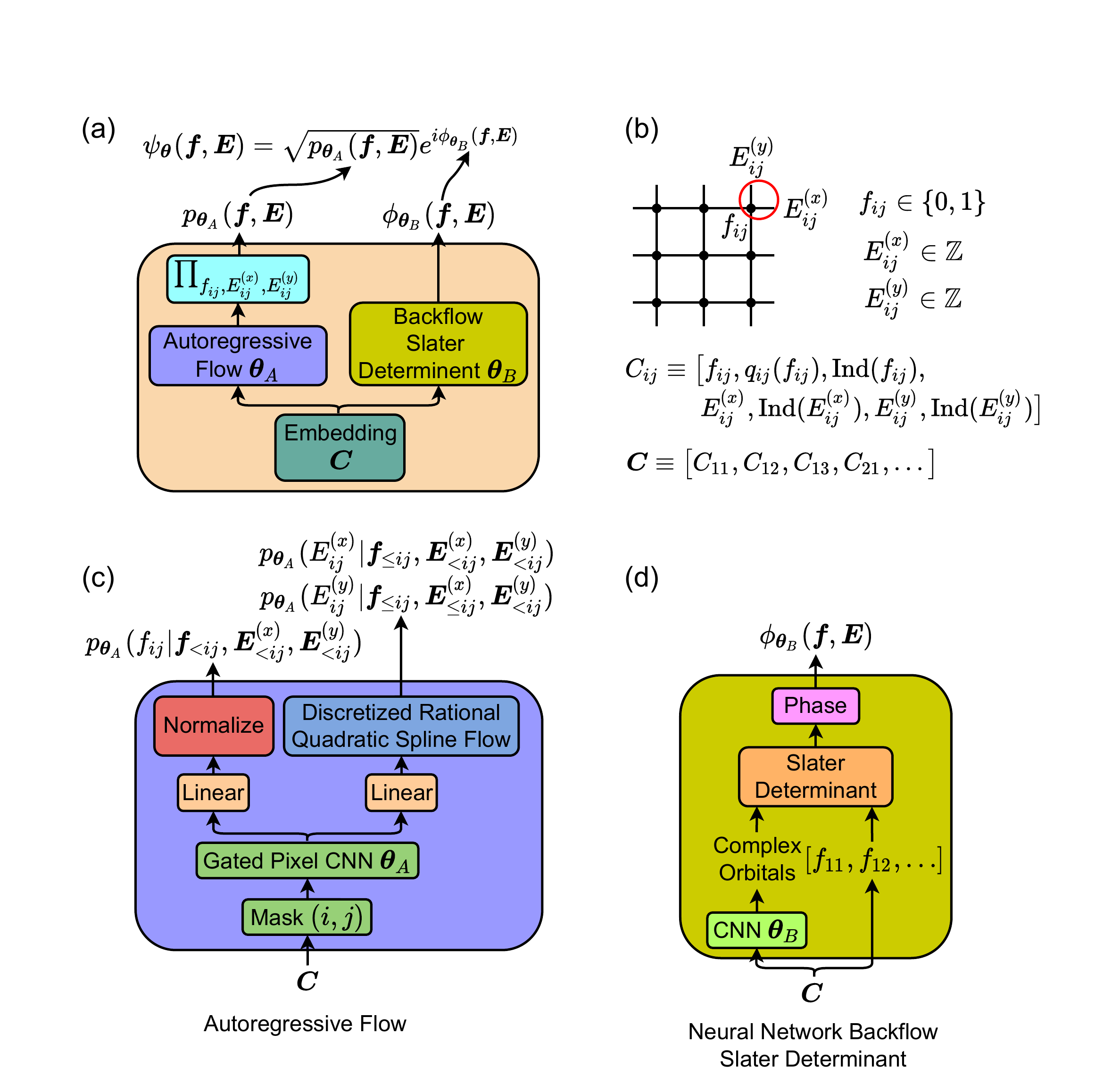}
    \caption{(a) General setup of the neural network. The neural network consists of three parts---the embedding; the autoregressive flow, which generates the probability distribution over the configurations; and the neural network backflow Slater determinant, which produces the phase of the configuration. (b) The embedding of the neural network which maps a configuration of the fermions and gauge fields into a vector $C_{ij}$.  In the configuration, there are two kinds of fields---the electric field (with values spanning all integers) and the fermion field with values 0 (no fermion) or 1 (a fermion). Inside the embedding vector, $q_{ij}$ is a function that gives the charge of $f_{ij}$, and $\text{Ind}$ is a function that outputs 1 if the unit cell $(i, j)$ contains the corresponding degree of freedom, and outputs 0 otherwise. (c) The autoregressive flow. The autoregressive flow evaluates three conditional probabilities for site $(i,j)$ using the embedded vectors $\bm C$. The embeded vectors are masked so that only the information necessary to generate the conditional probabilities is passed into the gated pixelCNN.
    (d) Neural network backflow Slater determinant. The backflow Slater determinant takes as input $\bm C$ and runs them through a multi-layer CNN to obtain the complex orbitals. Then, the complex orbitals and the fermion configuration is used to form a Slater determinant. Afterwards, we define the phase of the wave function as the phase of the Slater determinant.
    }
    \label{fig:set_up}
\end{figure}

The neural network can be divided into three parts---the embedding part, the probability distribution part which generates $p_{\bm \theta_A}(\bm f, \bm E)$,and the phase part which generates $\phi_{\bm \theta_B}(\bm f, \bm E)$ (Fig.~\ref{fig:set_up}). 

\begin{figure*}[ht]
    \centering
    \includegraphics[width=\linewidth]{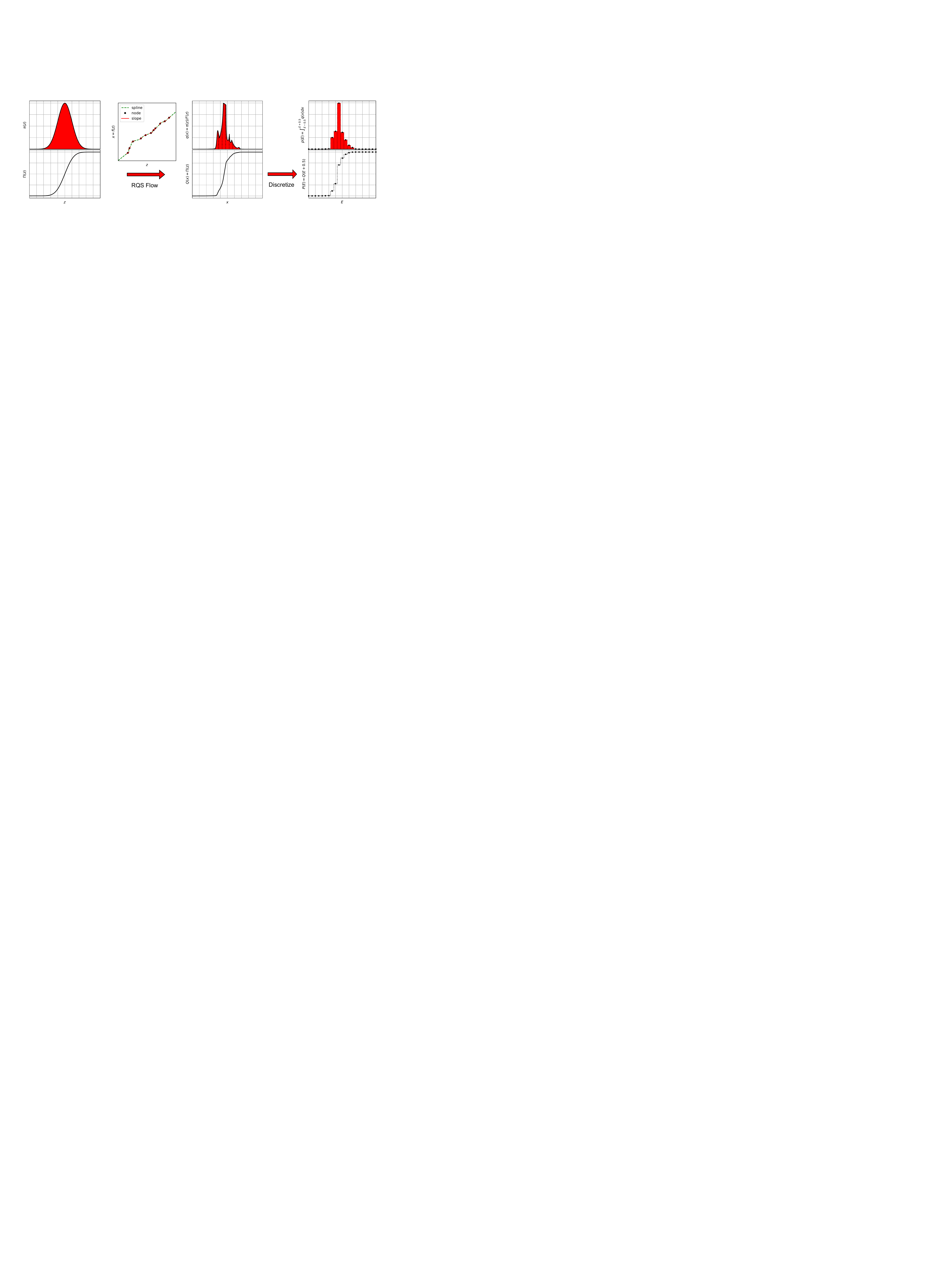}
    \caption{Illustration of the discretized rational quadratic spline (RQS) flow. We use a standard normal distribution as the prior probability distribution $\pi(z)$ (together with a cumulative distribution $\Pi(z)$). Then, we form a RQS using the outputs from the neural network as the nodes and the slopes. This RQS defines an invertible function $x=f(z)$, which induces a new probability distribution $q(x)=\pi(z)/f'(z)$ and a cumulative distribution $Q(x)=\Pi(z)$ with $z=f^{-1}(x)$. Afterwards, the continuous distribution is discretized to obtain a discrete distribution defined on integers (electric field $E$) by defining $p(E) = \int_{E-0.5}^{E+0.5} q(x) \dd x$ and $P(E)=Q(E+0.5)$.  Then $p(E)$ can be evaluate efficiently as $p(E)=P(E)-P(E-1) = Q(E+0.5) - Q(E-0.5) = \Pi(f^{-1}(E+0.5)) - \Pi(f^{-1}(E-0.5))$. The probability distribution $p(E)$ is the conditional distribution of the gauge field $E$ on a single link.
    }
    \label{fig:quadratic_spline}
\end{figure*}

\subsection{Embedding}

The embedding part (Fig.~\ref{fig:set_up} (a, b)) converts the field configurations into tensors of the shape $N_{\text{samples}}\times N_{\text{features}} \times (L_x+1) \times (L_y+1) $, where $L_x$ (or $L_y$) is the number of sites in the $x$-(or $y$-)direction. To be more specific, we define a fermion at location $(i, j)$, the horizontal electric field at location $(i, j)$, and the vertical electric field at location $(i, j)$ together as a unit cell and embed them into a vector of length  $N_{\text{features}} =7$. The elements of the vector are 
\begin{equation}
\begin{aligned}
    C_{ij} \equiv \bigg[&f_{ij}, q_{ij}(f_{ij}), \text{Ind}(f_{ij}), \\
    &E_{ij}^{(x)}, \text{Ind}(E_{ij}^{(x)}), E_{ij}^{(y)}, \text{Ind}(E_{ij}^{(y)})\bigg].
\end{aligned}
\end{equation}
Here the $q_{ij}(f_{ij}) = f_{ij} + [(-1)^{i+j}-1]/2$ measures the staggered charge at the site given the fermion configuration. In addition, we have the $\text{Ind}$ function which serves as an indicator that evaluates to 1 if the site or link exists inside the unit cell, and evaluates to 0 if the site or link does not exist inside the unit cell. This indicator function is needed because some unit cells consist of all three $f_{ij}$, $E_{ij}^{(x)}$, and $E_{ij}^{(y)}$, while other unit cells may only consist a subset of them. 
The embedding has no variational parameters. 

\subsection{Probability distribution with discretized flow}

We use the autoregressive flow (Fig.~\ref{fig:set_up} (a, c)) \cite{pmlr-v37-rezende15,https://doi.org/10.48550/arxiv.1705.07057,https://doi.org/10.48550/arxiv.1804.00779,  https://doi.org/10.48550/arxiv.2002.02547} to generate the probability part of the wave function. Specifically, the autoregressive flow takes as input the embedded configuration and masks them such that only the fermions and gauge fields that are used in the conditional probability distribution are included. This preserves the autoregressive structure of the neural network. Here, we use the notation $\bm X_{< ij}$ (or $\bm X_{\le ij}$) to indicate all $X$'s less than (or equal to) $ij$ in a fixed linear order of sites.  In practice, the mask is built inside the gated pixelCNN (see Appendix~\ref{app:dnn}). In addition, we only need to feed $\bm C$ once, to produce all conditional probability distributions in parallel. The $\bm C$ goes through a parameterized (with parameters $\bm \theta_A$)  gated pixelCNN (see Appendix~\ref{app:dnn}) and the output from the gated pixelCNN is fed into a linear layer. We use two different linear layers for fermions and gauge fields. For fermions, the linear layer outputs are considered as the unnormalized conditional probability distribution, which after normalization, becomes the conditional probability distribution for the fermions $p_{\bm \theta_A}(f_{ij}|\bm f_{<ij}, \bm E_{<ij}^{(x)}, \bm E_{<ij}^{(y)})$. In practice, we work in log space, so the output is log probability distribution and the normalization becomes a log\_softmax function. On the other hand, for gauge fields, the linear layer output is feed into a discretized rational quadratic spline (RQS) flow (Fig.~\ref{fig:quadratic_spline}) to produce the normalized conditional probability distribution of the gauge fields. Since the $U$(1) gauge fields in the electric field basis are integer, our flow is not the conventional flow in continuous space, but instead a discretized flow that produces a normalized distribution over the integer space.

The discretized RQS flow produces a (normalized) probability distribution by defining a flow from a prior distribution to an arbitrary continuous distribution before discretization. The general idea of the discretized flow is described as follows:
\begin{itemize}
    \item Start with a known prior probability distribution $\pi(z)$.
    \item Define an invertible function $x = f(z)$.
    \item Generate a continuous distribution over $x$ as $q(x) = \pi(f^{-1}(x))/f'(f^{-1}(x))$.
    \item Discretize the continuous distribution to obtain $p(E) = \int_{E-0.5}^{E+0.5} q(x) \dd x$.
\end{itemize}
Here, we choose the prior distribution to be the standard normal distribution, and notice that the conditional probability is in 1D, so we can work in the cumulative distribution and obtain $p(E)=\Pi(f^{-1}(E+0.5)) - \Pi(f^{-1}(E-0.5))$ efficiently and exactly, where $\Pi$ is the cumulative distribution of $\pi$.

This procedure defines a normalized discrete probability distribution over $\mathbb{Z}$, and by changing the transformation function $f$, we can in theory obtain any distribution. Here, we choose to parameterize $f$ using the RQS \cite{https://doi.org/10.48550/arxiv.1906.04032}. The RQS takes the outputs from the linear layer and use them to define a set of nodes $\{(z_i, x_i)\}$ that monotonically increases, and a set of derivatives ${\{d_i\}}$ for each node. Then, within the interval of node $i$ and node $i+1$, the function $f$ is defined as 
\begin{equation}
    f(z) = x_i + \frac{(x_{i+1} - x_i)[s \xi^2 + d_i \xi(1-\xi)]}{s + [d_{i+1} + d_i - 2s]\xi(1-\xi)},
\end{equation}
with $s = (x_{i+1}-x_i)/(z_{i+1} - z_i)$ and $\xi=(z - z_i)/(z_{i+1} - z_i)$. Outside the left most node or the right most node, the function is just a linear function with the derivative at the edge nodes.

The resulting probability distribution from the discretized RQS flow is then used as the conditional probability distribution for the gauge field at the current site $p_{\bm \theta_A}(E_{ij}^{(x)}|\bm f_{\le ij}, \bm E_{<ij}^{(x)}, \bm E_{<ij}^{(y)})$ and $p_{\bm \theta_A}(E_{ij}^{(y)}|\bm f_{\le ij}, \bm E_{\le ij}^{(x)}, \bm E_{<ij}^{(y)})$. 

To obtain the overal probability distribution for the whole configuration, we multiply all the conditional probability distribution together as
\begin{equation}
\begin{aligned}
    p_{\bm\theta_A}(\bm f, \bm E) = \prod_{ij} &p_{\bm\theta_A}(f_{ij}|\bm f_{<ij}, \bm E_{<ij}^{(x)}, \bm E_{<ij}^{(y)})\\
    &p_{\bm\theta_A}(E_{ij}^{(x)}|\bm f_{\le ij}, \bm E_{<ij}^{(x)}, \bm E_{<ij}^{(y)}) \\
    &p_{\bm\theta_A}(E_{ij}^{(y)}|\bm f_{\le ij}, \bm E_{\le ij}^{(x)}, \bm E_{<ij}^{(y)}).
\end{aligned}
\end{equation}
This procedure evaluates $p_{\bm\theta_A}$.    
The key approach to accomplish this exact sampling is to sequentially generate the conditional probability distributions and select the configuration for that site using this selection as part of the masked output for the next selections. For each step in the autoregressive procedure, the sampling can be done exactly from the discretized RQS. The details of the sampling is described in Appendix~\ref{app:sampling}.

\subsection{Phase construction with neural network backflow}
The phase of the wave function is produced by a neural network backflow Slater determinant \cite{Luo_2019} (Fig.~\ref{fig:set_up} (a, d)). Here the neural-network backflow generates configuration-dependent complex orbitals---i.e. the neural network takes as input the embedding ${\bm C}$ and outputs the $L_x \times L_y \times N_f$ array of single-particle orbitals $\chi_{\bm \theta_B}^{k}(i, j;{\bm C})$ for an $L_x$ by $L_y$ system with $N_f$ fermions, where $1 \leq k \leq N_f$, and $1 \leq i \leq L_x$, $1 \leq j \leq L_y$ are the position indices.  The neural network we use is a multi-layer CNN with parameters $\bm\theta_B$.

We then take the phase of the wave-function to be 

\begin{equation}
\label{Eq:SD}
\phi_{\bm\theta_B}(\bm f, \bm E) =  \arg
\mqty|
\chi_{\bm \theta_B}^{1}(i_1, j_1;{\bm C}) & \cdots& \chi_{\bm \theta_B}^{1}(i_{N_f}, j_{N_f};{\bm C}) \\ 
\vdots & \ddots & \vdots \\
\chi_{\bm \theta_B}^{N_f}(i_1, j_1;{\bm C}) & \cdots& \chi_{\bm \theta_B}^{N_f}(i_{N_f}, j_{N_f};{\bm C})|,
\end{equation}
where $(i_k, j_k)$ refers to the position of the $k$-th fermion.
Since we use complex numbers to form the Slater determinant, we are able to differentiate through the parameters.

\section{Variational Monte Carlo}\label{sec:vmc}
For the ground state optimization, we stochastically minimize the expectation of energy for a Hamiltonian $H$ and a normalized wave function $\ket{\psi_\theta}$ as
\begin{equation}
    \ev{\hat H}{\psi_\theta} \approx \frac{1}{N_s}\sum_{x \sim \abs{\psi_\theta}^2}^{N_s} \frac{\hat H\psi_\theta(x)}{\psi_\theta(x)} \equiv \frac{1}{N_s}\sum_{x \sim \abs{\psi_\theta}^2}^{N_s} E_\text{loc}(x),
\end{equation}
where $N_s$ is the batch size and the gradient is given by (see Ref.~\cite{https://doi.org/10.48550/arxiv.2101.07243} for derivation)
\begin{equation}
    \pdv{\theta}\ev{\hat H}{\psi_\theta} \approx \frac{2}{N_s}\sum_{x \sim \abs{\psi_\theta}^2}^{N_s} \real\left\{E_\text{loc}(x) \pdv{\theta} \log \psi_\theta^*(x)\right\}.
    \label{eq:variational_gradient}
\end{equation}
We further use variance control in the gradient formula by replacing $E_\text{loc}(x)$ with 
\begin{equation}
    E_\text{loc}'(x) = E_\text{loc}(x)-\frac{1}{N}\sum_{x \sim \abs{\psi_\theta}^2}^N E_\text{loc}(x).
\end{equation}
Compared to the conventional variational Monte Carlo approach, which often uses the Markov-chain Monte Carlo method, we can perform exact sampling to generate independent samples using the autoregressive models. We also apply transfer learning for optimization by first optimizing the small system and then using its neural network parameters as the initial parameters for larger system optimization. The details are provided in Appendix~\ref{app:hyper_opt}.

As a benchmark for testing our method, we first apply our approach to a pure gauge theory simulation, where $g_E=g$, $g_B=1/g$, $m=0$, $\kappa=0$ in Eq.~\ref{eq:ham}. We found that our method provides excellent agreement on different system sizes for both energy and string tension calculation.  The details of the simulation results are provided in the Appendix~\ref{app:pure}.

\section{Confinement and String Breaking}\label{sec:string}

\subsection{Zero density}
In this section, we show the charge confinement property of the Hamiltonian that manifests through the phenomena of string breaking. We can see string breaking by looking at the ground state of the system where we fix two static charges with opposite signs at two different locations. When the charges are close, an electric field line should connect the two charges, while when the charges are far apart, this electric field string breaks and a pair of mesons are formed.  This string-breaking behavior should be visible both in the density of field lines where we should explicitly see the string; as well as in the energy as a function of the distance between the static charges which should increase monotonically until the string breaks.

Consider the regime where $\kappa$ and $g_B$ are small; there is a competing energy between $H_E$ and $H_M$. Notice a length $L$ string costs energy $g_E^2 /2 \times L  + 2m$, while a pair of mesons costs energy $g_E^2 + 4m$. Therefore, a classical scaling analysis indicates string breaking happens if 

\begin{equation}
    g_E^2 + 4m < g_E^2 /2 \times L + 2m
\end{equation}
which implies $L \sim O(4m / g_E^2)$ \cite{Magnifico_2021}. This is consistent with the intuition that the lighter the mass is, the easier the string breaks to generate a pair of mesons to reduce energy. 

\begin{figure}[h]
    \centering
    \includegraphics[width=\linewidth]{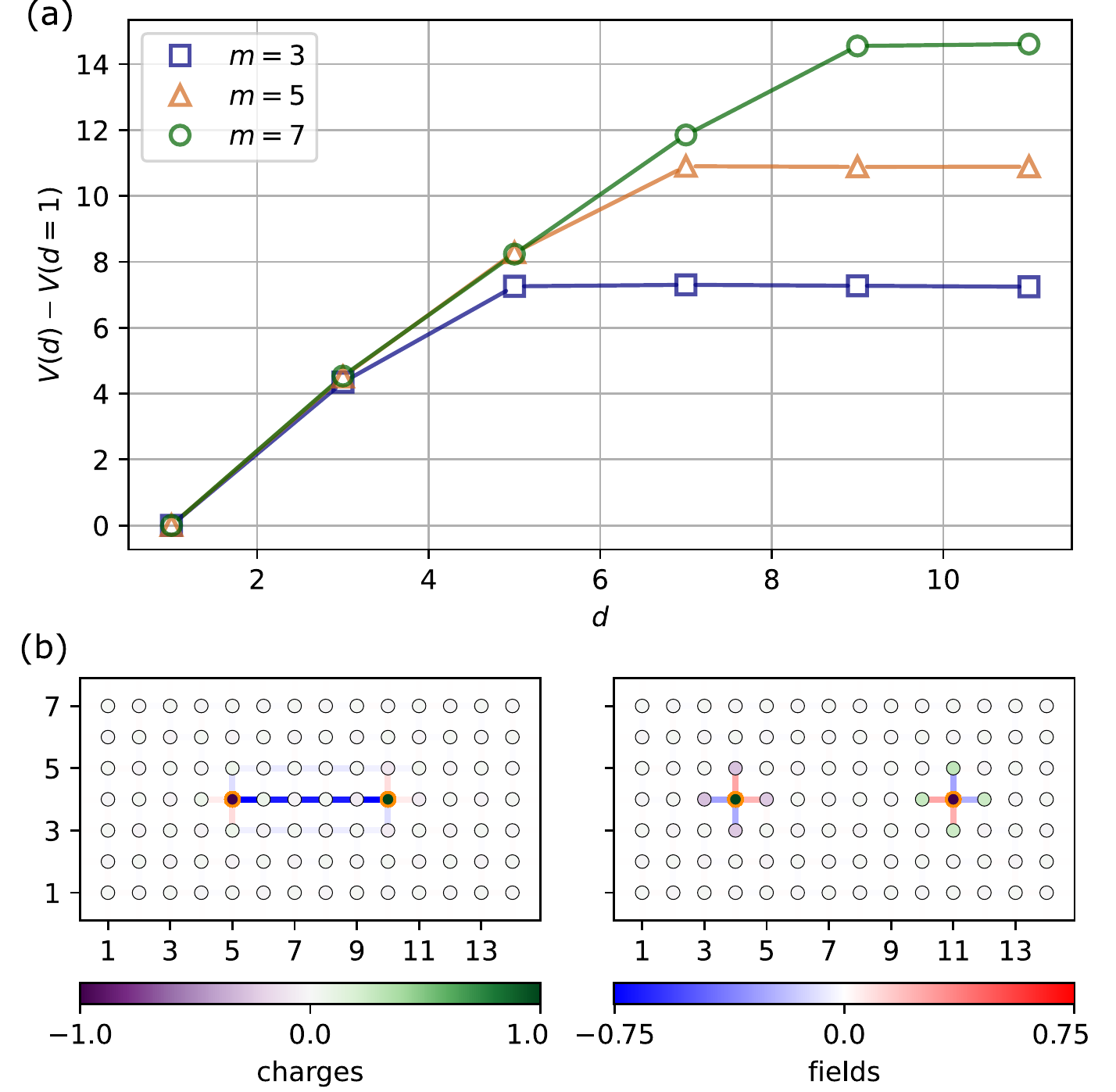}
    \caption{String breaking for ground state with static charges at zero density for a $14\times 7$ system. We put a pair of positive and negative static charges (surrounded with orange circles) at different distances. (a) Ground state energy as a function of distance between static charges $d$ for $g_E^2=4$. $g_B^2=2$ and $\kappa=1$. The energy shows a string breaking behavior, where the energy first increases linearly with the distance, then saturates to a value that depends on the mass $m$ of the fermion. (b) Electric field and charge density observables to illustrate the string breaking for $m=5$. 
    At a distance of $d=5$, the ground state has a string connecting the two static charges, whereas at a distance of $d=7$, the string breaks and new charges are created around the static charges to form mesons. The color scheme is the same as Fig.~\ref{fig:half_filling} (b) The neural network hyperparameters are listed in Appendix~\ref{app:hyper_opt}.
    }
    \label{fig:string_breaking}
\end{figure}

\begin{figure}[h]
    \centering
    \includegraphics[width=\linewidth]{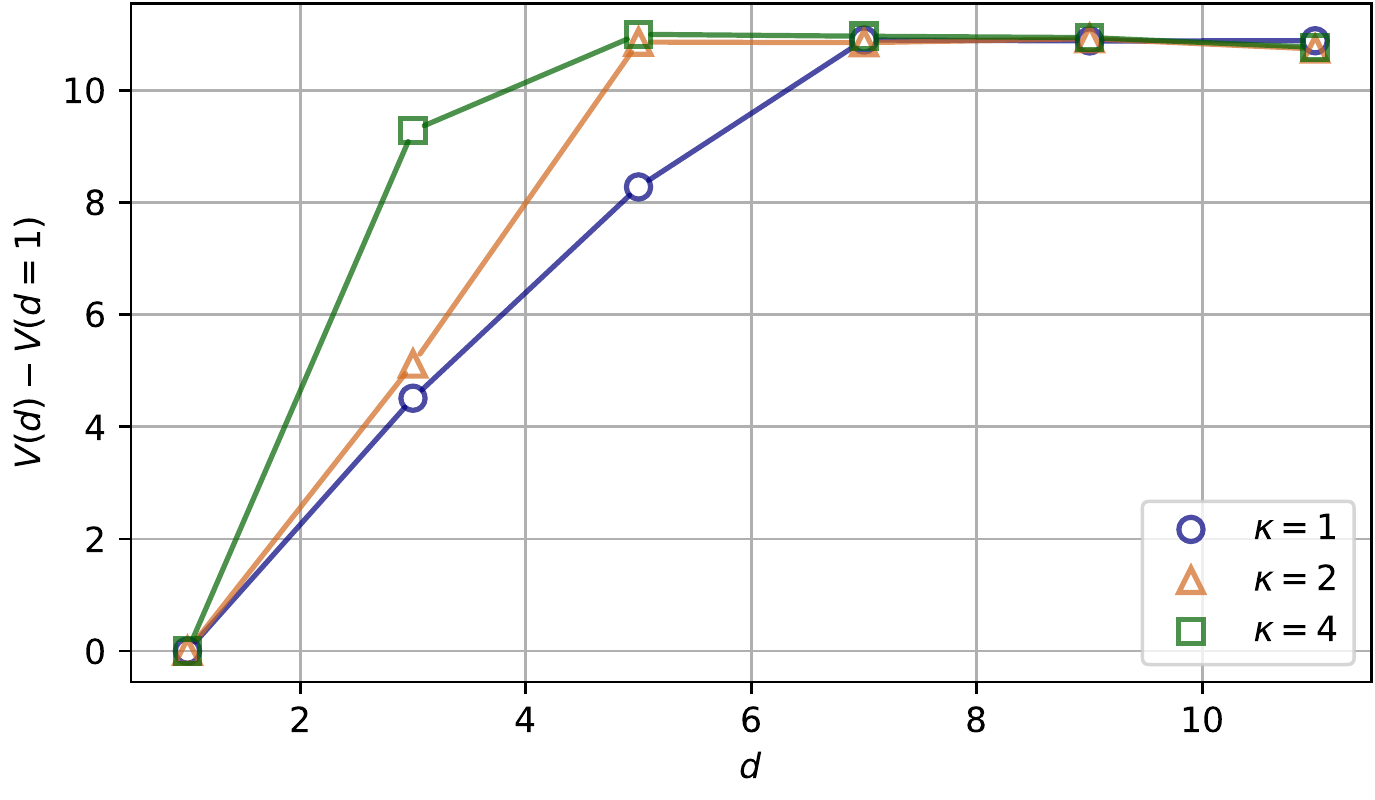}
    \caption{Ground state energy as a function of distance between static charges $d$ for $g_E^2=4$, $g_B^2=2$ and $m=5$ at zero density for a $14\times 7$ system. The energy shows a string breaking behavior, where the energy first increases linearly with the distance with a slope that depends on $\kappa$, then saturates to a value that depends on the mass $m$ of the fermion The neural network hyperparameters are listed in Appendix~\ref{app:hyper_opt}.
    }
    \label{fig:string_breaking_kappa}
\end{figure}

Here we present results for our simulations of Eq.~\ref{eq:ham} with both gauge and fermionic degrees of freedom. In Fig.~\ref{fig:string_breaking} (a) we show the energy as a function of distance between static charges for $m=3$, $m=5$ and $m=7$. We see that indeed the energy first increases largely linearly as the distance increases, corresponding to the electric field line being stretched. At large distance, the energy stays constant independent of distance. This corresponds to string breaking and meson formation. We also observe that the string breaking distance is longer for larger mass. This is expected since it is harder to form mesons for larger mass. In addition, we plot the electric field and charge density observables in Fig.~\ref{fig:string_breaking} (b) for $m=5$ and static charges at two distance before and after the string breaking. It is clear that when the distance is small, an electric field line is formed between the two charges, whereas when the distance is large, mesons are formed at each of the static charge.

We further study the quantum effect of fermion hopping on the string breaking phenomena. In Fig.~\ref{fig:string_breaking_kappa}, we vary $\kappa$ while fixing $g_E^2$, $g_B^2$ and $m$. We found that for larger $\kappa$, the energy increases faster and reaches the maximum value earlier as the charges are pulled apart. This means that, for larger $\kappa$, the string tension is higher before breaking and the string breaks at a shorter distance. We can understand the effect as follows. When the hopping coefficient $\kappa$ is large, the fermion hopping leads to larger gauge field fluctuations on the string, resulting in a high string tension. In addition, the fermion hopping also makes it easier to create charges, so mesons can form at a shorter distance and break the string earlier.

\subsection{Finite density}

\begin{figure}[h!]
    \centering
    \includegraphics[width=\linewidth]{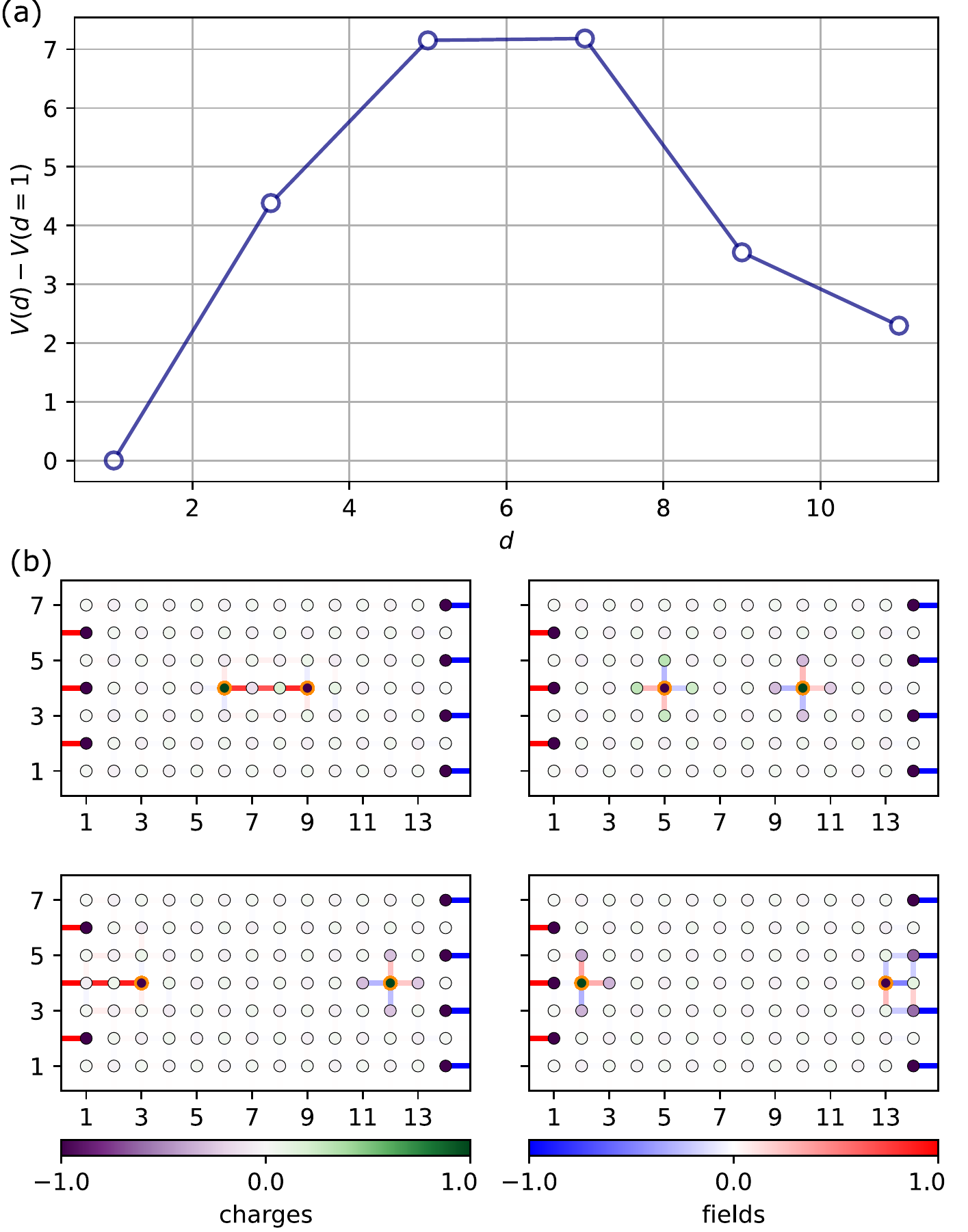}
    \caption{String breaking and boundary charge attachment for ground state with a total charge $Q=-7$ and $m=3$ and von Neumann boundary condition (fixed external fields) for a $14\times 7$ system. We put a pair of positive and negative static charges at different distances. (a) Ground state energy as a function of distance between static charges $d$ for $g_E^2=4$, $g_B^2=2$ and $\kappa=1$. (b) Electric field and charge density observables. The static charges are surrounded by orange circles. For small distance $d\le3$, the two charges are connected by a string, for intermediate distances $5\le d\le7$, the charges form mesons, and for large distances $d\ge 9$, one of the static charge forms a meson and the other charge creates a string that connects to the boundary. Here the color scheme is the same as Fig.~\ref{fig:half_filling} (b).
    }
    \label{fig:string_breaking_finite}
\end{figure}

In Fig.~\ref{fig:string_breaking_finite} (a) we show the energy as a function of distance between static charges for $m=3$ at finite density. In particular, we impose a total charge $Q=-7$ and impose a von Neumann boundary condition such that all the external field lines are on the short side of the system.  We see that, as in the case with zero density, the energy first increases as the distance increases. This corresponds to the electric field line being stretched and therefore increasing in energy. Then, for intermediate distance, the energy stays constant. This corresponds to string breaking and meson formation. Afterwards, we see the energy decreases again; this corresponds to the static charges forming a string with the boundary. 
In addition, we plot the electric field and charge density observables in Fig.~\ref{fig:string_breaking_finite} (b) for $m=3$ and static charges at four distances--- before breaking, after breaking, and two ways of connecting to boundary.

\section{Charge Crystal Phase to Vacuum Phase Transition}\label{sec:phase}

\subsection{Zero density}

\begin{figure}[h!]
    \centering
    \includegraphics[width=\linewidth]{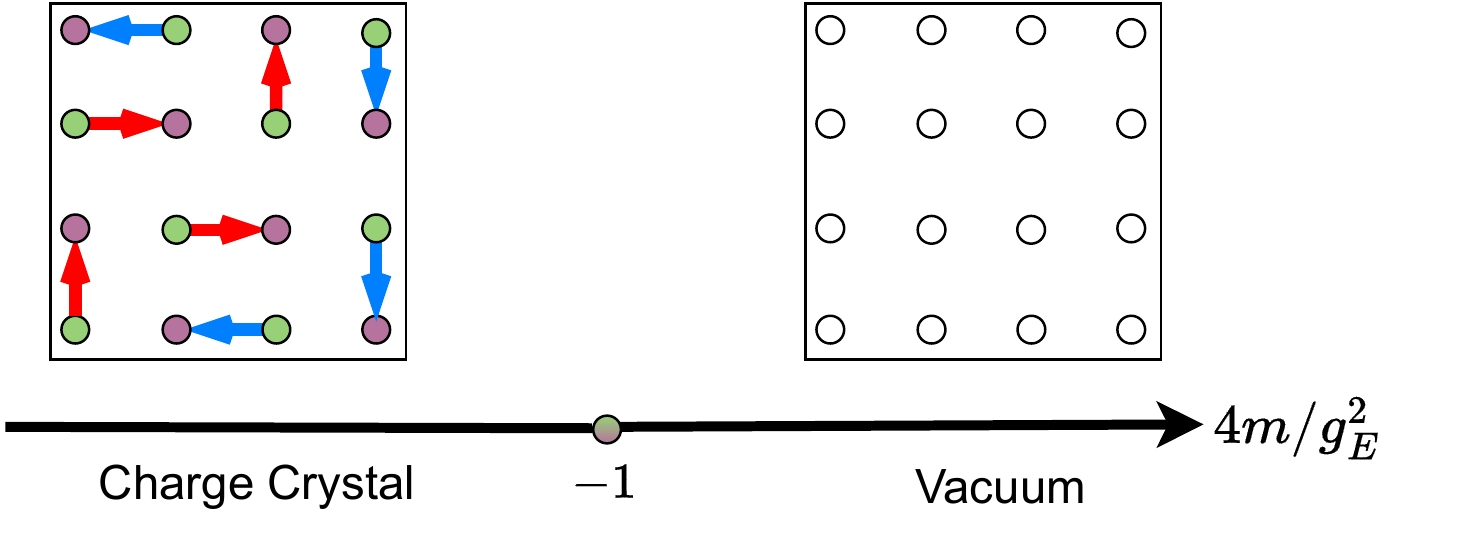}
    \caption{Illustration of the phase transition between charge crystal phase to vacuum phase in the semi-classical setting with $g_B^2=0$ and $\kappa=0$.
    }
    \label{fig:phase_transion_cartoon}
\end{figure}

The ground state of the Kogut--Susskind Hamiltonian undergoes a phase transition from the charge crystal phase to the vacuum phase. This can be understood classically as the competition between the electric term $\hat{H}_E$  and the mass term $\hat{H}_M$  of the Hamiltonian~\cite{PhysRevX.10.041040}.   Notice that when the fermions occupy an even site and leave the odd site empty (equivalently there is a pair of positive and negative charges on those sites), they have a gauge field energy of $g_E^2/2 + 2m$; alternatively when the fermions occupy an odd site leaving the even site empty (equivalently a vacuum), they 
contribute no additional energy.  Therefore, for large negative masses, it is energetically more favorable to create pairs of positive and negative charges, thereby producing charge crystals, but for positive mass, it is energetically more favorable for the system to stay in vacuum. The phase diagram (without $\hat{H}_B$ and $\hat{H}_K$) is illustrated in Fig.~\ref{fig:phase_transion_cartoon}.

\begin{figure}[h!]
    \centering
    \includegraphics[width=\linewidth]{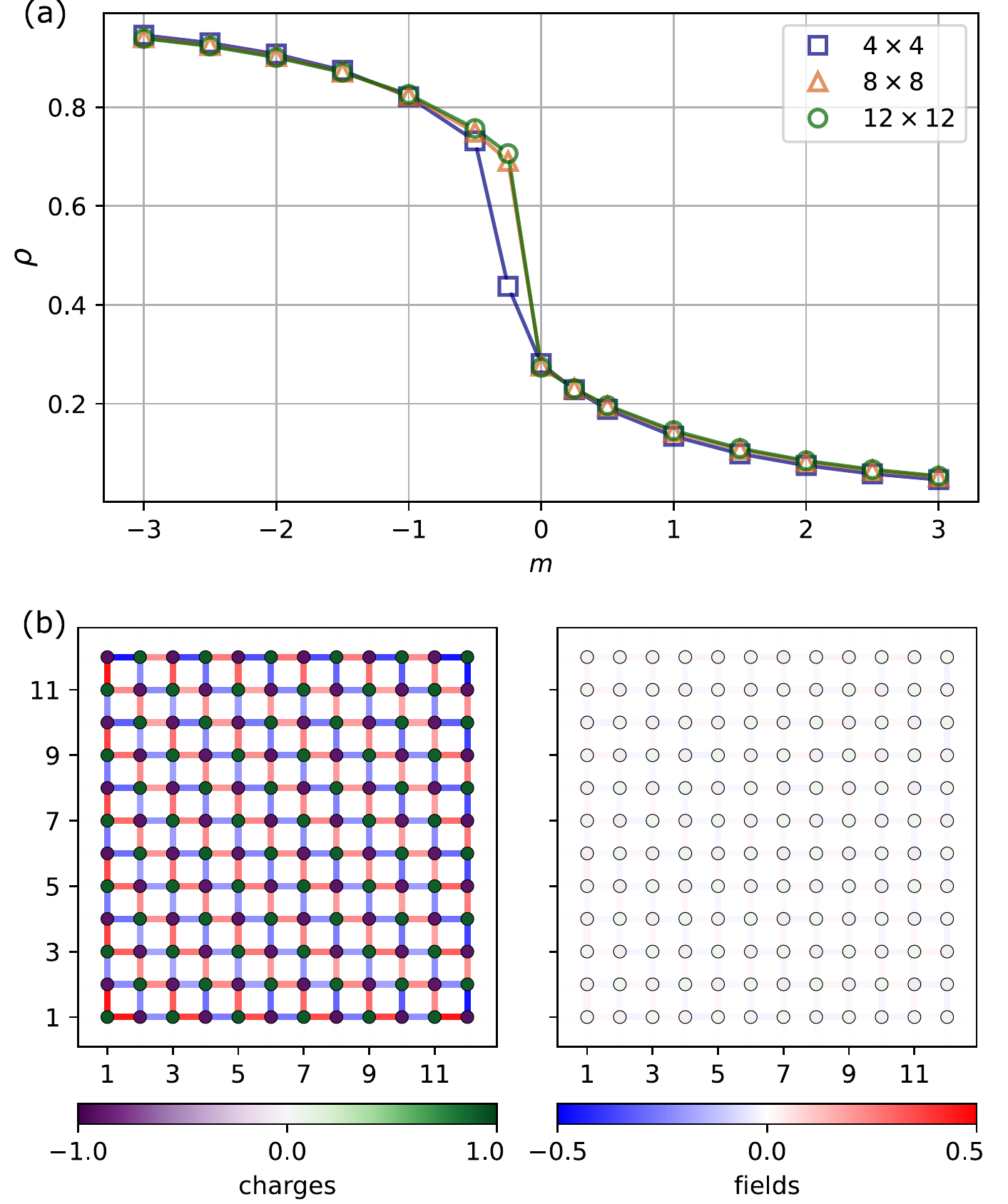}
    \caption{Phase transition at zero density with $g_E^2=2$, $g_B^2=4$ and $\kappa=1$. (a) Average particle density $\rho$ as a function of $m$ at zero density for different system sizes. We observe that a phase transition occurs from the charge crystal phase to the vacuum phase, and the transition gets sharper as the system size increases. (b) Electric field and charge density observable of the field for (left) $m=-2$ and (right) $m=3$ for $12\times 12$ system.  Here, purple represents negative charge while green represents positive charge. Red means the electric field points to the positive direction (right or upward) and blue means the electric field points to the negative direction (left or downward). It is clear that for $m=-2$, the system is in the charge crystal phase, where pairs of positive and negative charges are created and gauge fields are created to connect them, and for $m=3$, the system appears in the vacuum phase, with particle density and gauge field strength close to zero. The neural network hyperparameters are listed in Appendix~\ref{app:hyper_opt}.
    }
    \label{fig:half_filling}
\end{figure}

In Fig.~\ref{fig:half_filling} (a), we plot the average particle density $\rho\equiv1/N\sum_n \qty{\frac{1}{2}\qty[(-1)^n + 1] \ev{\hat \psi_n^\dagger \hat \psi_n} - \frac{1}{2}\qty[(-1)^n - 1] \ev{\hat\psi_n \hat\psi_n^\dagger}}$ which counts both positive and negative charges in zero density for different systems sizes. It can be seen that the phase transition happens between $m=-0.25$ and $m=0$. In Fig.~\ref{fig:half_filling} (b) we show the electric field and charge density observables for both the charge crystal phase and the vacuum phase. From the figure, it can be seen that in the charge crystal phase, positive charges (green) and negative charges (purple) are created with electric field lines connecting them due to Gauss's law. In the vacuum phase, however, everything stays at zero.

When the Hamiltonian only involves $\hat{H}_E$ and $\hat{H}_M$, the two terms commute with each other and the phase transition is first order as Fig.~\ref{fig:phase_transion_cartoon} shows. For the full Hamiltonian simulations with $\hat{H}_B$ and $\hat{H}_K$, the previous tensor network simulation in  3+1D QED~\cite{Magnifico_2021} with spin-1 representation, it is observed that the transition is second order. In our simulation (Fig.~\ref{fig:half_filling}), we see that continuous gauge degree of freedom may sharpen the order of phase transition, which could be weakly second order or first order. In Fig.~\ref{fig:cv_charge_density} in Appendix~\ref{app:cv}, we perform exact diagonalization on $2 \times 2$ lattice. We increase the cutoff for both the $\mathbb{Z}_N$ and the quantum link model, which both approach to the $U(1)$ theory at infinite cutoff, and observe the transition consistently gets sharper as cutoff increases.

\subsection{Finite density}
\begin{figure}[h!]
    \centering
    \includegraphics[width=\linewidth]{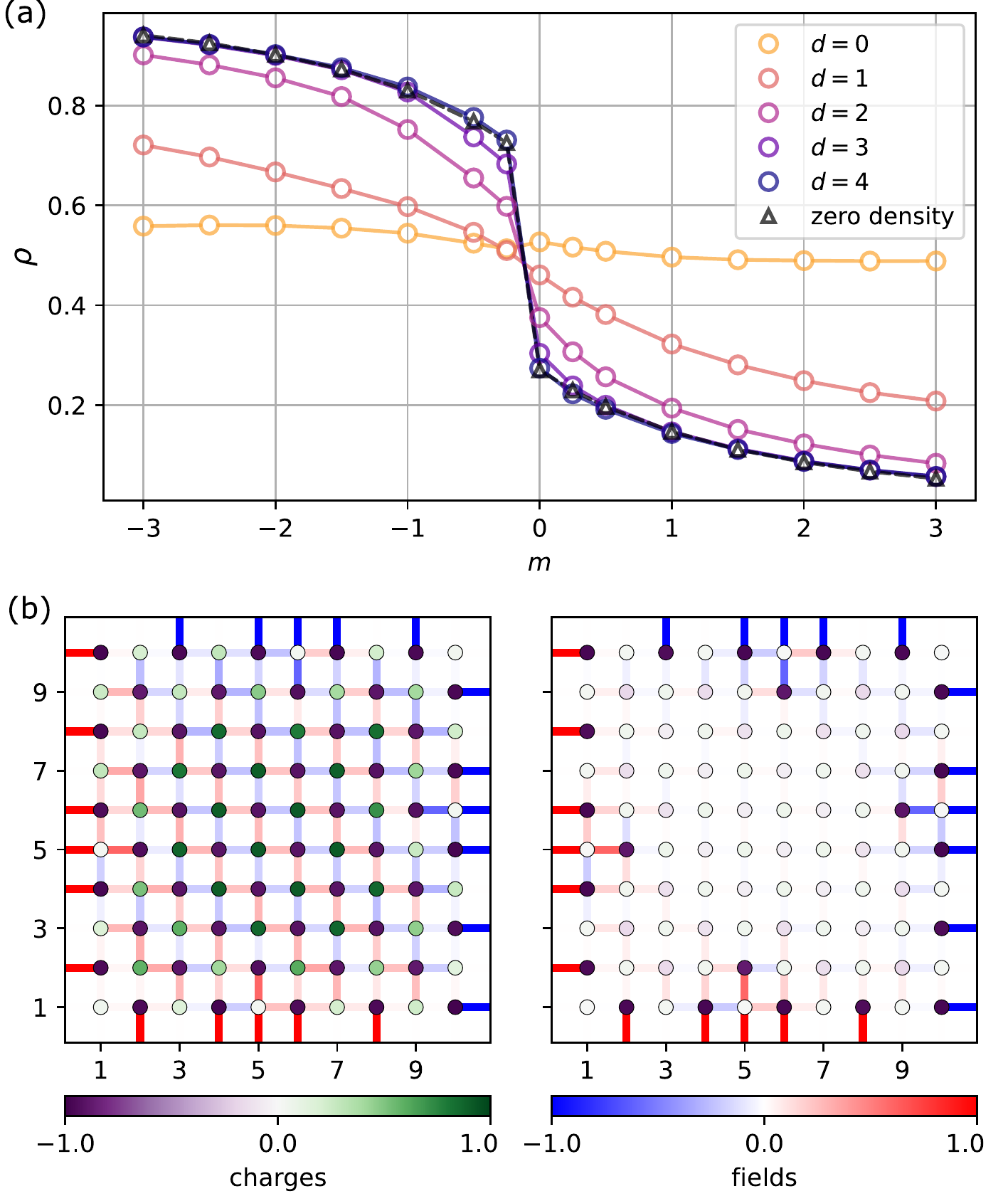}
    \caption{Finite density simulation with $g_E^2=2$, $g_B^2=4$ and $\kappa=1$. (a) Average surface particle density $\rho$ as a function of $m$ at difference distance from the boundary $d$ for $10\times 10$ finite density system with a total charge of $Q=-22$ and a von Neumann boundary condition (fixed external field lines). For the zero density curve, we average over the $6\times6$ bulk of a $12\times12$ system.  
    We see that the particle density approaches the zero density particle density as we go deeper into the bulk. 
    (b) Electric field and charge density observable of the field for (left) $m=-2$ and (right) $m=3$.
    The bulk of the systems also shares similar phenomena as the zero density case in Fig.~\ref{fig:half_filling}.
    The boundary of the systems remains the same regardless of the mass. 
    Here the color scheme is the same as Fig.~\ref{fig:half_filling} (b). 
    The neural network hyperparameters are listed in Appendix~\ref{app:hyper_opt}.
    }
    \label{fig:finite_density}
\end{figure}

In this section, we consider the Kogut-Susskind Hamiltonian at finite density.  We show the existence of a phase separation between the bulk and boundary in the vacuum phase; and demonstrate the net charge in the charge crystal phase fails to penetrate into the bulk potentially also leading to phase separation.
Here, we work with a $10\times10$ system and choose a net charge of $-22$. 
In Fig.~\ref{fig:finite_density} (a), we plot the surface particle density at different distances from the boundary. We find that the surface particle density is almost constant independent of $m$ at the surface, but approaches the zero density particle density deep in the bulk. Similar to zero density, we also plot the electric field and charge density observables in Fig.~\ref{fig:finite_density} (b). From the figure, we observe that the bulk undergoes a possible phase transition from the charge crystal phase to the vacuum phase, while the surface remains intact. This indicates a possible phase separation where the density deep in the bulk of the system recovers the particle density at zero density and the extra charges reside on the surface of the system.  

However, a previous study~\cite{PhysRevX.10.041040} on the quantum link model with spin-1 gauge field truncation and $g_B^2=0$ suggests a different result. It is found that all the extra charges are pushed to the boundary in the vacuum phase, while charges formed in the charge crystal phase are delocalized and free to move inside the system, resulting in a phase separation in the vaccuum phase and no phase separation in the charge crystal phase. Since Ref.~\onlinecite{PhysRevX.10.041040} only studies $g_B^2=0$ while our data in Fig.~\ref{fig:finite_density} shows $g_B^2=4$, we believe the magnetic interaction plays an important role on the possible phase separation in the charge crystal phase.

\begin{figure}[h!]
    \centering
    \includegraphics[width=\linewidth]{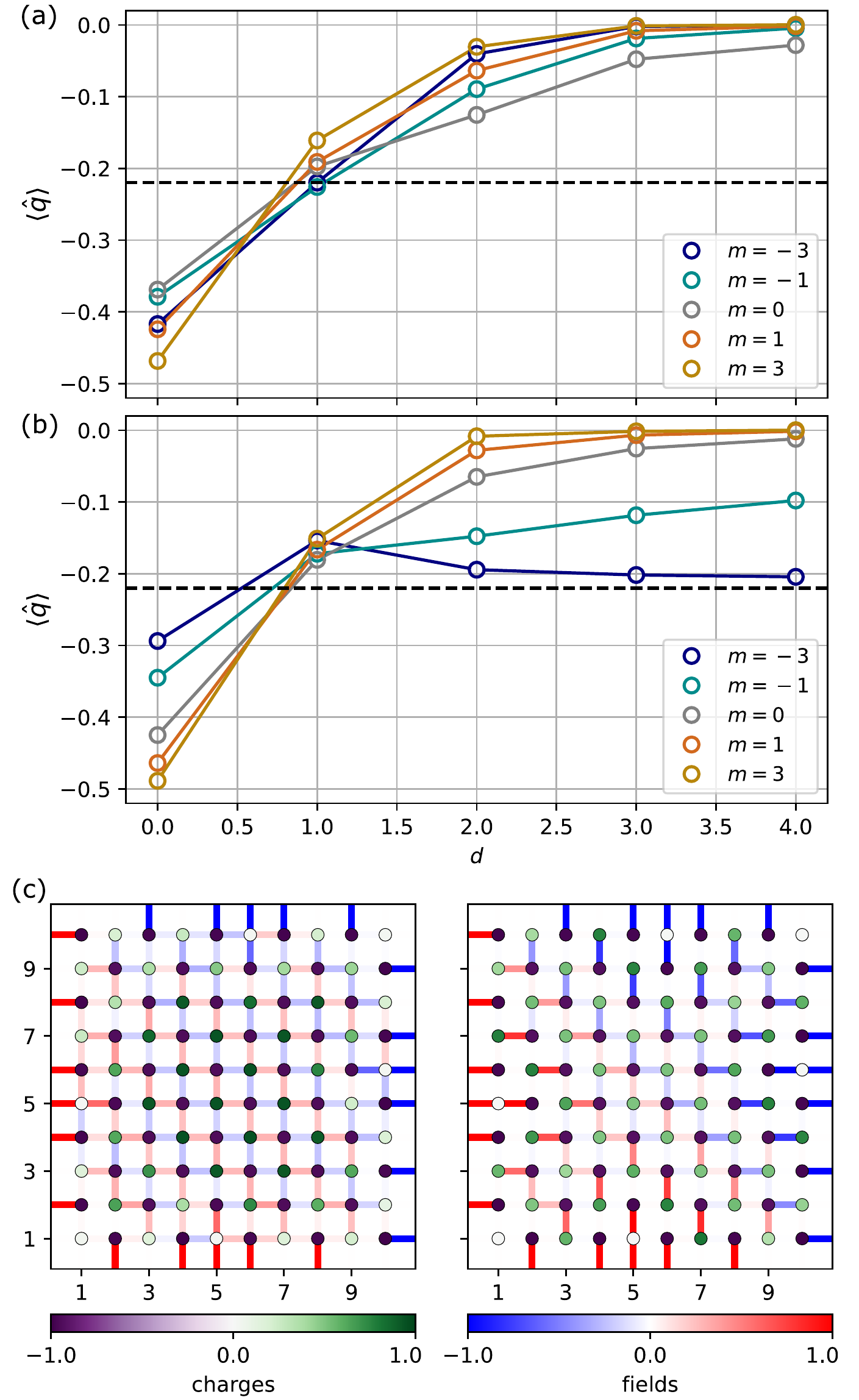}
    \caption{Net charge penetration blocking at finite density with $g_E^2=2$ and $\kappa=1$ at $10\times 10$ finite density system with a total charge of $Q=-22$ and von Neumann boundary condition. (a, b) Average surface charge density $\ev{\hat q}$ as a function of distance from the boundary $d$ with (a) $g_B^2=4$ and (b) $g_B^2=0$ for different $m$ . The black dashed line represents the overall average charge density of $-0.22$. We observe that with large $g_B^2$, the charge density decays to zero as we go into the bulk, while for $g_B^2=0$, the charge density penetrates into the bulk for negative $m$. (c) Electric field and charge density observables of the field for (both) $m=-3$, (left) $g_B^2=4$ and (right) $g_B^2=0$. The positive charges (green) cluster in the bulk when $g_B^2$ is large, resulting in a neutral bulk, while they distribute more evenly when $g_B^2=0$, resulting in a net negative charge throughout the system. Here the color scheme is the same as Fig.~\ref{fig:half_filling} (b). The neural network hyperparameters are listed in Appendix~\ref{app:hyper_opt}.
    }
    \label{fig:charge_penetration}
\end{figure}

Therefore, we further investigate the net charge penetration behavior for different strengths of magnetic interactions. We plot the average surface charge density $\ev{\hat q} \equiv 1/N \sum_n \ev{\hat q_n}$ for two sets of parameters---$g_E^2=2$, $g_B^2=4$, $\kappa=-1$ (Fig~\ref{fig:charge_penetration} (a)) and $g_E^2=2$, $g_B^2=0$, $\kappa=-1$ (Fig~\ref{fig:charge_penetration} (b)). It can be seen that, when $g_B^2$ is large, the surface charge density decays to zero in the bulk regardless of $m$. When $g_B^2=0$, however, the surface charge density decays to zero in the bulk when $m\ge 0$, while it remains finite in the bulk when $m < 0$. This indicates that the net charge density is able to penetrate into the bulk in the charge crystal phase when there is no magnetic interaction, but the net charge penetration is blocked under large magnetic interaction. We also plot the electric field and charge density observables for $m=-3$ with the two different $g_B^2$'s at each site in Fig~\ref{fig:charge_penetration} (c). We observe that, for large $g_B^2$, the extra negative charges (purple) reside on the boundary while the positive charges (green) are pushed into the bulk, creating a neutral net charge in the bulk, whereas for $g_B^2=0$, the positive charges are more uniformly distributed inside the system, and thus the net negative charge penetrates throughout the system.

Our results in Fig.~\ref{fig:charge_penetration} show similar phenomena to Ref.~\onlinecite{PhysRevX.10.041040} when the magnetic term is absent finding a phase separation in the vacuum phase, but no phase separation in the charge crystal phase. However, for large $g_B^2$, the vacuum phase separation remains, but new phenomenon arises in the charge crystal phase. In this case, the net charge penetration is blocked and a possible phase separation appears. This phenomenon could be related to the effect that, to reduce the magnetic energy, it is necessary to create superpositions of the gauge field. It is preferable to have the positive charges in the bulk instead of on the boundary
as the larger number of links in the bulk support a higher degree of quantum fluctuation.

\section{Magnetic Phase Transition with Dynamical Fermions}\label{sec:topo_phase}

In this section, we investigate the effect of the magnetic term $\hat H_B$ when dynamical fermions exist. 
We consider the competition between the Fermion kinetic energy $\hat H_K$ and the magnetic energy $\hat H_B$, which could be tuned by the coupling parameter ratio $\kappa / g_B^2$. We fix $g_E^2=0.01$, $g_B^2=1$ and $m=0.01$ and vary $\kappa$ in the simulations. We apply our neural network to simulate large systems up to $8 \times 8$ and measure the following two observables to study the physics: the average plaquette value 
\begin{equation}
\begin{aligned}
\langle\cos \phi^P\rangle &\equiv \frac{1}{L^2}\sum_{i, j}\langle \cos \phi_{i,j}^P\rangle
= \frac{1}{2L^2}\sum_{i, j}\langle \hat P_{i,j}+ \hat P_{i,j}^\dagger\rangle.
\end{aligned}
\end{equation}
and the average nearest neighbor plaquette correlation
\begin{equation}
\begin{aligned}
&\mathcal{C}(\phi^P)\\
&\equiv\frac{1}{2L(L-1)}\sum_{\ev{i_1, j_1; i_2, j_2}}\langle \sin \phi_{i_1, j_1}^P \sin \phi_{i_2, j_2}^P \rangle  \\
&= -\frac{1}{8L(L-1)}\sum_{\ev{i_1, j_1; i_2, j_2}}\langle (\hat P_{i_1, j_1}- \hat P_{i_1, j_1}^\dagger)(\hat P_{i_2, j_2}-\hat P_{i_2, j_2}^\dagger) \rangle,
\end{aligned}
\end{equation} where $\ev{i_1, j_1; i_2, j_2}$ refers to nearest-neighbor sites, and $\phi^P_{i,j}$ is the $U(1)$ variable that corresponds to the magnetic flux and the eigenvalues $e^{i \phi^P_{i,j}}$ of the plaquette operator $\hat P_{i, j}$.

\begin{figure}[h!]
    \centering
    \includegraphics[width=\linewidth]{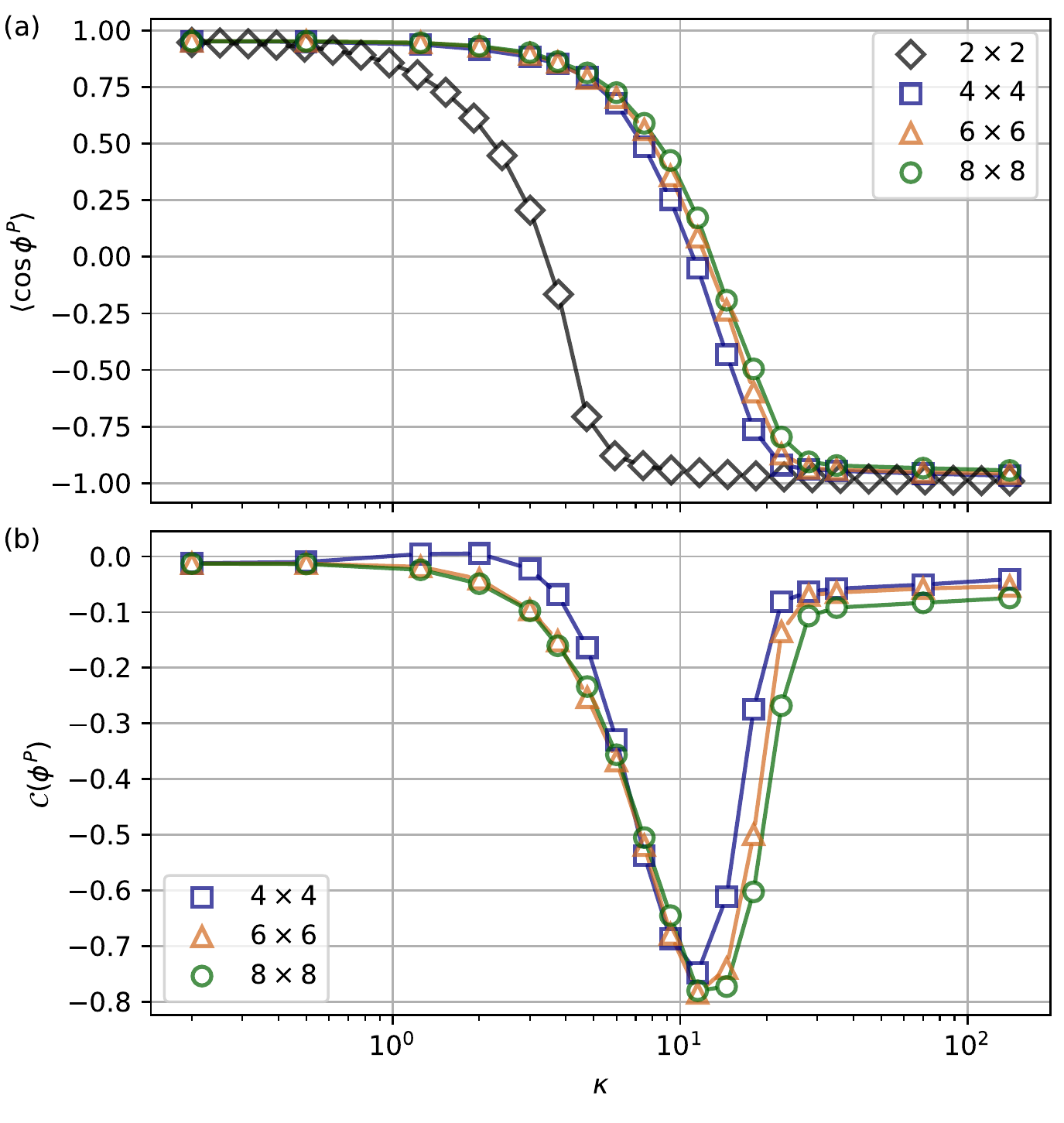}
    \caption{Magnetic effect with dynamical matter at zero density for $g_E^2=0.01$, $g_B^2=1$ and $m=0.01$ of different system sizes with open boundary condition. (a) Average plaquette value $\ev{\cos \phi^P}$ as a function of $\kappa$. The average plaquette values drops from $1$ to $-1$ as $\kappa$ increases. The $2\times2$ (one plaquette) result is calculated using exact diagonalization and all other results are from our neural network algorithm. (b) Average nearest neighbor plaquette correlation $\mathcal{C}(\phi^P)$ as a function of $\kappa$. As $\kappa$ increases, the average nearest neighbor plaquette correlation first decreases from 0 to negative values; then it goes back up again.  The neural network hyperparameters are listed in Appendix~\ref{app:hyper_opt}.
    }
    \label{fig:half_filling_topo}
\end{figure}

In Fig.~\ref{fig:half_filling_topo}, we plot (a) the average plaquette value and (b) the average plaquette correlation for different system sizes with open boundary condition in zero density. We observe that as the fermion kinetic energy coupling $\kappa$ increases, the average plaquette expectation goes from $1$ to $-1$, which indicates the magentic flux changes from 0 to $\pi$. $\mathcal{C}(\phi^P)$ measures the plaquette correlation, where zero indicates that the magnetic flux is $0$ or $\pi$ while negative values indicates the magnetic flux forms a staggered pattern. It is shown that for both small and large $\kappa$ the correlation is close to 0 corresponding to the magnetic flux being at 0 or $\pi$; 
at intermediate $\kappa$ a staggered flux at angles between $0$ and $\pi$ appears.

To further understand the underlying mechanism, we consider the following model without the interaction from the electric field $\hat H_E.$
\begin{equation}\label{eq:topo_ham}
    \hat{H} = \hat H_B + \hat H_M + \hat H_K,
\end{equation}
Notice that $\hat H_B$, $\hat H_M$ and $\hat H_K$ commute with each other, so that the Hamiltonian can be simultaneously diagonalized, and analytically solved under periodic boundary conditions. Therefore, we can write the energy of the Hamiltonian as $E_\text{magnetic} + E_\text{fermion}$, where
\begin{equation}
    \begin{aligned}
    E_{\text{magnetic}} &= -g_B^2 L^2 \cos \phi, \\
    E_{\text{fermion}} &= \sum_{i=0}^{L/2}\sum_{j=0}^L E^{-}\left(\frac{2 \pi i}{ L}, \frac{2 \pi j}{ L}, \phi, \phi_x, \phi_y\right),
    \end{aligned}
\end{equation}
with 
\begin{align}
\begin{split}
    E^{\pm}&(k_x, k_y, \phi, \phi_x, \phi_y) = \pm2 \kappa \\
    &\Bigg[\cos^2 \left(k_x-\frac{\phi_x}{2}\right)+
    \cos^2 \left(k_y-\frac{\phi_y}{2}\right)\\
    &+2\cos \frac{\phi}{2}\cos \left(k_x-\frac{\phi_x}{2}\right) \cos \left(k_y-\frac{\phi_y}{2}\right)+\frac{m^2}{4\kappa^2}\Bigg]^{1/2}.
\end{split}
\end{align}

Here $\phi \equiv \phi_{1, 1}^P \in (-\pi,\pi]$ is the $U(1)$ variable related to the flux over a particular plaquette and $\phi_x$ and $\phi_y$ are variables related to the links (see Appendix~\ref{app:analytic} for details). While this solution does not take into account the effect of Gauss's law, we can construct the correct solution by projecting it into the correct Gauss's law sector, which preserves both the energy and plaquette values (see Appendix~\ref{app:analytic}).

From the analytical construction, it turns out there is a competition between $E_\text{magnet}$ and $E_\text{fermion}$. When $\kappa \ll g_B^2$, the magnetic term dominates, and the system prefers $\phi=0$. When $\kappa \gg g_B^2$, the fermion term dominates, and it leads to $\phi=\pi$ . In the intermediate regime, $\phi\ne0,\pi$. Due to the staggered fermions and because the total flux must be 0, the system permits a staggered flux with adjacent plaquettes having fluxes with opposite signs, that breaks time-reverse symmetry. Therefore, the system admits three phases---the zero-flux phase, the stagger-flux phase, and the $\pi$-flux phase as illustrated in Fig.~\ref{fig:mag_tran} (see Appendix~\ref{app:analytic} for more details). 
With non-zero electric field interaction $\hat H_E$, we believe the three phases may still exist, because they differ in their symmetries and $\hat H_E$ term does not break the time reverse symmetry of the system.
The data from the neural network wave-functions at non-zero $g_E^2$ still suggests three separate phases (possibly with a higher order transition) although from the numerical results it's not possible to rule out the possibility that the $g_E^2=0$ transition turns into a crossover.  
In Appendix~\ref{app:magge}, we show that as we decrease $g_E^2$, the neural network result approaches the analytical calculation with $g_E^2=0$, giving evidence that the neural network results are of high quality.

\begin{figure}[h!]
    \centering
    \includegraphics[width=\linewidth]{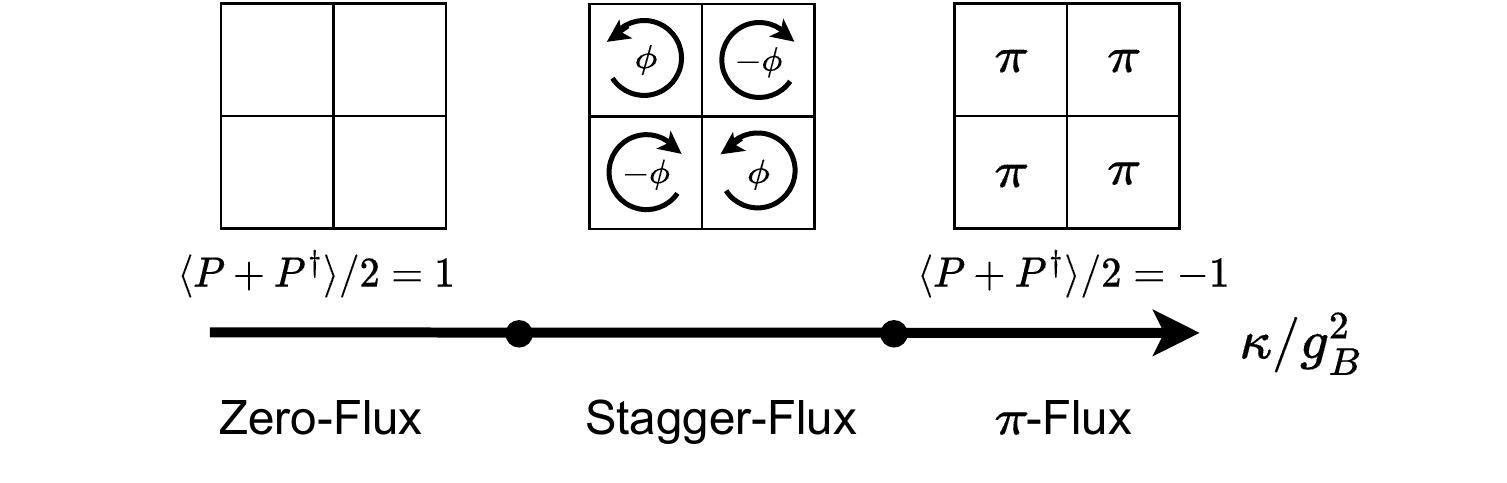}
    \caption{Illustration of the magnetic phase transition with $g_E^2=0$. Notice that the stagger-flux phase has a degeneracy with the exchange $\phi \rightarrow -\phi$.
    }
    \label{fig:mag_tran}
\end{figure}

We note that recently Ref.~\onlinecite{PRXQuantum.2.030334} has considered 
dynamical matter and magnetic fields for a single 
plaquette with a spin-$1$ representation on the gauge field in the context of quantum computation. 
They find that $\langle \cos \phi^P \rangle$ on one plaquette jumps sharply as a function of $\kappa/g_B^2$; this might indicate that if a phase transition exists it would be first order. 
However, our data shows that $\langle \cos \phi^P \rangle$ varies continuously indicating that if the phase transition exists, it is likely to be second or higher order. We believe this discrepancy in the order of the apparent phase transition may depend on the truncation of the gauge field. In Appendix~\ref{app:cv}, we further study this effect by measuring how the continuity of the observable depends on the level of truncating the gauge field.

\section{Conclusions}\label{sec:con}

In this work, we have demonstrated a novel and efficient neural network, Gauge-Fermion FlowNet, for simulating 2+1D lattice compact QED with finite density dynamical fermions. To our knowledge, it is the first construction of a variational quantum state that simultaneously encodes $U$(1) gauge degrees of freedom without cutoff and dynamical fermionic degrees of freedom while fulfilling the gauge symmetries. Our approach is free of a sign problem and works in the finite density regime. 

We apply our Gauge-Fermion FlowNet to simulate the string breaking phenomena related to confinement with different fermion density and hopping amplitudes. We study the phase transition from the charge crystal phase to the vacuum phase at zero density, as well as the phase separation and the net charge penetration blocking effect at finite density. In addition, we use both the neural network and analytical method to investigate the lesser known magnetic phase transition, demonstrating the ability to discover new physics using our approach. With the recent interests in simulating 2+1D lattice QED on quantum computers~\cite{haase2021resource,osborne2022large,PRXQuantum.2.030334,zohar2013simulating,PhysRevLett.127.130504}, our results also provide insights for exploring new phenomena with quantum devices.

Our approach opens up new opportunities for simulating gauge theories coupled to dynamical matter. One natural extension is to apply the method to higher dimensions like 3+1D QED and study different phases and real-time dynamics of the models. Another interesting direction is to further study different abelian theories such as the Abelian-Higgs model and QED with multiple fermion species. It will also be important to generalize the approach to nonabelian gauge theories and investigate QCD physics.

\section{Acknowledgements}
The authors are grateful for insightful suggestions from Yizhuang You, and acknowledge helpful discussion with Phiala Shanahan, Michael DeMarco, Lena Funcke, James Stokes, Ho Tat Lam, Hersh Singh, Hart Goldman, and Ruben Verresen. ZC acknowledges DARPA 134371-5113608 award. ZC and DL acknowledge support from the NSF AI Institute for Artificial Intelligence and Fundamental Interactions (IAIFI). BKC acknowledge support from the Department of Energy grant DOE DESC0020165. This material is based upon work supported by the U.S. Department of Energy, Office of Science, National Quantum Information Science Research Centers, Co-design Center for Quantum Advantage (C2QA) under contract number DE-SC0012704.

\bibliography{reference,reference_ml}

\begin{thebibliography}{71}%
\makeatletter
\providecommand \@ifxundefined [1]{%
 \@ifx{#1\undefined}
}%
\providecommand \@ifnum [1]{%
 \ifnum #1\expandafter \@firstoftwo
 \else \expandafter \@secondoftwo
 \fi
}%
\providecommand \@ifx [1]{%
 \ifx #1\expandafter \@firstoftwo
 \else \expandafter \@secondoftwo
 \fi
}%
\providecommand \natexlab [1]{#1}%
\providecommand \enquote  [1]{``#1''}%
\providecommand \bibnamefont  [1]{#1}%
\providecommand \bibfnamefont [1]{#1}%
\providecommand \citenamefont [1]{#1}%
\providecommand \href@noop [0]{\@secondoftwo}%
\providecommand \href [0]{\begingroup \@sanitize@url \@href}%
\providecommand \@href[1]{\@@startlink{#1}\@@href}%
\providecommand \@@href[1]{\endgroup#1\@@endlink}%
\providecommand \@sanitize@url [0]{\catcode `\\12\catcode `\$12\catcode
  `\&12\catcode `\#12\catcode `\^12\catcode `\_12\catcode `\%12\relax}%
\providecommand \@@startlink[1]{}%
\providecommand \@@endlink[0]{}%
\providecommand \url  [0]{\begingroup\@sanitize@url \@url }%
\providecommand \@url [1]{\endgroup\@href {#1}{\urlprefix }}%
\providecommand \urlprefix  [0]{URL }%
\providecommand \Eprint [0]{\href }%
\providecommand \doibase [0]{https://doi.org/}%
\providecommand \selectlanguage [0]{\@gobble}%
\providecommand \bibinfo  [0]{\@secondoftwo}%
\providecommand \bibfield  [0]{\@secondoftwo}%
\providecommand \translation [1]{[#1]}%
\providecommand \BibitemOpen [0]{}%
\providecommand \bibitemStop [0]{}%
\providecommand \bibitemNoStop [0]{.\EOS\space}%
\providecommand \EOS [0]{\spacefactor3000\relax}%
\providecommand \BibitemShut  [1]{\csname bibitem#1\endcsname}%
\let\auto@bib@innerbib\@empty
\bibitem [{\citenamefont {Kitaev}(2003)}]{Kitaev_2003}%
  \BibitemOpen
  \bibfield  {author} {\bibinfo {author} {\bibfnamefont {A.}~\bibnamefont
  {Kitaev}},\ }\bibfield  {title} {\bibinfo {title} {Fault-tolerant quantum
  computation by anyons},\ }\href
  {https://doi.org/10.1016/s0003-4916(02)00018-0} {\bibfield  {journal}
  {\bibinfo  {journal} {Annals of Physics}\ }\textbf {\bibinfo {volume}
  {303}},\ \bibinfo {pages} {2–30} (\bibinfo {year} {2003})}\BibitemShut
  {NoStop}%
\bibitem [{\citenamefont {Hamma}\ \emph {et~al.}(2005)\citenamefont {Hamma},
  \citenamefont {Zanardi},\ and\ \citenamefont {Wen}}]{Hamma_2005}%
  \BibitemOpen
  \bibfield  {author} {\bibinfo {author} {\bibfnamefont {A.}~\bibnamefont
  {Hamma}}, \bibinfo {author} {\bibfnamefont {P.}~\bibnamefont {Zanardi}},\
  and\ \bibinfo {author} {\bibfnamefont {X.-G.}\ \bibnamefont {Wen}},\
  }\bibfield  {title} {\bibinfo {title} {String and membrane condensation on
  three-dimensional lattices},\ }\bibfield  {journal} {\bibinfo  {journal}
  {Physical Review B}\ }\textbf {\bibinfo {volume} {72}},\ \href
  {https://doi.org/10.1103/physrevb.72.035307} {10.1103/physrevb.72.035307}
  (\bibinfo {year} {2005})\BibitemShut {NoStop}%
\bibitem [{\citenamefont {Vijay}\ \emph {et~al.}(2016)\citenamefont {Vijay},
  \citenamefont {Haah},\ and\ \citenamefont {Fu}}]{Vijay_2016}%
  \BibitemOpen
  \bibfield  {author} {\bibinfo {author} {\bibfnamefont {S.}~\bibnamefont
  {Vijay}}, \bibinfo {author} {\bibfnamefont {J.}~\bibnamefont {Haah}},\ and\
  \bibinfo {author} {\bibfnamefont {L.}~\bibnamefont {Fu}},\ }\bibfield
  {title} {\bibinfo {title} {Fracton topological order, generalized lattice
  gauge theory, and duality},\ }\bibfield  {journal} {\bibinfo  {journal}
  {Physical Review B}\ }\textbf {\bibinfo {volume} {94}},\ \href
  {https://doi.org/10.1103/physrevb.94.235157} {10.1103/physrevb.94.235157}
  (\bibinfo {year} {2016})\BibitemShut {NoStop}%
\bibitem [{\citenamefont {Rebbi}(1983)}]{rebbi_1983}%
  \BibitemOpen
  \bibfield  {author} {\bibinfo {author} {\bibfnamefont {C.}~\bibnamefont
  {Rebbi}},\ }\href@noop {} {\emph {\bibinfo {title} {Lattice gauge theories
  and Monte Carlo simulations}}}\ (\bibinfo  {publisher} {World Scientific},\
  \bibinfo {year} {1983})\BibitemShut {NoStop}%
\bibitem [{\citenamefont {Xu}\ \emph {et~al.}(2019)\citenamefont {Xu},
  \citenamefont {Qi}, \citenamefont {Zhang}, \citenamefont {Assaad},
  \citenamefont {Xu},\ and\ \citenamefont {Meng}}]{Xu_2019}%
  \BibitemOpen
  \bibfield  {author} {\bibinfo {author} {\bibfnamefont {X.~Y.}\ \bibnamefont
  {Xu}}, \bibinfo {author} {\bibfnamefont {Y.}~\bibnamefont {Qi}}, \bibinfo
  {author} {\bibfnamefont {L.}~\bibnamefont {Zhang}}, \bibinfo {author}
  {\bibfnamefont {F.~F.}\ \bibnamefont {Assaad}}, \bibinfo {author}
  {\bibfnamefont {C.}~\bibnamefont {Xu}},\ and\ \bibinfo {author}
  {\bibfnamefont {Z.~Y.}\ \bibnamefont {Meng}},\ }\bibfield  {title} {\bibinfo
  {title} {Monte~carlo study of lattice compact quantum electrodynamics with
  fermionic matter: The parent state of quantum phases},\ }\bibfield  {journal}
  {\bibinfo  {journal} {Physical Review X}\ }\textbf {\bibinfo {volume} {9}},\
  \href {https://doi.org/10.1103/physrevx.9.021022} {10.1103/physrevx.9.021022}
  (\bibinfo {year} {2019})\BibitemShut {NoStop}%
\bibitem [{\citenamefont {Bañuls}\ \emph {et~al.}(2020)\citenamefont
  {Bañuls}, \citenamefont {Blatt}, \citenamefont {Catani}, \citenamefont
  {Celi}, \citenamefont {Cirac}, \citenamefont {Dalmonte}, \citenamefont
  {Fallani}, \citenamefont {Jansen}, \citenamefont {Lewenstein}, \citenamefont
  {Montangero},\ and\ \citenamefont {et~al.}}]{Ba_uls_2020}%
  \BibitemOpen
  \bibfield  {author} {\bibinfo {author} {\bibfnamefont {M.~C.}\ \bibnamefont
  {Bañuls}}, \bibinfo {author} {\bibfnamefont {R.}~\bibnamefont {Blatt}},
  \bibinfo {author} {\bibfnamefont {J.}~\bibnamefont {Catani}}, \bibinfo
  {author} {\bibfnamefont {A.}~\bibnamefont {Celi}}, \bibinfo {author}
  {\bibfnamefont {J.~I.}\ \bibnamefont {Cirac}}, \bibinfo {author}
  {\bibfnamefont {M.}~\bibnamefont {Dalmonte}}, \bibinfo {author}
  {\bibfnamefont {L.}~\bibnamefont {Fallani}}, \bibinfo {author} {\bibfnamefont
  {K.}~\bibnamefont {Jansen}}, \bibinfo {author} {\bibfnamefont
  {M.}~\bibnamefont {Lewenstein}}, \bibinfo {author} {\bibfnamefont
  {S.}~\bibnamefont {Montangero}},\ and\ \bibinfo {author} {\bibnamefont
  {et~al.}},\ }\bibfield  {title} {\bibinfo {title} {Simulating lattice gauge
  theories within quantum technologies},\ }\bibfield  {journal} {\bibinfo
  {journal} {The European Physical Journal D}\ }\textbf {\bibinfo {volume}
  {74}},\ \href {https://doi.org/10.1140/epjd/e2020-100571-8}
  {10.1140/epjd/e2020-100571-8} (\bibinfo {year} {2020})\BibitemShut {NoStop}%
\bibitem [{\citenamefont {Emonts}\ \emph {et~al.}(2020)\citenamefont {Emonts},
  \citenamefont {Ba{\~n}uls}, \citenamefont {Cirac},\ and\ \citenamefont
  {Zohar}}]{emonts2020variational}%
  \BibitemOpen
  \bibfield  {author} {\bibinfo {author} {\bibfnamefont {P.}~\bibnamefont
  {Emonts}}, \bibinfo {author} {\bibfnamefont {M.~C.}\ \bibnamefont
  {Ba{\~n}uls}}, \bibinfo {author} {\bibfnamefont {I.}~\bibnamefont {Cirac}},\
  and\ \bibinfo {author} {\bibfnamefont {E.}~\bibnamefont {Zohar}},\ }\bibfield
   {title} {\bibinfo {title} {Variational monte carlo simulation with tensor
  networks of a pure z 3 gauge theory in (2+ 1) d},\ }\href@noop {} {\bibfield
  {journal} {\bibinfo  {journal} {Physical Review D}\ }\textbf {\bibinfo
  {volume} {102}},\ \bibinfo {pages} {074501} (\bibinfo {year}
  {2020})}\BibitemShut {NoStop}%
\bibitem [{\citenamefont {Hashizume}\ \emph {et~al.}(2022)\citenamefont
  {Hashizume}, \citenamefont {Halimeh}, \citenamefont {Hauke},\ and\
  \citenamefont {Banerjee}}]{qlm_2dphase}%
  \BibitemOpen
  \bibfield  {author} {\bibinfo {author} {\bibfnamefont {T.}~\bibnamefont
  {Hashizume}}, \bibinfo {author} {\bibfnamefont {J.~C.}\ \bibnamefont
  {Halimeh}}, \bibinfo {author} {\bibfnamefont {P.}~\bibnamefont {Hauke}},\
  and\ \bibinfo {author} {\bibfnamefont {D.}~\bibnamefont {Banerjee}},\
  }\bibfield  {title} {\bibinfo {title} {{Ground-state phase diagram of quantum
  link electrodynamics in $(2+1)$-d}},\ }\href
  {https://doi.org/10.21468/SciPostPhys.13.2.017} {\bibfield  {journal}
  {\bibinfo  {journal} {SciPost Phys.}\ }\textbf {\bibinfo {volume} {13}},\
  \bibinfo {pages} {017} (\bibinfo {year} {2022})}\BibitemShut {NoStop}%
\bibitem [{\citenamefont {Emonts}\ \emph {et~al.}(2022)\citenamefont {Emonts},
  \citenamefont {Kelman}, \citenamefont {Borla}, \citenamefont {Moroz},
  \citenamefont {Gazit},\ and\ \citenamefont {Zohar}}]{emonts2022finding}%
  \BibitemOpen
  \bibfield  {author} {\bibinfo {author} {\bibfnamefont {P.}~\bibnamefont
  {Emonts}}, \bibinfo {author} {\bibfnamefont {A.}~\bibnamefont {Kelman}},
  \bibinfo {author} {\bibfnamefont {U.}~\bibnamefont {Borla}}, \bibinfo
  {author} {\bibfnamefont {S.}~\bibnamefont {Moroz}}, \bibinfo {author}
  {\bibfnamefont {S.}~\bibnamefont {Gazit}},\ and\ \bibinfo {author}
  {\bibfnamefont {E.}~\bibnamefont {Zohar}},\ }\bibfield  {title} {\bibinfo
  {title} {Finding the ground state of a lattice gauge theory with fermionic
  tensor networks: a 2+ 1d $\mathbb{Z}_2$ demonstration},\ }\href@noop {}
  {\bibfield  {journal} {\bibinfo  {journal} {arXiv preprint arXiv:2211.00023}\
  } (\bibinfo {year} {2022})}\BibitemShut {NoStop}%
\bibitem [{\citenamefont {Zohar}\ and\ \citenamefont
  {Cirac}(2018)}]{zohar2018combining}%
  \BibitemOpen
  \bibfield  {author} {\bibinfo {author} {\bibfnamefont {E.}~\bibnamefont
  {Zohar}}\ and\ \bibinfo {author} {\bibfnamefont {J.~I.}\ \bibnamefont
  {Cirac}},\ }\bibfield  {title} {\bibinfo {title} {Combining tensor networks
  with monte carlo methods for lattice gauge theories},\ }\href@noop {}
  {\bibfield  {journal} {\bibinfo  {journal} {Physical Review D}\ }\textbf
  {\bibinfo {volume} {97}},\ \bibinfo {pages} {034510} (\bibinfo {year}
  {2018})}\BibitemShut {NoStop}%
\bibitem [{\citenamefont {Felser}\ \emph {et~al.}(2020)\citenamefont {Felser},
  \citenamefont {Silvi}, \citenamefont {Collura},\ and\ \citenamefont
  {Montangero}}]{PhysRevX.10.041040}%
  \BibitemOpen
  \bibfield  {author} {\bibinfo {author} {\bibfnamefont {T.}~\bibnamefont
  {Felser}}, \bibinfo {author} {\bibfnamefont {P.}~\bibnamefont {Silvi}},
  \bibinfo {author} {\bibfnamefont {M.}~\bibnamefont {Collura}},\ and\ \bibinfo
  {author} {\bibfnamefont {S.}~\bibnamefont {Montangero}},\ }\bibfield  {title}
  {\bibinfo {title} {Two-dimensional quantum-link lattice quantum
  electrodynamics at finite density},\ }\href
  {https://doi.org/10.1103/PhysRevX.10.041040} {\bibfield  {journal} {\bibinfo
  {journal} {Phys. Rev. X}\ }\textbf {\bibinfo {volume} {10}},\ \bibinfo
  {pages} {041040} (\bibinfo {year} {2020})}\BibitemShut {NoStop}%
\bibitem [{\citenamefont {Magnifico}\ \emph {et~al.}(2021)\citenamefont
  {Magnifico}, \citenamefont {Felser}, \citenamefont {Silvi},\ and\
  \citenamefont {Montangero}}]{Magnifico_2021}%
  \BibitemOpen
  \bibfield  {author} {\bibinfo {author} {\bibfnamefont {G.}~\bibnamefont
  {Magnifico}}, \bibinfo {author} {\bibfnamefont {T.}~\bibnamefont {Felser}},
  \bibinfo {author} {\bibfnamefont {P.}~\bibnamefont {Silvi}},\ and\ \bibinfo
  {author} {\bibfnamefont {S.}~\bibnamefont {Montangero}},\ }\bibfield  {title}
  {\bibinfo {title} {Lattice quantum electrodynamics in (3+1)-dimensions at
  finite density with tensor networks},\ }\bibfield  {journal} {\bibinfo
  {journal} {Nature Communications}\ }\textbf {\bibinfo {volume} {12}},\ \href
  {https://doi.org/10.1038/s41467-021-23646-3} {10.1038/s41467-021-23646-3}
  (\bibinfo {year} {2021})\BibitemShut {NoStop}%
\bibitem [{\citenamefont {Mazzola}\ \emph {et~al.}(2021)\citenamefont
  {Mazzola}, \citenamefont {Mathis}, \citenamefont {Mazzola},\ and\
  \citenamefont {Tavernelli}}]{mazzola2021gauge}%
  \BibitemOpen
  \bibfield  {author} {\bibinfo {author} {\bibfnamefont {G.}~\bibnamefont
  {Mazzola}}, \bibinfo {author} {\bibfnamefont {S.~V.}\ \bibnamefont {Mathis}},
  \bibinfo {author} {\bibfnamefont {G.}~\bibnamefont {Mazzola}},\ and\ \bibinfo
  {author} {\bibfnamefont {I.}~\bibnamefont {Tavernelli}},\ }\bibfield  {title}
  {\bibinfo {title} {Gauge-invariant quantum circuits for u (1) and yang-mills
  lattice gauge theories},\ }\href@noop {} {\bibfield  {journal} {\bibinfo
  {journal} {Physical Review Research}\ }\textbf {\bibinfo {volume} {3}},\
  \bibinfo {pages} {043209} (\bibinfo {year} {2021})}\BibitemShut {NoStop}%
\bibitem [{\citenamefont {Haase}\ \emph {et~al.}(2021)\citenamefont {Haase},
  \citenamefont {Dellantonio}, \citenamefont {Celi}, \citenamefont {Paulson},
  \citenamefont {Kan}, \citenamefont {Jansen},\ and\ \citenamefont
  {Muschik}}]{haase2021resource}%
  \BibitemOpen
  \bibfield  {author} {\bibinfo {author} {\bibfnamefont {J.~F.}\ \bibnamefont
  {Haase}}, \bibinfo {author} {\bibfnamefont {L.}~\bibnamefont {Dellantonio}},
  \bibinfo {author} {\bibfnamefont {A.}~\bibnamefont {Celi}}, \bibinfo {author}
  {\bibfnamefont {D.}~\bibnamefont {Paulson}}, \bibinfo {author} {\bibfnamefont
  {A.}~\bibnamefont {Kan}}, \bibinfo {author} {\bibfnamefont {K.}~\bibnamefont
  {Jansen}},\ and\ \bibinfo {author} {\bibfnamefont {C.~A.}\ \bibnamefont
  {Muschik}},\ }\bibfield  {title} {\bibinfo {title} {A resource efficient
  approach for quantum and classical simulations of gauge theories in particle
  physics},\ }\href@noop {} {\bibfield  {journal} {\bibinfo  {journal}
  {Quantum}\ }\textbf {\bibinfo {volume} {5}},\ \bibinfo {pages} {393}
  (\bibinfo {year} {2021})}\BibitemShut {NoStop}%
\bibitem [{\citenamefont {Paulson}\ \emph {et~al.}(2021)\citenamefont
  {Paulson}, \citenamefont {Dellantonio}, \citenamefont {Haase}, \citenamefont
  {Celi}, \citenamefont {Kan}, \citenamefont {Jena}, \citenamefont {Kokail},
  \citenamefont {van Bijnen}, \citenamefont {Jansen}, \citenamefont {Zoller},\
  and\ \citenamefont {Muschik}}]{PRXQuantum.2.030334}%
  \BibitemOpen
  \bibfield  {author} {\bibinfo {author} {\bibfnamefont {D.}~\bibnamefont
  {Paulson}}, \bibinfo {author} {\bibfnamefont {L.}~\bibnamefont
  {Dellantonio}}, \bibinfo {author} {\bibfnamefont {J.~F.}\ \bibnamefont
  {Haase}}, \bibinfo {author} {\bibfnamefont {A.}~\bibnamefont {Celi}},
  \bibinfo {author} {\bibfnamefont {A.}~\bibnamefont {Kan}}, \bibinfo {author}
  {\bibfnamefont {A.}~\bibnamefont {Jena}}, \bibinfo {author} {\bibfnamefont
  {C.}~\bibnamefont {Kokail}}, \bibinfo {author} {\bibfnamefont
  {R.}~\bibnamefont {van Bijnen}}, \bibinfo {author} {\bibfnamefont
  {K.}~\bibnamefont {Jansen}}, \bibinfo {author} {\bibfnamefont
  {P.}~\bibnamefont {Zoller}},\ and\ \bibinfo {author} {\bibfnamefont {C.~A.}\
  \bibnamefont {Muschik}},\ }\bibfield  {title} {\bibinfo {title} {Simulating
  2d effects in lattice gauge theories on a quantum computer},\ }\href
  {https://doi.org/10.1103/PRXQuantum.2.030334} {\bibfield  {journal} {\bibinfo
   {journal} {PRX Quantum}\ }\textbf {\bibinfo {volume} {2}},\ \bibinfo {pages}
  {030334} (\bibinfo {year} {2021})}\BibitemShut {NoStop}%
\bibitem [{\citenamefont {Nguyen}\ \emph {et~al.}(2022)\citenamefont {Nguyen},
  \citenamefont {Tran}, \citenamefont {Zhu}, \citenamefont {Green},
  \citenamefont {Alderete}, \citenamefont {Davoudi},\ and\ \citenamefont
  {Linke}}]{PRXQuantum.3.020324}%
  \BibitemOpen
  \bibfield  {author} {\bibinfo {author} {\bibfnamefont {N.~H.}\ \bibnamefont
  {Nguyen}}, \bibinfo {author} {\bibfnamefont {M.~C.}\ \bibnamefont {Tran}},
  \bibinfo {author} {\bibfnamefont {Y.}~\bibnamefont {Zhu}}, \bibinfo {author}
  {\bibfnamefont {A.~M.}\ \bibnamefont {Green}}, \bibinfo {author}
  {\bibfnamefont {C.~H.}\ \bibnamefont {Alderete}}, \bibinfo {author}
  {\bibfnamefont {Z.}~\bibnamefont {Davoudi}},\ and\ \bibinfo {author}
  {\bibfnamefont {N.~M.}\ \bibnamefont {Linke}},\ }\bibfield  {title} {\bibinfo
  {title} {Digital quantum simulation of the schwinger model and symmetry
  protection with trapped ions},\ }\href
  {https://doi.org/10.1103/PRXQuantum.3.020324} {\bibfield  {journal} {\bibinfo
   {journal} {PRX Quantum}\ }\textbf {\bibinfo {volume} {3}},\ \bibinfo {pages}
  {020324} (\bibinfo {year} {2022})}\BibitemShut {NoStop}%
\bibitem [{\citenamefont {Luo}\ \emph {et~al.}(2020)\citenamefont {Luo},
  \citenamefont {Shen}, \citenamefont {Highman}, \citenamefont {Clark},
  \citenamefont {DeMarco}, \citenamefont {El-Khadra},\ and\ \citenamefont
  {Gadway}}]{luo2020framework}%
  \BibitemOpen
  \bibfield  {author} {\bibinfo {author} {\bibfnamefont {D.}~\bibnamefont
  {Luo}}, \bibinfo {author} {\bibfnamefont {J.}~\bibnamefont {Shen}}, \bibinfo
  {author} {\bibfnamefont {M.}~\bibnamefont {Highman}}, \bibinfo {author}
  {\bibfnamefont {B.~K.}\ \bibnamefont {Clark}}, \bibinfo {author}
  {\bibfnamefont {B.}~\bibnamefont {DeMarco}}, \bibinfo {author} {\bibfnamefont
  {A.~X.}\ \bibnamefont {El-Khadra}},\ and\ \bibinfo {author} {\bibfnamefont
  {B.}~\bibnamefont {Gadway}},\ }\bibfield  {title} {\bibinfo {title}
  {Framework for simulating gauge theories with dipolar spin systems},\
  }\href@noop {} {\bibfield  {journal} {\bibinfo  {journal} {Physical Review
  A}\ }\textbf {\bibinfo {volume} {102}},\ \bibinfo {pages} {032617} (\bibinfo
  {year} {2020})}\BibitemShut {NoStop}%
\bibitem [{\citenamefont {Zhou}\ \emph {et~al.}(2022)\citenamefont {Zhou},
  \citenamefont {Su}, \citenamefont {Halimeh}, \citenamefont {Ott},
  \citenamefont {Sun}, \citenamefont {Hauke}, \citenamefont {Yang},
  \citenamefont {Yuan}, \citenamefont {Berges},\ and\ \citenamefont
  {Pan}}]{zhou2022thermalization}%
  \BibitemOpen
  \bibfield  {author} {\bibinfo {author} {\bibfnamefont {Z.-Y.}\ \bibnamefont
  {Zhou}}, \bibinfo {author} {\bibfnamefont {G.-X.}\ \bibnamefont {Su}},
  \bibinfo {author} {\bibfnamefont {J.~C.}\ \bibnamefont {Halimeh}}, \bibinfo
  {author} {\bibfnamefont {R.}~\bibnamefont {Ott}}, \bibinfo {author}
  {\bibfnamefont {H.}~\bibnamefont {Sun}}, \bibinfo {author} {\bibfnamefont
  {P.}~\bibnamefont {Hauke}}, \bibinfo {author} {\bibfnamefont
  {B.}~\bibnamefont {Yang}}, \bibinfo {author} {\bibfnamefont {Z.-S.}\
  \bibnamefont {Yuan}}, \bibinfo {author} {\bibfnamefont {J.}~\bibnamefont
  {Berges}},\ and\ \bibinfo {author} {\bibfnamefont {J.-W.}\ \bibnamefont
  {Pan}},\ }\bibfield  {title} {\bibinfo {title} {Thermalization dynamics of a
  gauge theory on a quantum simulator},\ }\href@noop {} {\bibfield  {journal}
  {\bibinfo  {journal} {Science}\ }\textbf {\bibinfo {volume} {377}},\ \bibinfo
  {pages} {311} (\bibinfo {year} {2022})}\BibitemShut {NoStop}%
\bibitem [{\citenamefont {Yang}\ \emph {et~al.}(2020)\citenamefont {Yang},
  \citenamefont {Sun}, \citenamefont {Ott}, \citenamefont {Wang}, \citenamefont
  {Zache}, \citenamefont {Halimeh}, \citenamefont {Yuan}, \citenamefont
  {Hauke},\ and\ \citenamefont {Pan}}]{yang2020observation}%
  \BibitemOpen
  \bibfield  {author} {\bibinfo {author} {\bibfnamefont {B.}~\bibnamefont
  {Yang}}, \bibinfo {author} {\bibfnamefont {H.}~\bibnamefont {Sun}}, \bibinfo
  {author} {\bibfnamefont {R.}~\bibnamefont {Ott}}, \bibinfo {author}
  {\bibfnamefont {H.-Y.}\ \bibnamefont {Wang}}, \bibinfo {author}
  {\bibfnamefont {T.~V.}\ \bibnamefont {Zache}}, \bibinfo {author}
  {\bibfnamefont {J.~C.}\ \bibnamefont {Halimeh}}, \bibinfo {author}
  {\bibfnamefont {Z.-S.}\ \bibnamefont {Yuan}}, \bibinfo {author}
  {\bibfnamefont {P.}~\bibnamefont {Hauke}},\ and\ \bibinfo {author}
  {\bibfnamefont {J.-W.}\ \bibnamefont {Pan}},\ }\bibfield  {title} {\bibinfo
  {title} {Observation of gauge invariance in a 71-site bose--hubbard quantum
  simulator},\ }\href@noop {} {\bibfield  {journal} {\bibinfo  {journal}
  {Nature}\ }\textbf {\bibinfo {volume} {587}},\ \bibinfo {pages} {392}
  (\bibinfo {year} {2020})}\BibitemShut {NoStop}%
\bibitem [{\citenamefont {Osborne}\ \emph {et~al.}(2022)\citenamefont
  {Osborne}, \citenamefont {McCulloch}, \citenamefont {Yang}, \citenamefont
  {Hauke},\ and\ \citenamefont {Halimeh}}]{osborne2022large}%
  \BibitemOpen
  \bibfield  {author} {\bibinfo {author} {\bibfnamefont {J.}~\bibnamefont
  {Osborne}}, \bibinfo {author} {\bibfnamefont {I.~P.}\ \bibnamefont
  {McCulloch}}, \bibinfo {author} {\bibfnamefont {B.}~\bibnamefont {Yang}},
  \bibinfo {author} {\bibfnamefont {P.}~\bibnamefont {Hauke}},\ and\ \bibinfo
  {author} {\bibfnamefont {J.~C.}\ \bibnamefont {Halimeh}},\ }\bibfield
  {title} {\bibinfo {title} {Large-scale $2+ 1$ d u(1) gauge theory with
  dynamical matter in a cold-atom quantum simulator},\ }\href@noop {}
  {\bibfield  {journal} {\bibinfo  {journal} {arXiv preprint arXiv:2211.01380}\
  } (\bibinfo {year} {2022})}\BibitemShut {NoStop}%
\bibitem [{\citenamefont {Kan}\ \emph {et~al.}(2021)\citenamefont {Kan},
  \citenamefont {Funcke}, \citenamefont {K{\"u}hn}, \citenamefont
  {Dellantonio}, \citenamefont {Zhang}, \citenamefont {Haase}, \citenamefont
  {Muschik},\ and\ \citenamefont {Jansen}}]{kan2021investigating}%
  \BibitemOpen
  \bibfield  {author} {\bibinfo {author} {\bibfnamefont {A.}~\bibnamefont
  {Kan}}, \bibinfo {author} {\bibfnamefont {L.}~\bibnamefont {Funcke}},
  \bibinfo {author} {\bibfnamefont {S.}~\bibnamefont {K{\"u}hn}}, \bibinfo
  {author} {\bibfnamefont {L.}~\bibnamefont {Dellantonio}}, \bibinfo {author}
  {\bibfnamefont {J.}~\bibnamefont {Zhang}}, \bibinfo {author} {\bibfnamefont
  {J.~F.}\ \bibnamefont {Haase}}, \bibinfo {author} {\bibfnamefont {C.~A.}\
  \bibnamefont {Muschik}},\ and\ \bibinfo {author} {\bibfnamefont
  {K.}~\bibnamefont {Jansen}},\ }\bibfield  {title} {\bibinfo {title}
  {Investigating a (3+ 1) d topological $\theta$-term in the hamiltonian
  formulation of lattice gauge theories for quantum and classical
  simulations},\ }\href@noop {} {\bibfield  {journal} {\bibinfo  {journal}
  {Physical Review D}\ }\textbf {\bibinfo {volume} {104}},\ \bibinfo {pages}
  {034504} (\bibinfo {year} {2021})}\BibitemShut {NoStop}%
\bibitem [{\citenamefont {Klco}\ \emph {et~al.}(2020)\citenamefont {Klco},
  \citenamefont {Savage},\ and\ \citenamefont {Stryker}}]{PhysRevD.101.074512}%
  \BibitemOpen
  \bibfield  {author} {\bibinfo {author} {\bibfnamefont {N.}~\bibnamefont
  {Klco}}, \bibinfo {author} {\bibfnamefont {M.~J.}\ \bibnamefont {Savage}},\
  and\ \bibinfo {author} {\bibfnamefont {J.~R.}\ \bibnamefont {Stryker}},\
  }\bibfield  {title} {\bibinfo {title} {Su(2) non-abelian gauge field theory
  in one dimension on digital quantum computers},\ }\href
  {https://doi.org/10.1103/PhysRevD.101.074512} {\bibfield  {journal} {\bibinfo
   {journal} {Phys. Rev. D}\ }\textbf {\bibinfo {volume} {101}},\ \bibinfo
  {pages} {074512} (\bibinfo {year} {2020})}\BibitemShut {NoStop}%
\bibitem [{\citenamefont {Ciavarella}\ \emph {et~al.}(2021)\citenamefont
  {Ciavarella}, \citenamefont {Klco},\ and\ \citenamefont
  {Savage}}]{PhysRevD.103.094501}%
  \BibitemOpen
  \bibfield  {author} {\bibinfo {author} {\bibfnamefont {A.}~\bibnamefont
  {Ciavarella}}, \bibinfo {author} {\bibfnamefont {N.}~\bibnamefont {Klco}},\
  and\ \bibinfo {author} {\bibfnamefont {M.~J.}\ \bibnamefont {Savage}},\
  }\bibfield  {title} {\bibinfo {title} {Trailhead for quantum simulation of
  su(3) yang-mills lattice gauge theory in the local multiplet basis},\ }\href
  {https://doi.org/10.1103/PhysRevD.103.094501} {\bibfield  {journal} {\bibinfo
   {journal} {Phys. Rev. D}\ }\textbf {\bibinfo {volume} {103}},\ \bibinfo
  {pages} {094501} (\bibinfo {year} {2021})}\BibitemShut {NoStop}%
\bibitem [{\citenamefont {Davoudi}\ \emph {et~al.}(2021)\citenamefont
  {Davoudi}, \citenamefont {Raychowdhury},\ and\ \citenamefont
  {Shaw}}]{PhysRevD.104.074505}%
  \BibitemOpen
  \bibfield  {author} {\bibinfo {author} {\bibfnamefont {Z.}~\bibnamefont
  {Davoudi}}, \bibinfo {author} {\bibfnamefont {I.}~\bibnamefont
  {Raychowdhury}},\ and\ \bibinfo {author} {\bibfnamefont {A.}~\bibnamefont
  {Shaw}},\ }\bibfield  {title} {\bibinfo {title} {Search for efficient
  formulations for hamiltonian simulation of non-abelian lattice gauge
  theories},\ }\href {https://doi.org/10.1103/PhysRevD.104.074505} {\bibfield
  {journal} {\bibinfo  {journal} {Phys. Rev. D}\ }\textbf {\bibinfo {volume}
  {104}},\ \bibinfo {pages} {074505} (\bibinfo {year} {2021})}\BibitemShut
  {NoStop}%
\bibitem [{\citenamefont {Zohar}\ \emph {et~al.}(2013)\citenamefont {Zohar},
  \citenamefont {Cirac},\ and\ \citenamefont {Reznik}}]{zohar2013simulating}%
  \BibitemOpen
  \bibfield  {author} {\bibinfo {author} {\bibfnamefont {E.}~\bibnamefont
  {Zohar}}, \bibinfo {author} {\bibfnamefont {J.~I.}\ \bibnamefont {Cirac}},\
  and\ \bibinfo {author} {\bibfnamefont {B.}~\bibnamefont {Reznik}},\
  }\bibfield  {title} {\bibinfo {title} {Simulating (2+ 1)-dimensional lattice
  qed with dynamical matter using ultracold atoms},\ }\href@noop {} {\bibfield
  {journal} {\bibinfo  {journal} {Physical review letters}\ }\textbf {\bibinfo
  {volume} {110}},\ \bibinfo {pages} {055302} (\bibinfo {year}
  {2013})}\BibitemShut {NoStop}%
\bibitem [{\citenamefont {Luo}\ \emph {et~al.}(2021{\natexlab{a}})\citenamefont
  {Luo}, \citenamefont {Carleo}, \citenamefont {Clark},\ and\ \citenamefont
  {Stokes}}]{PhysRevLett.127.276402}%
  \BibitemOpen
  \bibfield  {author} {\bibinfo {author} {\bibfnamefont {D.}~\bibnamefont
  {Luo}}, \bibinfo {author} {\bibfnamefont {G.}~\bibnamefont {Carleo}},
  \bibinfo {author} {\bibfnamefont {B.~K.}\ \bibnamefont {Clark}},\ and\
  \bibinfo {author} {\bibfnamefont {J.}~\bibnamefont {Stokes}},\ }\bibfield
  {title} {\bibinfo {title} {Gauge equivariant neural networks for quantum
  lattice gauge theories},\ }\href
  {https://doi.org/10.1103/PhysRevLett.127.276402} {\bibfield  {journal}
  {\bibinfo  {journal} {Phys. Rev. Lett.}\ }\textbf {\bibinfo {volume} {127}},\
  \bibinfo {pages} {276402} (\bibinfo {year} {2021}{\natexlab{a}})}\BibitemShut
  {NoStop}%
\bibitem [{\citenamefont {Surace}\ \emph {et~al.}(2020)\citenamefont {Surace},
  \citenamefont {Mazza}, \citenamefont {Giudici}, \citenamefont {Lerose},
  \citenamefont {Gambassi},\ and\ \citenamefont
  {Dalmonte}}]{PhysRevX.10.021041}%
  \BibitemOpen
  \bibfield  {author} {\bibinfo {author} {\bibfnamefont {F.~M.}\ \bibnamefont
  {Surace}}, \bibinfo {author} {\bibfnamefont {P.~P.}\ \bibnamefont {Mazza}},
  \bibinfo {author} {\bibfnamefont {G.}~\bibnamefont {Giudici}}, \bibinfo
  {author} {\bibfnamefont {A.}~\bibnamefont {Lerose}}, \bibinfo {author}
  {\bibfnamefont {A.}~\bibnamefont {Gambassi}},\ and\ \bibinfo {author}
  {\bibfnamefont {M.}~\bibnamefont {Dalmonte}},\ }\bibfield  {title} {\bibinfo
  {title} {Lattice gauge theories and string dynamics in rydberg atom quantum
  simulators},\ }\href {https://doi.org/10.1103/PhysRevX.10.021041} {\bibfield
  {journal} {\bibinfo  {journal} {Phys. Rev. X}\ }\textbf {\bibinfo {volume}
  {10}},\ \bibinfo {pages} {021041} (\bibinfo {year} {2020})}\BibitemShut
  {NoStop}%
\bibitem [{\citenamefont {Favoni}\ \emph {et~al.}(2022)\citenamefont {Favoni},
  \citenamefont {Ipp}, \citenamefont {M{\"u}ller},\ and\ \citenamefont
  {Schuh}}]{favoni2022lattice}%
  \BibitemOpen
  \bibfield  {author} {\bibinfo {author} {\bibfnamefont {M.}~\bibnamefont
  {Favoni}}, \bibinfo {author} {\bibfnamefont {A.}~\bibnamefont {Ipp}},
  \bibinfo {author} {\bibfnamefont {D.~I.}\ \bibnamefont {M{\"u}ller}},\ and\
  \bibinfo {author} {\bibfnamefont {D.}~\bibnamefont {Schuh}},\ }\bibfield
  {title} {\bibinfo {title} {Lattice gauge equivariant convolutional neural
  networks},\ }\href@noop {} {\bibfield  {journal} {\bibinfo  {journal}
  {Physical Review Letters}\ }\textbf {\bibinfo {volume} {128}},\ \bibinfo
  {pages} {032003} (\bibinfo {year} {2022})}\BibitemShut {NoStop}%
\bibitem [{\citenamefont {Boyda}\ \emph {et~al.}(2021)\citenamefont {Boyda},
  \citenamefont {Kanwar}, \citenamefont {Racani{\`{e}}re}, \citenamefont
  {Rezende}, \citenamefont {Albergo}, \citenamefont {Cranmer}, \citenamefont
  {Hackett},\ and\ \citenamefont {Shanahan}}]{Boyda_2021}%
  \BibitemOpen
  \bibfield  {author} {\bibinfo {author} {\bibfnamefont {D.}~\bibnamefont
  {Boyda}}, \bibinfo {author} {\bibfnamefont {G.}~\bibnamefont {Kanwar}},
  \bibinfo {author} {\bibfnamefont {S.}~\bibnamefont {Racani{\`{e}}re}},
  \bibinfo {author} {\bibfnamefont {D.~J.}\ \bibnamefont {Rezende}}, \bibinfo
  {author} {\bibfnamefont {M.~S.}\ \bibnamefont {Albergo}}, \bibinfo {author}
  {\bibfnamefont {K.}~\bibnamefont {Cranmer}}, \bibinfo {author} {\bibfnamefont
  {D.~C.}\ \bibnamefont {Hackett}},\ and\ \bibinfo {author} {\bibfnamefont
  {P.~E.}\ \bibnamefont {Shanahan}},\ }\bibfield  {title} {\bibinfo {title}
  {Sampling using su(n) gauge equivariant flows},\ }\bibfield  {journal}
  {\bibinfo  {journal} {Physical Review D}\ }\textbf {\bibinfo {volume}
  {103}},\ \href {https://doi.org/10.1103/physrevd.103.074504}
  {10.1103/physrevd.103.074504} (\bibinfo {year} {2021})\BibitemShut {NoStop}%
\bibitem [{\citenamefont {Kanwar}\ \emph {et~al.}(2020)\citenamefont {Kanwar},
  \citenamefont {Albergo}, \citenamefont {Boyda}, \citenamefont {Cranmer},
  \citenamefont {Hackett}, \citenamefont {Racani\`ere}, \citenamefont
  {Rezende},\ and\ \citenamefont {Shanahan}}]{PhysRevLett.125.121601}%
  \BibitemOpen
  \bibfield  {author} {\bibinfo {author} {\bibfnamefont {G.}~\bibnamefont
  {Kanwar}}, \bibinfo {author} {\bibfnamefont {M.~S.}\ \bibnamefont {Albergo}},
  \bibinfo {author} {\bibfnamefont {D.}~\bibnamefont {Boyda}}, \bibinfo
  {author} {\bibfnamefont {K.}~\bibnamefont {Cranmer}}, \bibinfo {author}
  {\bibfnamefont {D.~C.}\ \bibnamefont {Hackett}}, \bibinfo {author}
  {\bibfnamefont {S.}~\bibnamefont {Racani\`ere}}, \bibinfo {author}
  {\bibfnamefont {D.~J.}\ \bibnamefont {Rezende}},\ and\ \bibinfo {author}
  {\bibfnamefont {P.~E.}\ \bibnamefont {Shanahan}},\ }\bibfield  {title}
  {\bibinfo {title} {Equivariant flow-based sampling for lattice gauge
  theory},\ }\href {https://doi.org/10.1103/PhysRevLett.125.121601} {\bibfield
  {journal} {\bibinfo  {journal} {Phys. Rev. Lett.}\ }\textbf {\bibinfo
  {volume} {125}},\ \bibinfo {pages} {121601} (\bibinfo {year}
  {2020})}\BibitemShut {NoStop}%
\bibitem [{\citenamefont {Albergo}\ \emph {et~al.}(2019)\citenamefont
  {Albergo}, \citenamefont {Kanwar},\ and\ \citenamefont
  {Shanahan}}]{albergo2019flow}%
  \BibitemOpen
  \bibfield  {author} {\bibinfo {author} {\bibfnamefont {M.~S.}\ \bibnamefont
  {Albergo}}, \bibinfo {author} {\bibfnamefont {G.}~\bibnamefont {Kanwar}},\
  and\ \bibinfo {author} {\bibfnamefont {P.~E.}\ \bibnamefont {Shanahan}},\
  }\bibfield  {title} {\bibinfo {title} {Flow-based generative models for
  markov chain monte carlo in lattice field theory},\ }\href@noop {} {\bibfield
   {journal} {\bibinfo  {journal} {Physical Review D}\ }\textbf {\bibinfo
  {volume} {100}},\ \bibinfo {pages} {034515} (\bibinfo {year}
  {2019})}\BibitemShut {NoStop}%
\bibitem [{\citenamefont {Hu}\ \emph {et~al.}(2020)\citenamefont {Hu},
  \citenamefont {Li}, \citenamefont {Wang},\ and\ \citenamefont
  {You}}]{PhysRevResearch.2.023369}%
  \BibitemOpen
  \bibfield  {author} {\bibinfo {author} {\bibfnamefont {H.-Y.}\ \bibnamefont
  {Hu}}, \bibinfo {author} {\bibfnamefont {S.-H.}\ \bibnamefont {Li}}, \bibinfo
  {author} {\bibfnamefont {L.}~\bibnamefont {Wang}},\ and\ \bibinfo {author}
  {\bibfnamefont {Y.-Z.}\ \bibnamefont {You}},\ }\bibfield  {title} {\bibinfo
  {title} {Machine learning holographic mapping by neural network
  renormalization group},\ }\href
  {https://doi.org/10.1103/PhysRevResearch.2.023369} {\bibfield  {journal}
  {\bibinfo  {journal} {Phys. Rev. Research}\ }\textbf {\bibinfo {volume}
  {2}},\ \bibinfo {pages} {023369} (\bibinfo {year} {2020})}\BibitemShut
  {NoStop}%
\bibitem [{\citenamefont {Abbott}\ \emph {et~al.}(2022)\citenamefont {Abbott},
  \citenamefont {Albergo}, \citenamefont {Boyda}, \citenamefont {Cranmer},
  \citenamefont {Hackett}, \citenamefont {Kanwar}, \citenamefont {Racanière},
  \citenamefont {Rezende}, \citenamefont {Romero-López}, \citenamefont
  {Shanahan}, \citenamefont {Tian},\ and\ \citenamefont
  {Urban}}]{https://doi.org/10.48550/arxiv.2207.08945}%
  \BibitemOpen
  \bibfield  {author} {\bibinfo {author} {\bibfnamefont {R.}~\bibnamefont
  {Abbott}}, \bibinfo {author} {\bibfnamefont {M.~S.}\ \bibnamefont {Albergo}},
  \bibinfo {author} {\bibfnamefont {D.}~\bibnamefont {Boyda}}, \bibinfo
  {author} {\bibfnamefont {K.}~\bibnamefont {Cranmer}}, \bibinfo {author}
  {\bibfnamefont {D.~C.}\ \bibnamefont {Hackett}}, \bibinfo {author}
  {\bibfnamefont {G.}~\bibnamefont {Kanwar}}, \bibinfo {author} {\bibfnamefont
  {S.}~\bibnamefont {Racanière}}, \bibinfo {author} {\bibfnamefont {D.~J.}\
  \bibnamefont {Rezende}}, \bibinfo {author} {\bibfnamefont {F.}~\bibnamefont
  {Romero-López}}, \bibinfo {author} {\bibfnamefont {P.~E.}\ \bibnamefont
  {Shanahan}}, \bibinfo {author} {\bibfnamefont {B.}~\bibnamefont {Tian}},\
  and\ \bibinfo {author} {\bibfnamefont {J.~M.}\ \bibnamefont {Urban}},\ }\href
  {https://doi.org/10.48550/ARXIV.2207.08945} {\bibinfo {title}
  {Gauge-equivariant flow models for sampling in lattice field theories with
  pseudofermions}} (\bibinfo {year} {2022})\BibitemShut {NoStop}%
\bibitem [{\citenamefont {Luo}\ \emph {et~al.}(2021{\natexlab{b}})\citenamefont
  {Luo}, \citenamefont {Chen}, \citenamefont {Hu}, \citenamefont {Zhao},
  \citenamefont {Hur},\ and\ \citenamefont
  {Clark}}]{https://doi.org/10.48550/arxiv.2101.07243}%
  \BibitemOpen
  \bibfield  {author} {\bibinfo {author} {\bibfnamefont {D.}~\bibnamefont
  {Luo}}, \bibinfo {author} {\bibfnamefont {Z.}~\bibnamefont {Chen}}, \bibinfo
  {author} {\bibfnamefont {K.}~\bibnamefont {Hu}}, \bibinfo {author}
  {\bibfnamefont {Z.}~\bibnamefont {Zhao}}, \bibinfo {author} {\bibfnamefont
  {V.~M.}\ \bibnamefont {Hur}},\ and\ \bibinfo {author} {\bibfnamefont {B.~K.}\
  \bibnamefont {Clark}},\ }\href {https://doi.org/10.48550/ARXIV.2101.07243}
  {\bibinfo {title} {Gauge invariant autoregressive neural networks for quantum
  lattice models}} (\bibinfo {year} {2021}{\natexlab{b}})\BibitemShut {NoStop}%
\bibitem [{\citenamefont {Luo}\ \emph {et~al.}(2022{\natexlab{a}})\citenamefont
  {Luo}, \citenamefont {Yuan}, \citenamefont {Stokes},\ and\ \citenamefont
  {Clark}}]{luo2022gauge}%
  \BibitemOpen
  \bibfield  {author} {\bibinfo {author} {\bibfnamefont {D.}~\bibnamefont
  {Luo}}, \bibinfo {author} {\bibfnamefont {S.}~\bibnamefont {Yuan}}, \bibinfo
  {author} {\bibfnamefont {J.}~\bibnamefont {Stokes}},\ and\ \bibinfo {author}
  {\bibfnamefont {B.~K.}\ \bibnamefont {Clark}},\ }\bibfield  {title} {\bibinfo
  {title} {Gauge equivariant neural networks for 2+ 1d u (1) gauge theory
  simulations in hamiltonian formulation},\ }\href@noop {} {\bibfield
  {journal} {\bibinfo  {journal} {arXiv preprint arXiv:2211.03198}\ } (\bibinfo
  {year} {2022}{\natexlab{a}})}\BibitemShut {NoStop}%
\bibitem [{\citenamefont {Carleo}\ and\ \citenamefont
  {Troyer}(2017)}]{doi:10.1126/science.aag2302}%
  \BibitemOpen
  \bibfield  {author} {\bibinfo {author} {\bibfnamefont {G.}~\bibnamefont
  {Carleo}}\ and\ \bibinfo {author} {\bibfnamefont {M.}~\bibnamefont
  {Troyer}},\ }\bibfield  {title} {\bibinfo {title} {Solving the quantum
  many-body problem with artificial neural networks},\ }\href
  {https://doi.org/10.1126/science.aag2302} {\bibfield  {journal} {\bibinfo
  {journal} {Science}\ }\textbf {\bibinfo {volume} {355}},\ \bibinfo {pages}
  {602} (\bibinfo {year} {2017})},\ \Eprint
  {https://arxiv.org/abs/https://www.science.org/doi/pdf/10.1126/science.aag2302}
  {https://www.science.org/doi/pdf/10.1126/science.aag2302} \BibitemShut
  {NoStop}%
\bibitem [{\citenamefont {Hibat-Allah}\ \emph
  {et~al.}(2020{\natexlab{a}})\citenamefont {Hibat-Allah}, \citenamefont
  {Ganahl}, \citenamefont {Hayward}, \citenamefont {Melko},\ and\ \citenamefont
  {Carrasquilla}}]{Hibat_Allah_2020}%
  \BibitemOpen
  \bibfield  {author} {\bibinfo {author} {\bibfnamefont {M.}~\bibnamefont
  {Hibat-Allah}}, \bibinfo {author} {\bibfnamefont {M.}~\bibnamefont {Ganahl}},
  \bibinfo {author} {\bibfnamefont {L.~E.}\ \bibnamefont {Hayward}}, \bibinfo
  {author} {\bibfnamefont {R.~G.}\ \bibnamefont {Melko}},\ and\ \bibinfo
  {author} {\bibfnamefont {J.}~\bibnamefont {Carrasquilla}},\ }\bibfield
  {title} {\bibinfo {title} {Recurrent neural network wave functions},\
  }\bibfield  {journal} {\bibinfo  {journal} {Physical Review Research}\
  }\textbf {\bibinfo {volume} {2}},\ \href
  {https://doi.org/10.1103/physrevresearch.2.023358}
  {10.1103/physrevresearch.2.023358} (\bibinfo {year}
  {2020}{\natexlab{a}})\BibitemShut {NoStop}%
\bibitem [{\citenamefont {Sharir}\ \emph {et~al.}(2020)\citenamefont {Sharir},
  \citenamefont {Levine}, \citenamefont {Wies}, \citenamefont {Carleo},\ and\
  \citenamefont {Shashua}}]{PhysRevLett.124.020503}%
  \BibitemOpen
  \bibfield  {author} {\bibinfo {author} {\bibfnamefont {O.}~\bibnamefont
  {Sharir}}, \bibinfo {author} {\bibfnamefont {Y.}~\bibnamefont {Levine}},
  \bibinfo {author} {\bibfnamefont {N.}~\bibnamefont {Wies}}, \bibinfo {author}
  {\bibfnamefont {G.}~\bibnamefont {Carleo}},\ and\ \bibinfo {author}
  {\bibfnamefont {A.}~\bibnamefont {Shashua}},\ }\bibfield  {title} {\bibinfo
  {title} {Deep autoregressive models for the efficient variational simulation
  of many-body quantum systems},\ }\href
  {https://doi.org/10.1103/PhysRevLett.124.020503} {\bibfield  {journal}
  {\bibinfo  {journal} {Phys. Rev. Lett.}\ }\textbf {\bibinfo {volume} {124}},\
  \bibinfo {pages} {020503} (\bibinfo {year} {2020})}\BibitemShut {NoStop}%
\bibitem [{\citenamefont {Irikura}\ and\ \citenamefont
  {Saito}(2020)}]{Irikura_2020}%
  \BibitemOpen
  \bibfield  {author} {\bibinfo {author} {\bibfnamefont {N.}~\bibnamefont
  {Irikura}}\ and\ \bibinfo {author} {\bibfnamefont {H.}~\bibnamefont
  {Saito}},\ }\bibfield  {title} {\bibinfo {title} {Neural-network quantum
  states at finite temperature},\ }\bibfield  {journal} {\bibinfo  {journal}
  {Physical Review Research}\ }\textbf {\bibinfo {volume} {2}},\ \href
  {https://doi.org/10.1103/physrevresearch.2.013284}
  {10.1103/physrevresearch.2.013284} (\bibinfo {year} {2020})\BibitemShut
  {NoStop}%
\bibitem [{\citenamefont {Lee}\ \emph {et~al.}(2021)\citenamefont {Lee},
  \citenamefont {Patil}, \citenamefont {Zhang},\ and\ \citenamefont
  {Hsieh}}]{PhysRevResearch.3.023095}%
  \BibitemOpen
  \bibfield  {author} {\bibinfo {author} {\bibfnamefont {C.~K.}\ \bibnamefont
  {Lee}}, \bibinfo {author} {\bibfnamefont {P.}~\bibnamefont {Patil}}, \bibinfo
  {author} {\bibfnamefont {S.}~\bibnamefont {Zhang}},\ and\ \bibinfo {author}
  {\bibfnamefont {C.~Y.}\ \bibnamefont {Hsieh}},\ }\bibfield  {title} {\bibinfo
  {title} {Neural-network variational quantum algorithm for simulating
  many-body dynamics},\ }\href
  {https://doi.org/10.1103/PhysRevResearch.3.023095} {\bibfield  {journal}
  {\bibinfo  {journal} {Phys. Rev. Research}\ }\textbf {\bibinfo {volume}
  {3}},\ \bibinfo {pages} {023095} (\bibinfo {year} {2021})}\BibitemShut
  {NoStop}%
\bibitem [{\citenamefont {Han}\ and\ \citenamefont
  {Hartnoll}(2020)}]{Han_2020}%
  \BibitemOpen
  \bibfield  {author} {\bibinfo {author} {\bibfnamefont {X.}~\bibnamefont
  {Han}}\ and\ \bibinfo {author} {\bibfnamefont {S.~A.}\ \bibnamefont
  {Hartnoll}},\ }\bibfield  {title} {\bibinfo {title} {Deep quantum geometry of
  matrices},\ }\bibfield  {journal} {\bibinfo  {journal} {Physical Review X}\
  }\textbf {\bibinfo {volume} {10}},\ \href
  {https://doi.org/10.1103/physrevx.10.011069} {10.1103/physrevx.10.011069}
  (\bibinfo {year} {2020})\BibitemShut {NoStop}%
\bibitem [{\citenamefont {Pfau}\ \emph {et~al.}(2020)\citenamefont {Pfau},
  \citenamefont {Spencer}, \citenamefont {Matthews},\ and\ \citenamefont
  {Foulkes}}]{ferminet}%
  \BibitemOpen
  \bibfield  {author} {\bibinfo {author} {\bibfnamefont {D.}~\bibnamefont
  {Pfau}}, \bibinfo {author} {\bibfnamefont {J.~S.}\ \bibnamefont {Spencer}},
  \bibinfo {author} {\bibfnamefont {A.~G. D.~G.}\ \bibnamefont {Matthews}},\
  and\ \bibinfo {author} {\bibfnamefont {W.~M.~C.}\ \bibnamefont {Foulkes}},\
  }\bibfield  {title} {\bibinfo {title} {Ab initio solution of the
  many-electron schr\"odinger equation with deep neural networks},\ }\href
  {https://doi.org/10.1103/PhysRevResearch.2.033429} {\bibfield  {journal}
  {\bibinfo  {journal} {Phys. Rev. Research}\ }\textbf {\bibinfo {volume}
  {2}},\ \bibinfo {pages} {033429} (\bibinfo {year} {2020})}\BibitemShut
  {NoStop}%
\bibitem [{\citenamefont {Choo}\ \emph {et~al.}(2019)\citenamefont {Choo},
  \citenamefont {Neupert},\ and\ \citenamefont {Carleo}}]{Choo_2019}%
  \BibitemOpen
  \bibfield  {author} {\bibinfo {author} {\bibfnamefont {K.}~\bibnamefont
  {Choo}}, \bibinfo {author} {\bibfnamefont {T.}~\bibnamefont {Neupert}},\ and\
  \bibinfo {author} {\bibfnamefont {G.}~\bibnamefont {Carleo}},\ }\bibfield
  {title} {\bibinfo {title} {Two-dimensional frustrated j1−j2 model studied
  with neural network quantum states},\ }\bibfield  {journal} {\bibinfo
  {journal} {Physical Review B}\ }\textbf {\bibinfo {volume} {100}},\ \href
  {https://doi.org/10.1103/physrevb.100.125124} {10.1103/physrevb.100.125124}
  (\bibinfo {year} {2019})\BibitemShut {NoStop}%
\bibitem [{\citenamefont {Hibat-Allah}\ \emph
  {et~al.}(2020{\natexlab{b}})\citenamefont {Hibat-Allah}, \citenamefont
  {Ganahl}, \citenamefont {Hayward}, \citenamefont {Melko},\ and\ \citenamefont
  {Carrasquilla}}]{rnn_wavefunction}%
  \BibitemOpen
  \bibfield  {author} {\bibinfo {author} {\bibfnamefont {M.}~\bibnamefont
  {Hibat-Allah}}, \bibinfo {author} {\bibfnamefont {M.}~\bibnamefont {Ganahl}},
  \bibinfo {author} {\bibfnamefont {L.~E.}\ \bibnamefont {Hayward}}, \bibinfo
  {author} {\bibfnamefont {R.~G.}\ \bibnamefont {Melko}},\ and\ \bibinfo
  {author} {\bibfnamefont {J.}~\bibnamefont {Carrasquilla}},\ }\bibfield
  {title} {\bibinfo {title} {Recurrent neural network wave functions},\ }\href
  {https://doi.org/10.1103/PhysRevResearch.2.023358} {\bibfield  {journal}
  {\bibinfo  {journal} {Phys. Rev. Research}\ }\textbf {\bibinfo {volume}
  {2}},\ \bibinfo {pages} {023358} (\bibinfo {year}
  {2020}{\natexlab{b}})}\BibitemShut {NoStop}%
\bibitem [{\citenamefont {Hermann}\ \emph {et~al.}(2019)\citenamefont
  {Hermann}, \citenamefont {Schätzle},\ and\ \citenamefont {Noé}}]{paulinet}%
  \BibitemOpen
  \bibfield  {author} {\bibinfo {author} {\bibfnamefont {J.}~\bibnamefont
  {Hermann}}, \bibinfo {author} {\bibfnamefont {Z.}~\bibnamefont {Schätzle}},\
  and\ \bibinfo {author} {\bibfnamefont {F.}~\bibnamefont {Noé}},\ }\href@noop
  {} {\bibinfo {title} {Deep neural network solution of the electronic
  schrödinger equation}} (\bibinfo {year} {2019}),\ \Eprint
  {https://arxiv.org/abs/1909.08423} {arXiv:1909.08423 [physics.comp-ph]}
  \BibitemShut {NoStop}%
\bibitem [{\citenamefont {Glasser}\ \emph {et~al.}(2018)\citenamefont
  {Glasser}, \citenamefont {Pancotti}, \citenamefont {August}, \citenamefont
  {Rodriguez},\ and\ \citenamefont {Cirac}}]{Glasser_2018}%
  \BibitemOpen
  \bibfield  {author} {\bibinfo {author} {\bibfnamefont {I.}~\bibnamefont
  {Glasser}}, \bibinfo {author} {\bibfnamefont {N.}~\bibnamefont {Pancotti}},
  \bibinfo {author} {\bibfnamefont {M.}~\bibnamefont {August}}, \bibinfo
  {author} {\bibfnamefont {I.~D.}\ \bibnamefont {Rodriguez}},\ and\ \bibinfo
  {author} {\bibfnamefont {J.~I.}\ \bibnamefont {Cirac}},\ }\bibfield  {title}
  {\bibinfo {title} {Neural-network quantum states, string-bond states, and
  chiral topological states},\ }\bibfield  {journal} {\bibinfo  {journal}
  {Physical Review X}\ }\textbf {\bibinfo {volume} {8}},\ \href
  {https://doi.org/10.1103/physrevx.8.011006} {10.1103/physrevx.8.011006}
  (\bibinfo {year} {2018})\BibitemShut {NoStop}%
\bibitem [{\citenamefont {Stokes}\ \emph {et~al.}(2020)\citenamefont {Stokes},
  \citenamefont {Moreno}, \citenamefont {Pnevmatikakis},\ and\ \citenamefont
  {Carleo}}]{Stokes_2020}%
  \BibitemOpen
  \bibfield  {author} {\bibinfo {author} {\bibfnamefont {J.}~\bibnamefont
  {Stokes}}, \bibinfo {author} {\bibfnamefont {J.~R.}\ \bibnamefont {Moreno}},
  \bibinfo {author} {\bibfnamefont {E.~A.}\ \bibnamefont {Pnevmatikakis}},\
  and\ \bibinfo {author} {\bibfnamefont {G.}~\bibnamefont {Carleo}},\
  }\bibfield  {title} {\bibinfo {title} {Phases of two-dimensional spinless
  lattice fermions with first-quantized deep neural-network quantum states},\
  }\bibfield  {journal} {\bibinfo  {journal} {Physical Review B}\ }\textbf
  {\bibinfo {volume} {102}},\ \href
  {https://doi.org/10.1103/physrevb.102.205122} {10.1103/physrevb.102.205122}
  (\bibinfo {year} {2020})\BibitemShut {NoStop}%
\bibitem [{\citenamefont {Nomura}\ \emph {et~al.}(2017)\citenamefont {Nomura},
  \citenamefont {Darmawan}, \citenamefont {Yamaji},\ and\ \citenamefont
  {Imada}}]{Nomura_2017}%
  \BibitemOpen
  \bibfield  {author} {\bibinfo {author} {\bibfnamefont {Y.}~\bibnamefont
  {Nomura}}, \bibinfo {author} {\bibfnamefont {A.~S.}\ \bibnamefont
  {Darmawan}}, \bibinfo {author} {\bibfnamefont {Y.}~\bibnamefont {Yamaji}},\
  and\ \bibinfo {author} {\bibfnamefont {M.}~\bibnamefont {Imada}},\ }\bibfield
   {title} {\bibinfo {title} {Restricted boltzmann machine learning for solving
  strongly correlated quantum systems},\ }\bibfield  {journal} {\bibinfo
  {journal} {Physical Review B}\ }\textbf {\bibinfo {volume} {96}},\ \href
  {https://doi.org/10.1103/physrevb.96.205152} {10.1103/physrevb.96.205152}
  (\bibinfo {year} {2017})\BibitemShut {NoStop}%
\bibitem [{\citenamefont {Martyn}\ \emph {et~al.}(2022)\citenamefont {Martyn},
  \citenamefont {Najafi},\ and\ \citenamefont {Luo}}]{martyn2022variational}%
  \BibitemOpen
  \bibfield  {author} {\bibinfo {author} {\bibfnamefont {J.~M.}\ \bibnamefont
  {Martyn}}, \bibinfo {author} {\bibfnamefont {K.}~\bibnamefont {Najafi}},\
  and\ \bibinfo {author} {\bibfnamefont {D.}~\bibnamefont {Luo}},\ }\bibfield
  {title} {\bibinfo {title} {Variational neural-network ansatz for continuum
  quantum field theory},\ }\href@noop {} {\bibfield  {journal} {\bibinfo
  {journal} {arXiv preprint arXiv:2212.00782}\ } (\bibinfo {year}
  {2022})}\BibitemShut {NoStop}%
\bibitem [{\citenamefont {Luo}\ and\ \citenamefont {Clark}(2019)}]{Luo_2019}%
  \BibitemOpen
  \bibfield  {author} {\bibinfo {author} {\bibfnamefont {D.}~\bibnamefont
  {Luo}}\ and\ \bibinfo {author} {\bibfnamefont {B.~K.}\ \bibnamefont
  {Clark}},\ }\bibfield  {title} {\bibinfo {title} {Backflow transformations
  via neural networks for quantum many-body wave functions},\ }\bibfield
  {journal} {\bibinfo  {journal} {Physical Review Letters}\ }\textbf {\bibinfo
  {volume} {122}},\ \href {https://doi.org/10.1103/physrevlett.122.226401}
  {10.1103/physrevlett.122.226401} (\bibinfo {year} {2019})\BibitemShut
  {NoStop}%
\bibitem [{\citenamefont {Xie}\ \emph {et~al.}(2021)\citenamefont {Xie},
  \citenamefont {Zhang},\ and\ \citenamefont {Wang}}]{xie2021ab}%
  \BibitemOpen
  \bibfield  {author} {\bibinfo {author} {\bibfnamefont {H.}~\bibnamefont
  {Xie}}, \bibinfo {author} {\bibfnamefont {L.}~\bibnamefont {Zhang}},\ and\
  \bibinfo {author} {\bibfnamefont {L.}~\bibnamefont {Wang}},\ }\bibfield
  {title} {\bibinfo {title} {Ab-initio study of interacting fermions at finite
  temperature with neural canonical transformation},\ }\href@noop {} {\bibfield
   {journal} {\bibinfo  {journal} {arXiv preprint arXiv:2105.08644}\ }
  (\bibinfo {year} {2021})}\BibitemShut {NoStop}%
\bibitem [{\citenamefont {Wang}\ \emph {et~al.}(2021)\citenamefont {Wang},
  \citenamefont {Chen}, \citenamefont {Luo}, \citenamefont {Zhao},
  \citenamefont {Hur},\ and\ \citenamefont {Clark}}]{wang2021spacetime}%
  \BibitemOpen
  \bibfield  {author} {\bibinfo {author} {\bibfnamefont {J.}~\bibnamefont
  {Wang}}, \bibinfo {author} {\bibfnamefont {Z.}~\bibnamefont {Chen}}, \bibinfo
  {author} {\bibfnamefont {D.}~\bibnamefont {Luo}}, \bibinfo {author}
  {\bibfnamefont {Z.}~\bibnamefont {Zhao}}, \bibinfo {author} {\bibfnamefont
  {V.~M.}\ \bibnamefont {Hur}},\ and\ \bibinfo {author} {\bibfnamefont {B.~K.}\
  \bibnamefont {Clark}},\ }\href@noop {} {\bibinfo {title} {Spacetime neural
  network for high dimensional quantum dynamics}} (\bibinfo {year} {2021}),\
  \Eprint {https://arxiv.org/abs/2108.02200} {arXiv:2108.02200
  [cond-mat.dis-nn]} \BibitemShut {NoStop}%
\bibitem [{\citenamefont {Astrakhantsev}\ \emph {et~al.}(2021)\citenamefont
  {Astrakhantsev}, \citenamefont {Westerhout}, \citenamefont {Tiwari},
  \citenamefont {Choo}, \citenamefont {Chen}, \citenamefont {Fischer},
  \citenamefont {Carleo},\ and\ \citenamefont {Neupert}}]{py2021}%
  \BibitemOpen
  \bibfield  {author} {\bibinfo {author} {\bibfnamefont {N.}~\bibnamefont
  {Astrakhantsev}}, \bibinfo {author} {\bibfnamefont {T.}~\bibnamefont
  {Westerhout}}, \bibinfo {author} {\bibfnamefont {A.}~\bibnamefont {Tiwari}},
  \bibinfo {author} {\bibfnamefont {K.}~\bibnamefont {Choo}}, \bibinfo {author}
  {\bibfnamefont {A.}~\bibnamefont {Chen}}, \bibinfo {author} {\bibfnamefont
  {M.~H.}\ \bibnamefont {Fischer}}, \bibinfo {author} {\bibfnamefont
  {G.}~\bibnamefont {Carleo}},\ and\ \bibinfo {author} {\bibfnamefont
  {T.}~\bibnamefont {Neupert}},\ }\bibfield  {title} {\bibinfo {title}
  {Broken-symmetry ground states of the heisenberg model on the pyrochlore
  lattice},\ }\bibfield  {journal} {\bibinfo  {journal} {Physical Review X}\
  }\textbf {\bibinfo {volume} {11}},\ \href
  {https://doi.org/10.1103/physrevx.11.041021} {10.1103/physrevx.11.041021}
  (\bibinfo {year} {2021})\BibitemShut {NoStop}%
\bibitem [{\citenamefont {Gutiérrez}\ and\ \citenamefont
  {Mendl}(2020)}]{gutierrez2020real}%
  \BibitemOpen
  \bibfield  {author} {\bibinfo {author} {\bibfnamefont {I.~L.}\ \bibnamefont
  {Gutiérrez}}\ and\ \bibinfo {author} {\bibfnamefont {C.~B.}\ \bibnamefont
  {Mendl}},\ }\href@noop {} {\bibinfo {title} {Real time evolution with
  neural-network quantum states}} (\bibinfo {year} {2020}),\ \Eprint
  {https://arxiv.org/abs/1912.08831} {arXiv:1912.08831 [cond-mat.dis-nn]}
  \BibitemShut {NoStop}%
\bibitem [{\citenamefont {Schmitt}\ and\ \citenamefont
  {Heyl}(2020)}]{Schmitt_2020}%
  \BibitemOpen
  \bibfield  {author} {\bibinfo {author} {\bibfnamefont {M.}~\bibnamefont
  {Schmitt}}\ and\ \bibinfo {author} {\bibfnamefont {M.}~\bibnamefont {Heyl}},\
  }\bibfield  {title} {\bibinfo {title} {Quantum many-body dynamics in two
  dimensions with artificial neural networks},\ }\bibfield  {journal} {\bibinfo
   {journal} {Physical Review Letters}\ }\textbf {\bibinfo {volume} {125}},\
  \href {https://doi.org/10.1103/physrevlett.125.100503}
  {10.1103/physrevlett.125.100503} (\bibinfo {year} {2020})\BibitemShut
  {NoStop}%
\bibitem [{\citenamefont {Vicentini}\ \emph {et~al.}(2019)\citenamefont
  {Vicentini}, \citenamefont {Biella}, \citenamefont {Regnault},\ and\
  \citenamefont {Ciuti}}]{Vicentini_2019}%
  \BibitemOpen
  \bibfield  {author} {\bibinfo {author} {\bibfnamefont {F.}~\bibnamefont
  {Vicentini}}, \bibinfo {author} {\bibfnamefont {A.}~\bibnamefont {Biella}},
  \bibinfo {author} {\bibfnamefont {N.}~\bibnamefont {Regnault}},\ and\
  \bibinfo {author} {\bibfnamefont {C.}~\bibnamefont {Ciuti}},\ }\bibfield
  {title} {\bibinfo {title} {Variational neural-network ansatz for steady
  states in open quantum systems},\ }\bibfield  {journal} {\bibinfo  {journal}
  {Physical Review Letters}\ }\textbf {\bibinfo {volume} {122}},\ \href
  {https://doi.org/10.1103/physrevlett.122.250503}
  {10.1103/physrevlett.122.250503} (\bibinfo {year} {2019})\BibitemShut
  {NoStop}%
\bibitem [{\citenamefont {Yoshioka}\ and\ \citenamefont
  {Hamazaki}(2019)}]{PhysRevB.99.214306}%
  \BibitemOpen
  \bibfield  {author} {\bibinfo {author} {\bibfnamefont {N.}~\bibnamefont
  {Yoshioka}}\ and\ \bibinfo {author} {\bibfnamefont {R.}~\bibnamefont
  {Hamazaki}},\ }\bibfield  {title} {\bibinfo {title} {Constructing neural
  stationary states for open quantum many-body systems},\ }\href
  {https://doi.org/10.1103/PhysRevB.99.214306} {\bibfield  {journal} {\bibinfo
  {journal} {Phys. Rev. B}\ }\textbf {\bibinfo {volume} {99}},\ \bibinfo
  {pages} {214306} (\bibinfo {year} {2019})}\BibitemShut {NoStop}%
\bibitem [{\citenamefont {Hartmann}\ and\ \citenamefont
  {Carleo}(2019)}]{PhysRevLett.122.250502}%
  \BibitemOpen
  \bibfield  {author} {\bibinfo {author} {\bibfnamefont {M.~J.}\ \bibnamefont
  {Hartmann}}\ and\ \bibinfo {author} {\bibfnamefont {G.}~\bibnamefont
  {Carleo}},\ }\bibfield  {title} {\bibinfo {title} {Neural-network approach to
  dissipative quantum many-body dynamics},\ }\href
  {https://doi.org/10.1103/PhysRevLett.122.250502} {\bibfield  {journal}
  {\bibinfo  {journal} {Phys. Rev. Lett.}\ }\textbf {\bibinfo {volume} {122}},\
  \bibinfo {pages} {250502} (\bibinfo {year} {2019})}\BibitemShut {NoStop}%
\bibitem [{\citenamefont {Nagy}\ and\ \citenamefont
  {Savona}(2019)}]{PhysRevLett.122.250501}%
  \BibitemOpen
  \bibfield  {author} {\bibinfo {author} {\bibfnamefont {A.}~\bibnamefont
  {Nagy}}\ and\ \bibinfo {author} {\bibfnamefont {V.}~\bibnamefont {Savona}},\
  }\bibfield  {title} {\bibinfo {title} {Variational quantum monte carlo method
  with a neural-network ansatz for open quantum systems},\ }\href
  {https://doi.org/10.1103/PhysRevLett.122.250501} {\bibfield  {journal}
  {\bibinfo  {journal} {Phys. Rev. Lett.}\ }\textbf {\bibinfo {volume} {122}},\
  \bibinfo {pages} {250501} (\bibinfo {year} {2019})}\BibitemShut {NoStop}%
\bibitem [{\citenamefont {Luo}\ \emph {et~al.}(2021{\natexlab{c}})\citenamefont
  {Luo}, \citenamefont {Chen}, \citenamefont {Hu}, \citenamefont {Zhao},
  \citenamefont {Hur},\ and\ \citenamefont {Clark}}]{luo_gauge_inv}%
  \BibitemOpen
  \bibfield  {author} {\bibinfo {author} {\bibfnamefont {D.}~\bibnamefont
  {Luo}}, \bibinfo {author} {\bibfnamefont {Z.}~\bibnamefont {Chen}}, \bibinfo
  {author} {\bibfnamefont {K.}~\bibnamefont {Hu}}, \bibinfo {author}
  {\bibfnamefont {Z.}~\bibnamefont {Zhao}}, \bibinfo {author} {\bibfnamefont
  {V.~M.}\ \bibnamefont {Hur}},\ and\ \bibinfo {author} {\bibfnamefont {B.~K.}\
  \bibnamefont {Clark}},\ }\href {https://doi.org/10.48550/ARXIV.2101.07243}
  {\bibinfo {title} {Gauge invariant autoregressive neural networks for quantum
  lattice models}} (\bibinfo {year} {2021}{\natexlab{c}})\BibitemShut {NoStop}%
\bibitem [{\citenamefont {Luo}\ \emph {et~al.}(2022{\natexlab{b}})\citenamefont
  {Luo}, \citenamefont {Chen}, \citenamefont {Carrasquilla},\ and\
  \citenamefont {Clark}}]{luo_povm}%
  \BibitemOpen
  \bibfield  {author} {\bibinfo {author} {\bibfnamefont {D.}~\bibnamefont
  {Luo}}, \bibinfo {author} {\bibfnamefont {Z.}~\bibnamefont {Chen}}, \bibinfo
  {author} {\bibfnamefont {J.}~\bibnamefont {Carrasquilla}},\ and\ \bibinfo
  {author} {\bibfnamefont {B.~K.}\ \bibnamefont {Clark}},\ }\bibfield  {title}
  {\bibinfo {title} {Autoregressive neural network for simulating open quantum
  systems via a probabilistic formulation},\ }\href
  {https://doi.org/10.1103/PhysRevLett.128.090501} {\bibfield  {journal}
  {\bibinfo  {journal} {Phys. Rev. Lett.}\ }\textbf {\bibinfo {volume} {128}},\
  \bibinfo {pages} {090501} (\bibinfo {year} {2022}{\natexlab{b}})}\BibitemShut
  {NoStop}%
\bibitem [{\citenamefont {Bender}\ \emph {et~al.}(2020)\citenamefont {Bender},
  \citenamefont {Emonts}, \citenamefont {Zohar},\ and\ \citenamefont
  {Cirac}}]{PhysRevResearch.2.043145}%
  \BibitemOpen
  \bibfield  {author} {\bibinfo {author} {\bibfnamefont {J.}~\bibnamefont
  {Bender}}, \bibinfo {author} {\bibfnamefont {P.}~\bibnamefont {Emonts}},
  \bibinfo {author} {\bibfnamefont {E.}~\bibnamefont {Zohar}},\ and\ \bibinfo
  {author} {\bibfnamefont {J.~I.}\ \bibnamefont {Cirac}},\ }\bibfield  {title}
  {\bibinfo {title} {Real-time dynamics in $2+1d$ compact qed using complex
  periodic gaussian states},\ }\href
  {https://doi.org/10.1103/PhysRevResearch.2.043145} {\bibfield  {journal}
  {\bibinfo  {journal} {Phys. Rev. Research}\ }\textbf {\bibinfo {volume}
  {2}},\ \bibinfo {pages} {043145} (\bibinfo {year} {2020})}\BibitemShut
  {NoStop}%
\bibitem [{\citenamefont {Rezende}\ and\ \citenamefont
  {Mohamed}(2015)}]{pmlr-v37-rezende15}%
  \BibitemOpen
  \bibfield  {author} {\bibinfo {author} {\bibfnamefont {D.}~\bibnamefont
  {Rezende}}\ and\ \bibinfo {author} {\bibfnamefont {S.}~\bibnamefont
  {Mohamed}},\ }\bibfield  {title} {\bibinfo {title} {Variational inference
  with normalizing flows},\ }in\ \href
  {https://proceedings.mlr.press/v37/rezende15.html} {\emph {\bibinfo
  {booktitle} {Proceedings of the 32nd International Conference on Machine
  Learning}}},\ \bibinfo {series} {Proceedings of Machine Learning Research},
  Vol.~\bibinfo {volume} {37},\ \bibinfo {editor} {edited by\ \bibinfo {editor}
  {\bibfnamefont {F.}~\bibnamefont {Bach}}\ and\ \bibinfo {editor}
  {\bibfnamefont {D.}~\bibnamefont {Blei}}}\ (\bibinfo  {publisher} {PMLR},\
  \bibinfo {address} {Lille, France},\ \bibinfo {year} {2015})\ pp.\ \bibinfo
  {pages} {1530--1538}\BibitemShut {NoStop}%
\bibitem [{\citenamefont {Papamakarios}\ \emph {et~al.}(2017)\citenamefont
  {Papamakarios}, \citenamefont {Pavlakou},\ and\ \citenamefont
  {Murray}}]{https://doi.org/10.48550/arxiv.1705.07057}%
  \BibitemOpen
  \bibfield  {author} {\bibinfo {author} {\bibfnamefont {G.}~\bibnamefont
  {Papamakarios}}, \bibinfo {author} {\bibfnamefont {T.}~\bibnamefont
  {Pavlakou}},\ and\ \bibinfo {author} {\bibfnamefont {I.}~\bibnamefont
  {Murray}},\ }\href {https://doi.org/10.48550/ARXIV.1705.07057} {\bibinfo
  {title} {Masked autoregressive flow for density estimation}} (\bibinfo {year}
  {2017})\BibitemShut {NoStop}%
\bibitem [{\citenamefont {Huang}\ \emph {et~al.}(2018)\citenamefont {Huang},
  \citenamefont {Krueger}, \citenamefont {Lacoste},\ and\ \citenamefont
  {Courville}}]{https://doi.org/10.48550/arxiv.1804.00779}%
  \BibitemOpen
  \bibfield  {author} {\bibinfo {author} {\bibfnamefont {C.-W.}\ \bibnamefont
  {Huang}}, \bibinfo {author} {\bibfnamefont {D.}~\bibnamefont {Krueger}},
  \bibinfo {author} {\bibfnamefont {A.}~\bibnamefont {Lacoste}},\ and\ \bibinfo
  {author} {\bibfnamefont {A.}~\bibnamefont {Courville}},\ }\href
  {https://doi.org/10.48550/ARXIV.1804.00779} {\bibinfo {title} {Neural
  autoregressive flows}} (\bibinfo {year} {2018})\BibitemShut {NoStop}%
\bibitem [{\citenamefont {Nielsen}\ and\ \citenamefont
  {Winther}(2020)}]{https://doi.org/10.48550/arxiv.2002.02547}%
  \BibitemOpen
  \bibfield  {author} {\bibinfo {author} {\bibfnamefont {D.}~\bibnamefont
  {Nielsen}}\ and\ \bibinfo {author} {\bibfnamefont {O.}~\bibnamefont
  {Winther}},\ }\href {https://doi.org/10.48550/ARXIV.2002.02547} {\bibinfo
  {title} {Closing the dequantization gap: Pixelcnn as a single-layer flow}}
  (\bibinfo {year} {2020})\BibitemShut {NoStop}%
\bibitem [{\citenamefont {Durkan}\ \emph {et~al.}(2019)\citenamefont {Durkan},
  \citenamefont {Bekasov}, \citenamefont {Murray},\ and\ \citenamefont
  {Papamakarios}}]{https://doi.org/10.48550/arxiv.1906.04032}%
  \BibitemOpen
  \bibfield  {author} {\bibinfo {author} {\bibfnamefont {C.}~\bibnamefont
  {Durkan}}, \bibinfo {author} {\bibfnamefont {A.}~\bibnamefont {Bekasov}},
  \bibinfo {author} {\bibfnamefont {I.}~\bibnamefont {Murray}},\ and\ \bibinfo
  {author} {\bibfnamefont {G.}~\bibnamefont {Papamakarios}},\ }\href
  {https://doi.org/10.48550/ARXIV.1906.04032} {\bibinfo {title} {Neural spline
  flows}} (\bibinfo {year} {2019})\BibitemShut {NoStop}%
\bibitem [{\citenamefont {Ott}\ \emph {et~al.}(2021)\citenamefont {Ott},
  \citenamefont {Zache}, \citenamefont {Jendrzejewski},\ and\ \citenamefont
  {Berges}}]{PhysRevLett.127.130504}%
  \BibitemOpen
  \bibfield  {author} {\bibinfo {author} {\bibfnamefont {R.}~\bibnamefont
  {Ott}}, \bibinfo {author} {\bibfnamefont {T.~V.}\ \bibnamefont {Zache}},
  \bibinfo {author} {\bibfnamefont {F.}~\bibnamefont {Jendrzejewski}},\ and\
  \bibinfo {author} {\bibfnamefont {J.}~\bibnamefont {Berges}},\ }\bibfield
  {title} {\bibinfo {title} {Scalable cold-atom quantum simulator for
  two-dimensional qed},\ }\href
  {https://doi.org/10.1103/PhysRevLett.127.130504} {\bibfield  {journal}
  {\bibinfo  {journal} {Phys. Rev. Lett.}\ }\textbf {\bibinfo {volume} {127}},\
  \bibinfo {pages} {130504} (\bibinfo {year} {2021})}\BibitemShut {NoStop}%
\bibitem [{\citenamefont {Oord}\ \emph {et~al.}(2016)\citenamefont {Oord},
  \citenamefont {Kalchbrenner}, \citenamefont {Vinyals}, \citenamefont
  {Espeholt}, \citenamefont {Graves},\ and\ \citenamefont
  {Kavukcuoglu}}]{https://doi.org/10.48550/arxiv.1606.05328}%
  \BibitemOpen
  \bibfield  {author} {\bibinfo {author} {\bibfnamefont {A.~v.~d.}\
  \bibnamefont {Oord}}, \bibinfo {author} {\bibfnamefont {N.}~\bibnamefont
  {Kalchbrenner}}, \bibinfo {author} {\bibfnamefont {O.}~\bibnamefont
  {Vinyals}}, \bibinfo {author} {\bibfnamefont {L.}~\bibnamefont {Espeholt}},
  \bibinfo {author} {\bibfnamefont {A.}~\bibnamefont {Graves}},\ and\ \bibinfo
  {author} {\bibfnamefont {K.}~\bibnamefont {Kavukcuoglu}},\ }\href
  {https://doi.org/10.48550/ARXIV.1606.05328} {\bibinfo {title} {Conditional
  image generation with pixelcnn decoders}} (\bibinfo {year}
  {2016})\BibitemShut {NoStop}%
\bibitem [{\citenamefont {Paszke}\ \emph {et~al.}(2019)\citenamefont {Paszke},
  \citenamefont {Gross}, \citenamefont {Massa}, \citenamefont {Lerer},
  \citenamefont {Bradbury}, \citenamefont {Chanan}, \citenamefont {Killeen},
  \citenamefont {Lin}, \citenamefont {Gimelshein}, \citenamefont {Antiga},
  \citenamefont {Desmaison}, \citenamefont {Kopf}, \citenamefont {Yang},
  \citenamefont {DeVito}, \citenamefont {Raison}, \citenamefont {Tejani},
  \citenamefont {Chilamkurthy}, \citenamefont {Steiner}, \citenamefont {Fang},
  \citenamefont {Bai},\ and\ \citenamefont {Chintala}}]{NEURIPS2019_9015}%
  \BibitemOpen
  \bibfield  {author} {\bibinfo {author} {\bibfnamefont {A.}~\bibnamefont
  {Paszke}}, \bibinfo {author} {\bibfnamefont {S.}~\bibnamefont {Gross}},
  \bibinfo {author} {\bibfnamefont {F.}~\bibnamefont {Massa}}, \bibinfo
  {author} {\bibfnamefont {A.}~\bibnamefont {Lerer}}, \bibinfo {author}
  {\bibfnamefont {J.}~\bibnamefont {Bradbury}}, \bibinfo {author}
  {\bibfnamefont {G.}~\bibnamefont {Chanan}}, \bibinfo {author} {\bibfnamefont
  {T.}~\bibnamefont {Killeen}}, \bibinfo {author} {\bibfnamefont
  {Z.}~\bibnamefont {Lin}}, \bibinfo {author} {\bibfnamefont {N.}~\bibnamefont
  {Gimelshein}}, \bibinfo {author} {\bibfnamefont {L.}~\bibnamefont {Antiga}},
  \bibinfo {author} {\bibfnamefont {A.}~\bibnamefont {Desmaison}}, \bibinfo
  {author} {\bibfnamefont {A.}~\bibnamefont {Kopf}}, \bibinfo {author}
  {\bibfnamefont {E.}~\bibnamefont {Yang}}, \bibinfo {author} {\bibfnamefont
  {Z.}~\bibnamefont {DeVito}}, \bibinfo {author} {\bibfnamefont
  {M.}~\bibnamefont {Raison}}, \bibinfo {author} {\bibfnamefont
  {A.}~\bibnamefont {Tejani}}, \bibinfo {author} {\bibfnamefont
  {S.}~\bibnamefont {Chilamkurthy}}, \bibinfo {author} {\bibfnamefont
  {B.}~\bibnamefont {Steiner}}, \bibinfo {author} {\bibfnamefont
  {L.}~\bibnamefont {Fang}}, \bibinfo {author} {\bibfnamefont {J.}~\bibnamefont
  {Bai}},\ and\ \bibinfo {author} {\bibfnamefont {S.}~\bibnamefont
  {Chintala}},\ }\bibfield  {title} {\bibinfo {title} {Pytorch: An imperative
  style, high-performance deep learning library},\ }in\ \href
  {http://papers.neurips.cc/paper/9015-pytorch-an-imperative-style-high-performance-deep-learning-library.pdf}
  {\emph {\bibinfo {booktitle} {Advances in Neural Information Processing
  Systems 32}}},\ \bibinfo {editor} {edited by\ \bibinfo {editor}
  {\bibfnamefont {H.}~\bibnamefont {Wallach}}, \bibinfo {editor} {\bibfnamefont
  {H.}~\bibnamefont {Larochelle}}, \bibinfo {editor} {\bibfnamefont
  {A.}~\bibnamefont {Beygelzimer}}, \bibinfo {editor} {\bibfnamefont
  {F.}~\bibnamefont {d\textquotesingle Alch\'{e}-Buc}}, \bibinfo {editor}
  {\bibfnamefont {E.}~\bibnamefont {Fox}},\ and\ \bibinfo {editor}
  {\bibfnamefont {R.}~\bibnamefont {Garnett}}}\ (\bibinfo  {publisher} {Curran
  Associates, Inc.},\ \bibinfo {year} {2019})\ pp.\ \bibinfo {pages}
  {8024--8035}\BibitemShut {NoStop}%
\bibitem [{\citenamefont {Kingma}\ and\ \citenamefont
  {Ba}(2014)}]{kingma2014method}%
  \BibitemOpen
  \bibfield  {author} {\bibinfo {author} {\bibfnamefont {D.~P.}\ \bibnamefont
  {Kingma}}\ and\ \bibinfo {author} {\bibfnamefont {J.}~\bibnamefont {Ba}},\
  }\href {http://arxiv.org/abs/1412.6980} {\bibinfo {title} {Adam: A method for
  stochastic optimization}} (\bibinfo {year} {2014}),\ \bibinfo {note} {cite
  arxiv:1412.6980Comment: Published as a conference paper at the 3rd
  International Conference for Learning Representations, San Diego,
  2015}\BibitemShut {NoStop}%
\end{thebibliography}%

\clearpage

\begin{center}
	\noindent\textbf{Appendix}
	\bigskip
		
	\noindent\textbf{\large{Gauge Invariant Autoregressive Flow Neural Networks for Quantum Electrodynamics at Finite Density}}
		
\end{center}

\appendix

\renewcommand\thefigure{A\arabic{figure}}  
\renewcommand\thetable{A\arabic{table}}  
\setcounter{figure}{0}  
\setcounter{table}{0}

\section{Details of Neural Network Architectures} \label{app:dnn}

In this section, we describe the details of the neural network architectures. 

\begin{figure}[h!]
    \centering
    \includegraphics[width=0.85\linewidth]{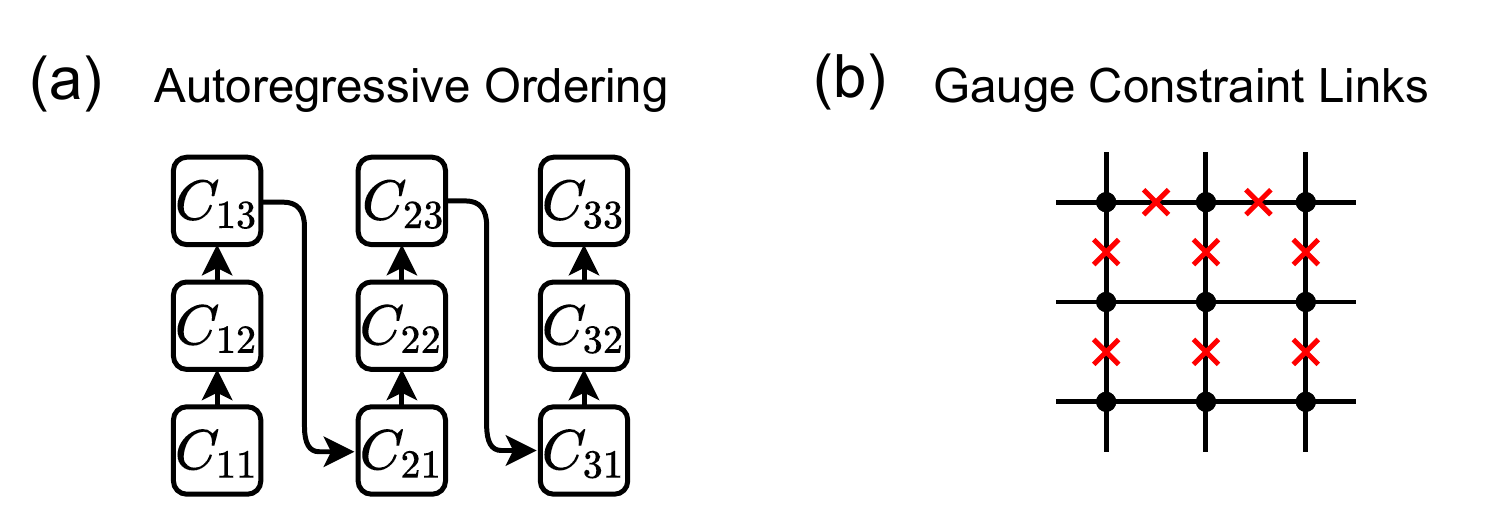}
    \caption{(a) Autoregressive ordering of the neural network. (b) Gauge constrained links. The crosses correspond to links where the gauge constrains the value of the field completely. 
    }
    \label{fig:order_gauge}
\end{figure}

Since we are constructing the probability distribution autoregressively, we need to choose a certain order of the neural network. In this work, we choose an zig-zag order across unit cells $C_{ij}$ as described in Fig.~\ref{fig:order_gauge} (a). Within each unit cell, the ordering is $f_{ij}\rightarrow E_{ij}^{(x)} \rightarrow E_{ij}^{(y)}$.

In addition, we impose the gauge symmetry to construct a gauge-invariant neural network. The gauge symmetry says the sum of gauge field around each vertex should be equal to the charge at the vertex. Following the autoregressive ordering, each time we encounter a constrained gauge field that is determined by the other three fields and the charge at a specific vertex, we don't run it though the neural network, but define $p(E_{ij}^{(x) \text{ or } (y)}) = 1$ if it satisfies the gauge symmetry and $p(E_{ij}^{(x) \text{ or } (y)}) = 0$ if it does not satisfy the gauge symmetry. The constrained gauge fields are illustrated in Fig.~\ref{fig:order_gauge} (b).

\begin{figure}[h!]
    \centering
    \includegraphics[width=\linewidth]{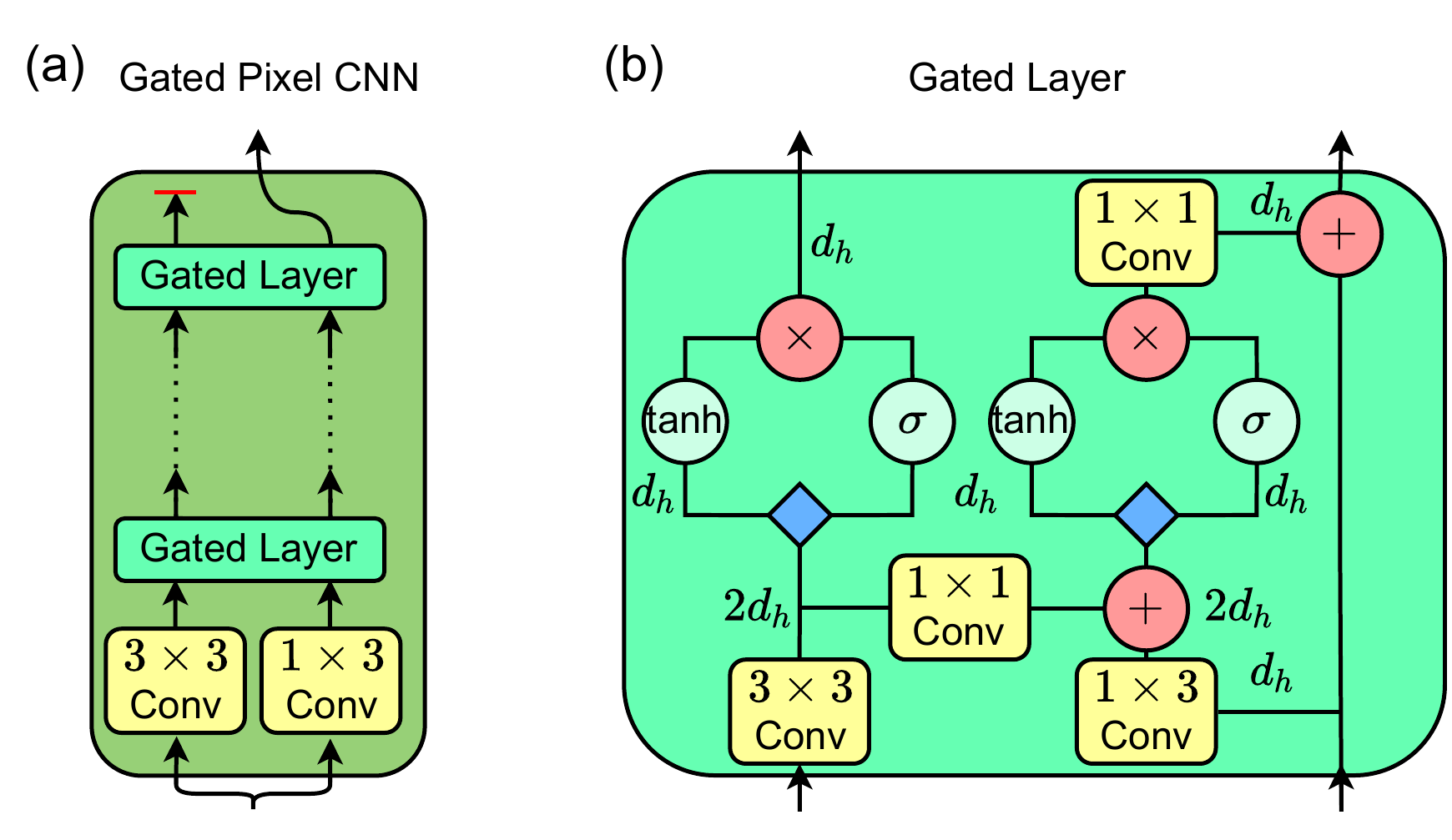}
    \caption{(a) Setup of gated pixelCNN and (b) setup of the gated layer as in Ref.~ \onlinecite{https://doi.org/10.48550/arxiv.1606.05328}.
    }
    \label{fig:gated_pixel_cnn_figure}
\end{figure}

Now, we go to the details of the gated pixelCNN. The gated pixelCNN is shown in Fig~\ref{fig:gated_pixel_cnn_figure} (a), with the gated layer defined in Fig~\ref{fig:gated_pixel_cnn_figure} (b). Here, we follow the implementation in Ref.~\onlinecite{https://doi.org/10.48550/arxiv.1606.05328}. The input of the gated pixelCNN is first put into two different convolutions to obtain the input of the gated layers. For the last gated layer, we ignore the left output and put the right output through two linear layers to obtain the conditional hidden vectors for fermions and electric fields. Inside each gated layer, there are two branches. This specific design allows the gated pixelCNN to capture a large area of perception avoiding the blind spots.

\begin{figure}[h!]
    \centering
    \includegraphics[width=\linewidth]{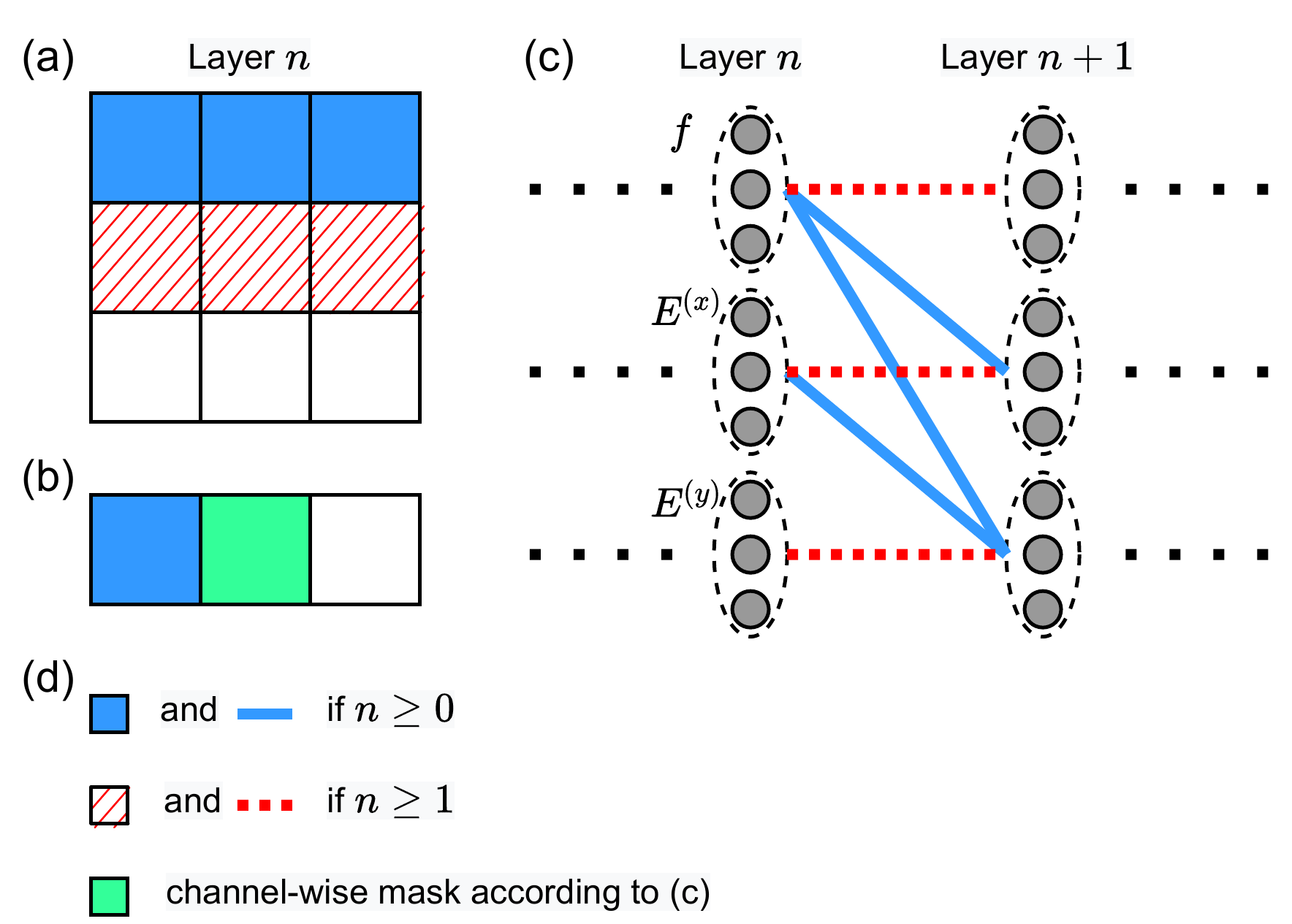}
    \caption{Illustration of the convolution mask. (a) Convolution mask for left branch. We keep the weight in the blue boxes for all the layers and keep the weight in the red boxes for layers other than the zeroth layer. (b) Convolution mask for right branch. We keep the weight in the blue box for all the layers and keep the weight in the green box with a channel-wise mask according to Fig.~(c). (c) Channel wise mask for Fig.~(b). We keep the blue connections for all layers and keep the red connections for layers other than the zeroth layer. 
    }
    \label{fig:mask}
\end{figure}

As described in the main text, when calculating the conditional probability distributions, we don't input $C_{\le ij}$ for each $(i, j)$, but apply appropriate masks to each convolution filters in the gated pixelCNN. The masks are described in Fig.~\ref{fig:mask}. Here, the left branch uses the mask in Fig.~\ref{fig:mask} (a) and the right branch uses the mask in Fig.~\ref{fig:mask} (b). For mask (a), since we use the channel dimension to encode the fermions and electric fields within each unit cell, an additional channel-wise mask is applied for the center point. Notice that the mask on layer zero (the layer before the gated layer) is slightly different from the other layers as Fig.~\ref{fig:mask} shows. These masks ensures the neural network always perceives the correct portion of the $C_{ij}$'s to ensure the autoregressive structure of the neural network while allowing as a large perception area as possible.

\begin{figure}[h!]
    \centering
    \includegraphics[width=\linewidth]{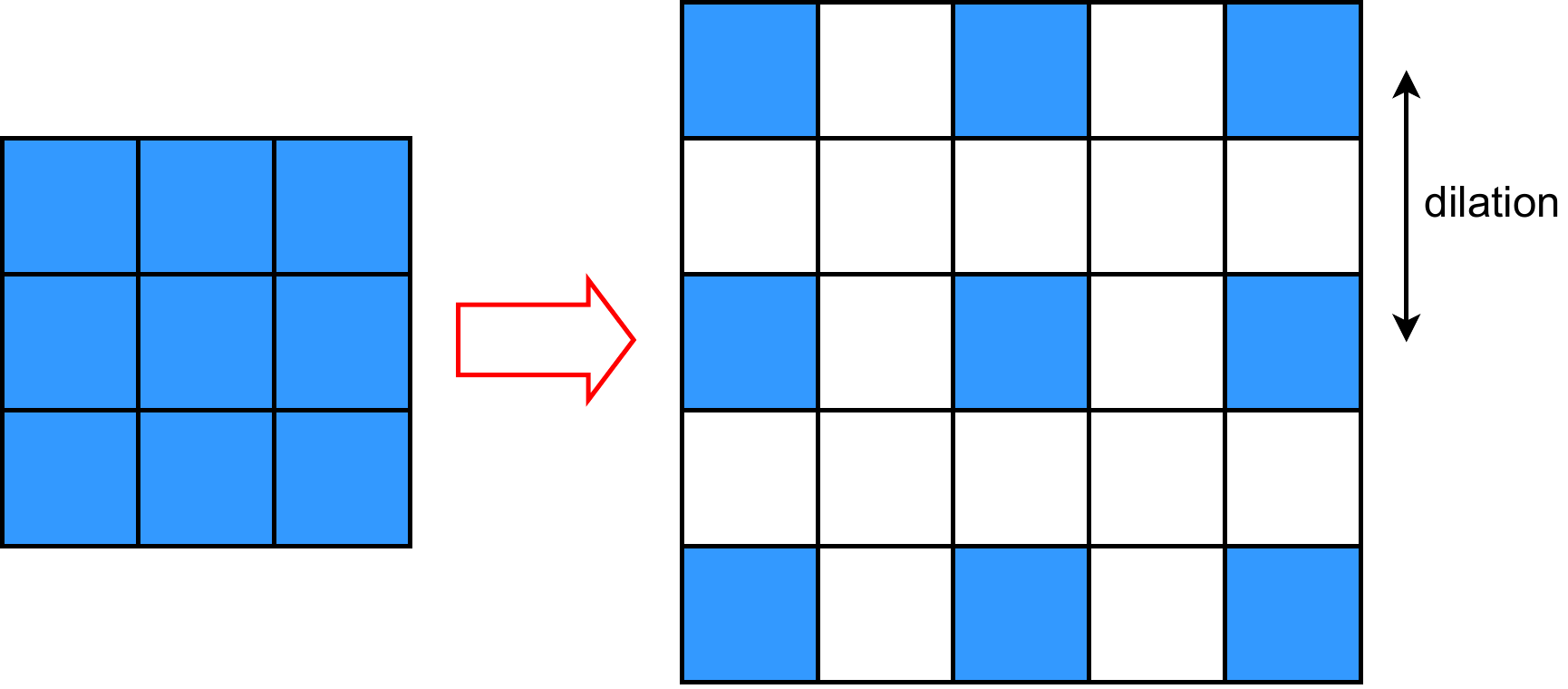}
    \caption{Illustration of the dilation of the convolution filter. The weights of the filter is only applied to the blue pixels.
    }
    \label{fig:dilation}
\end{figure}

To capture long range correlations, we further apply dilations to the convolutional filter as described by Fig.~\ref{fig:dilation}. 

\section{Details of Hyperparameters and Optimization} \label{app:hyper_opt}
In this work, the gated PixelCNN has a hidden dimension of 36 and consists of 7 gated layers. The gated layers have dilations 1, 2, 1, 4, 1, 2, and 1 for layers 1 through 7 respectively. The CNN for Slater determinant has a hidden dimension of 24 and 7 layers with the same dilation in each layer as the gated pixelCNN.

The neural network parameters are initialized randomly with PyTorch's \cite{NEURIPS2019_9015} default initialization scheme. The prior distribution of the autoregressive flow is chosen to be the standard normal distribution with a mean of 0 and a standard deviation of 1.  We use PyTorch's automatic differentiation to compute the energy derivative with respect to the parameters. For the neural network backflow Slater determinant (Sec.~\ref{sec:phase}), we implement a custom backward function by taking the imaginary part of the derivative of log\_determinant function.

For optimization, we use Adam optimizer \cite{kingma2014method} with an initial learning rate of $0.003$. The learning rate is halved at iterations 800, 1200, 1800, and 2500. For the phase transition result (Sec.~\ref{sec:phase} and Sec.~\ref{sec:topo_phase}), we use the transfer learning technique, where we first train the neural network on small systems before moving on to large systems. Previous work \cite{https://doi.org/10.48550/arxiv.2101.07243} suggests that using transfer learning allows the neural network to be trained faster. More specifically, we start with a random initialization of the neural network and train it on the $4\times4$ systems for 4000 iterations; then, we transfer the neural network to $8\times8$ system for 2700 iterations. For results with large system sizes, we continue to train the neural network on $10\times10$ system for 2000 iterations and on $12\times12$ system for another 1600 iterations. For the string breaking study (\ref{sec:string}), however, transfer learning is not applicable so we train a random initialization of the neural network for 2500 iterations.

\section{Sampling from the Wave Function}\label{app:sampling}

In this section, we describe how to sample from the wave function $(\bm f, \bm E) \sim \abs{\psi(\bm f, \bm E)}^2=p(\bm f, \bm E)$. To sample from the wave function, we don't need the phase of the wave function (i.e. the neural network backflow Slater Determinant), so we only focus on the probability part of the neural network. 

Suppose an overall probability distribution is defined as the product of conditional distributions as
\begin{equation}
    p(\bm x) = p(x_1) p(x_2|x_1) p(x_3|x_1, x_2) \dots p(x_N|\bm{x}_{<N}),
\end{equation}
then, we can sample from this probability distribution sequentially as 
\begin{enumerate}
    \item Sample $x_1$ from $p(x_1)$.
    \item Sample $x_2$ from $p(x_2 | x_1)$ by plugging in the sampled $x_1$.
    \item Sample $x_i$ from $p(x_i | \bm{x}_{<i})$ by plugging in all the sampled $\bm{x}_{<i}$ until all the $x_i$'s are sampled.
\end{enumerate}
Because our probability distribution is defined from conditional probability distributions $p_{\bm\theta_A}(f_{ij}|\bm f_{<ij}, \bm E_{<ij}^{(x)}, \bm E_{<ij}^{(y)})$, $p_{\bm\theta_A}(E_{ij}^{(x)}|\bm f_{\le ij}, \bm E_{<ij}^{(x)}, \bm E_{<ij}^{(y)})$, and $p_{\bm\theta_A}(E_{ij}^{(y)}|\bm f_{\le ij}, \bm E_{\le ij}^{(x)}, \bm E_{<ij}^{(y)})$, we can use the same procedure to sample from the probability distribution by following the conditioning order as defined in Fig.~\ref{fig:order_gauge} (a).

The conditional probability distribution for fermions is just a binary distribution $p(f=0)$ and $p(f=1)$, which is easy to sample from. To sample from the conditional probability distribution for gauge fields, we first sample from the prior distribution $z\sim \pi(z)$. Then, apply the transformation function $x = f(z)$. Afterwards, we round $x$ to the nearest integer $E = \text{round}(x)$. It is easy to verify that this procedure generates a sample from the conditional distribution. 

Although the samples are generated sequentially for each site, it can be made parallel and independent for a large sample size. In addition, it does not suffer from the auto correlation time of MCMC sampling. Therefore, the autoregressive sampling method is very efficient.

\section{Pure Gauge Theory}\label{app:pure}

\begin{figure}[h!]
    \centering
    \includegraphics[width=\linewidth]{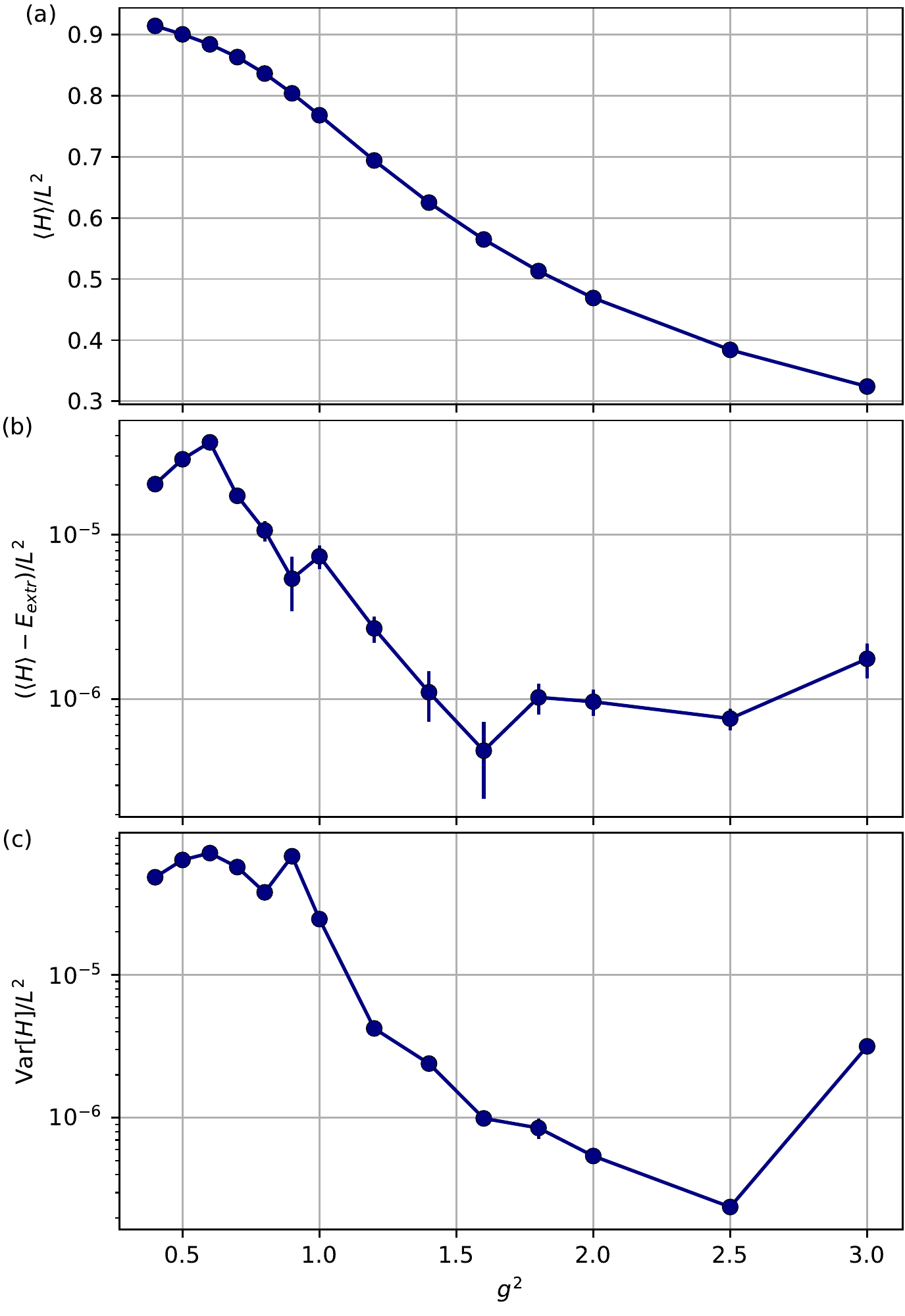}
    \caption{Pure gauge theory calculations of system size $12 \times 12$  (a) Energy per plaquette, (b) energy error per plaquette where we use the variance extrapolation to estimate the energy of the true ground state , and (c) variance per plaquette. 
    }
    \label{fig:energy_vs_g}
\end{figure}

In the pure gauge theory, the $H_M$ term and $H_K$ term are both 0, and we use the convention $g_E^2 = 1/g_B^2 = g^2$. We test our neural network on a one plaquette system, and find the error in energy is consistently within $\num{1e-5}$ compared to the exact diagonalization with gauge field cutoff 1000. We then test our neural network on $12\times12$ system. The results are shown in Fig.~\ref{fig:energy_vs_g}. We perform variance extrapolation to estimate the energy of the true ground state and calculate the per plaquette error of the neural network energy in Fig.~\ref{fig:energy_vs_g} (b). It is clear that the per plaquette error (below $\num{3e-5}$) agrees with the error we obtained for the one plaquette system.

\begin{figure}[h!]
    \centering
    \includegraphics[width=\linewidth]{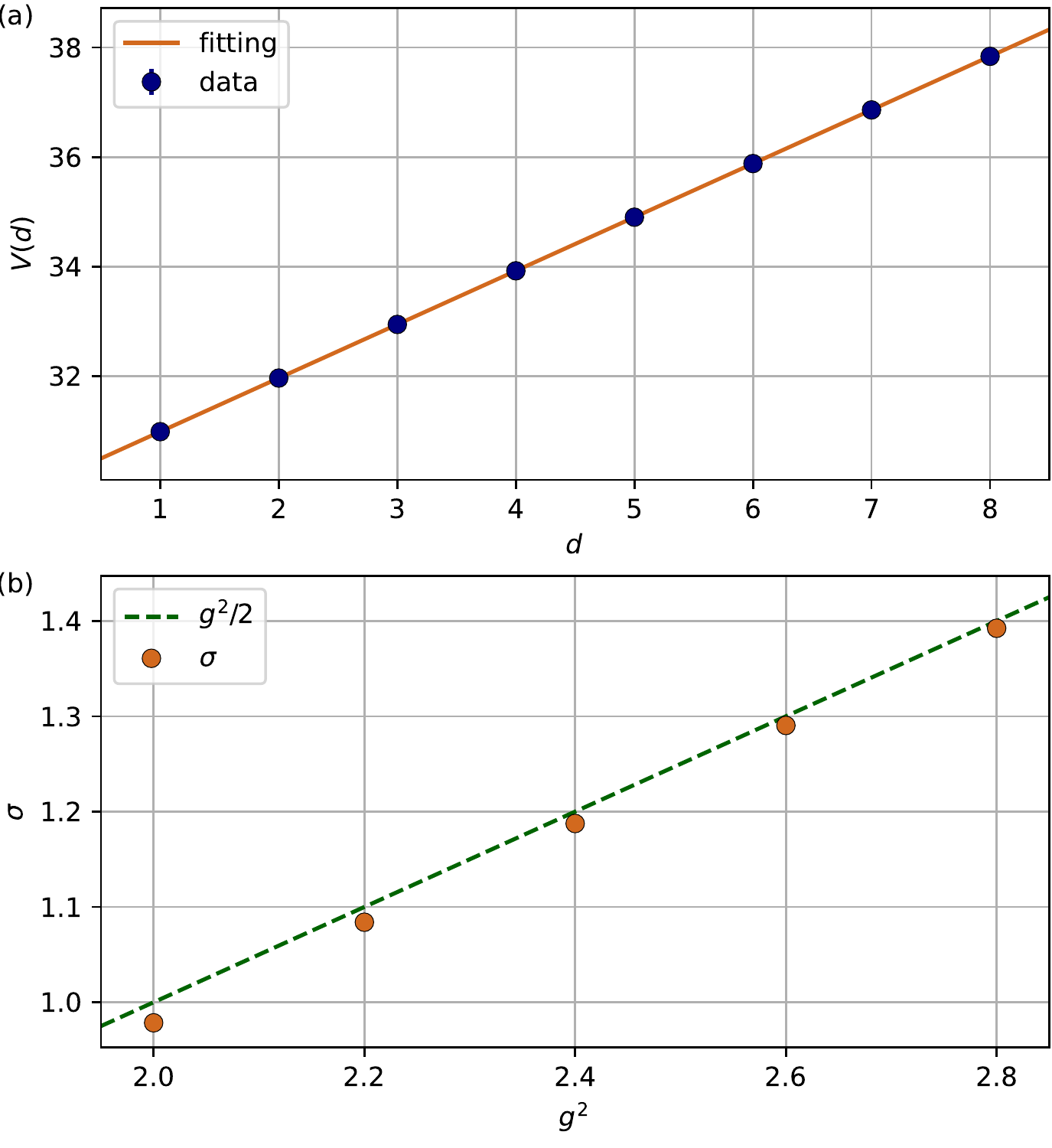}
    \caption{Static charge energy for pure gauge theory and the string tension $\sigma$ fitting. (a) Energy vs distance between static charges for $g^2=2$ (blue points). The orange line is fitted with $V(d)=\sigma d + a \log d + c$. (b) $\sigma$ vs $g^2$. $\sigma$ approaches $g^2 /2$ (green dash line) for large $g$ which is consistent with the theoretical prediction.
    }
    \label{fig:static_charge}
\end{figure}

In addition, we study the case with static charges. The results are shown in Fig.~\ref{fig:static_charge}. Here, we put two static charges at a distance $d$ apart and fit the ground state energy as a function of $d$ as $V(d)=\sigma d + a \log d + c$ (Fig.~\ref{fig:static_charge} (a)). It is predicted that for large $g^2$, the fitting parameter $\sigma$ should approach $g^2/2$. In, fig.~\ref{fig:static_charge} (b) we show that our neural network results agree with the theoretical prediction.

\begin{figure}[h!]
    \centering
    \includegraphics[width=\linewidth]{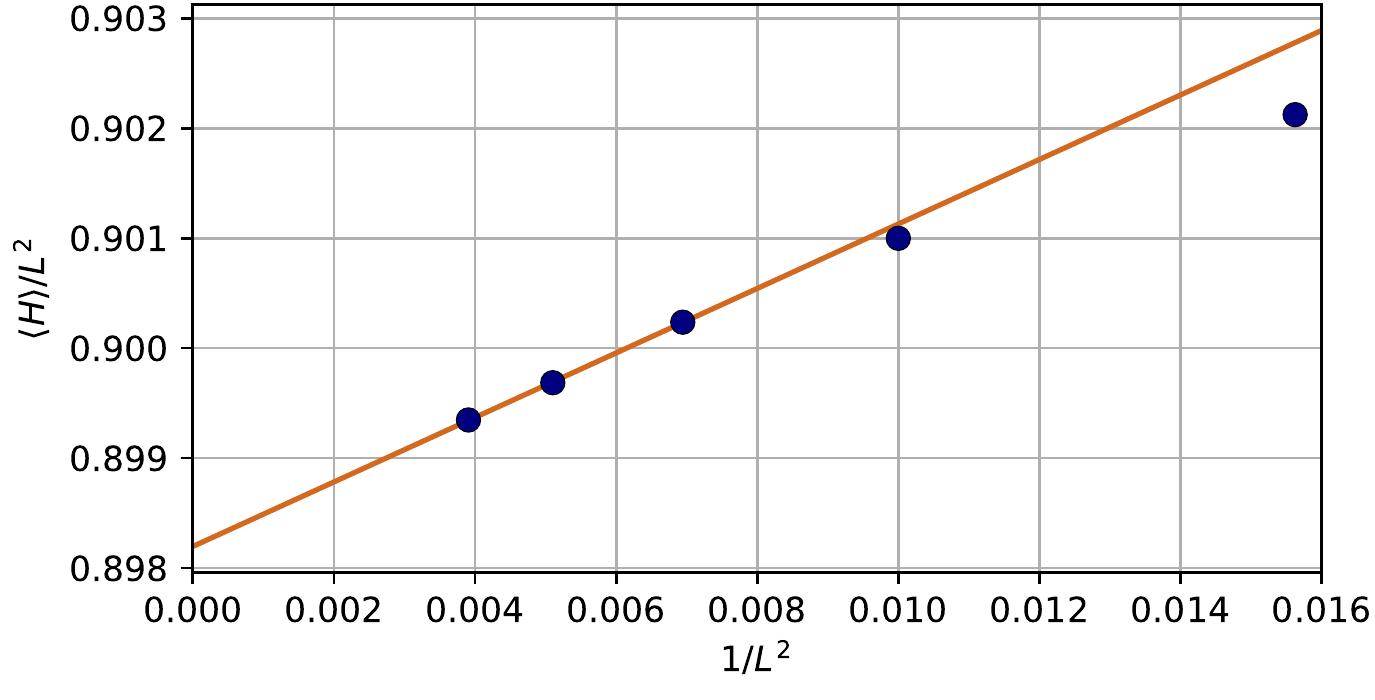}
    \caption{Infinite system size energy extrapolation for $g^2=1/2$. Here, we use a linear fit for the three points with largest system sizes to estimate the infinite system size energy. The extrapolated energy per site is 0.89819, which is consistent with the result of 0.89815 digitized from Ref.~\onlinecite{PhysRevResearch.2.043145} (with periodic boundary condition) with a relative difference of $4\times 10^{-5}$
    }
    \label{fig:extrapolation}
\end{figure}

Moreover, we simulate the ground state energy for $g^2=1/2$ for different system sizes, and extrapolate the ground state energy to the infinite system size limit in Fig.~\ref{fig:extrapolation}. We find that our infinite system size extrapolation from open boundary condition agrees with the extrapolation of Ref.~\onlinecite{PhysRevResearch.2.043145} from periodic boundary conditions up to 4 decimal places.

\section{Additional Data for Gauge Field Coupled to Fermions}

\begin{figure}[h!]
    \centering
    \includegraphics[width=\linewidth]{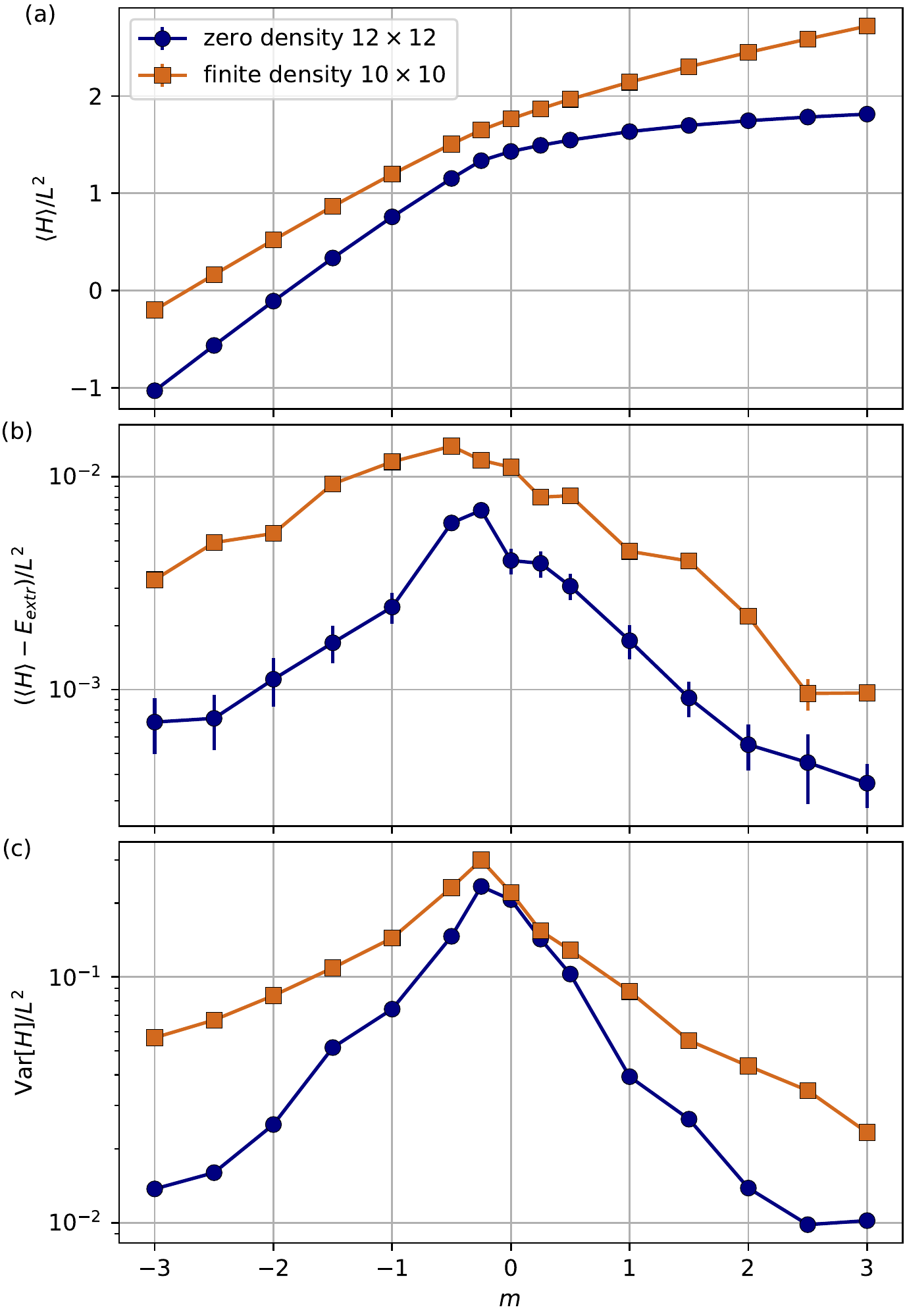}
    \caption{Zero density and finite density calculation. (a) Energy per site, (b) energy error per site, and (c) variance per site. We use the variance extrapolation to estimate the energy of the true ground state and use it to calculate the error for the neural network. The error is smaller than $2 \times 10^{-2}$ for all $m$'s. 
    }
    \label{fig:fermion}
\end{figure}

For the gauge field coupled to fermions, we test our neural network on one plaquette system, and find the error in energy to be consistently within $\num{1.5e-3}$ compared to exact diagonalization with gauge field cutoff of $\pm 25$. In addition, in Fig.~\ref{fig:fermion}, we show the energy per site, energy error per site, and variance per site for the $12\times12$ with zero density, and $10\times10$ system with finite density. We use variance extrapolation to estimate the true ground state energy and calculate the energy error per site. 

\section{Effect of Continuous Variable}\label{app:cv}

\begin{figure}[ht!]
    \centering
    \includegraphics[width=\linewidth]{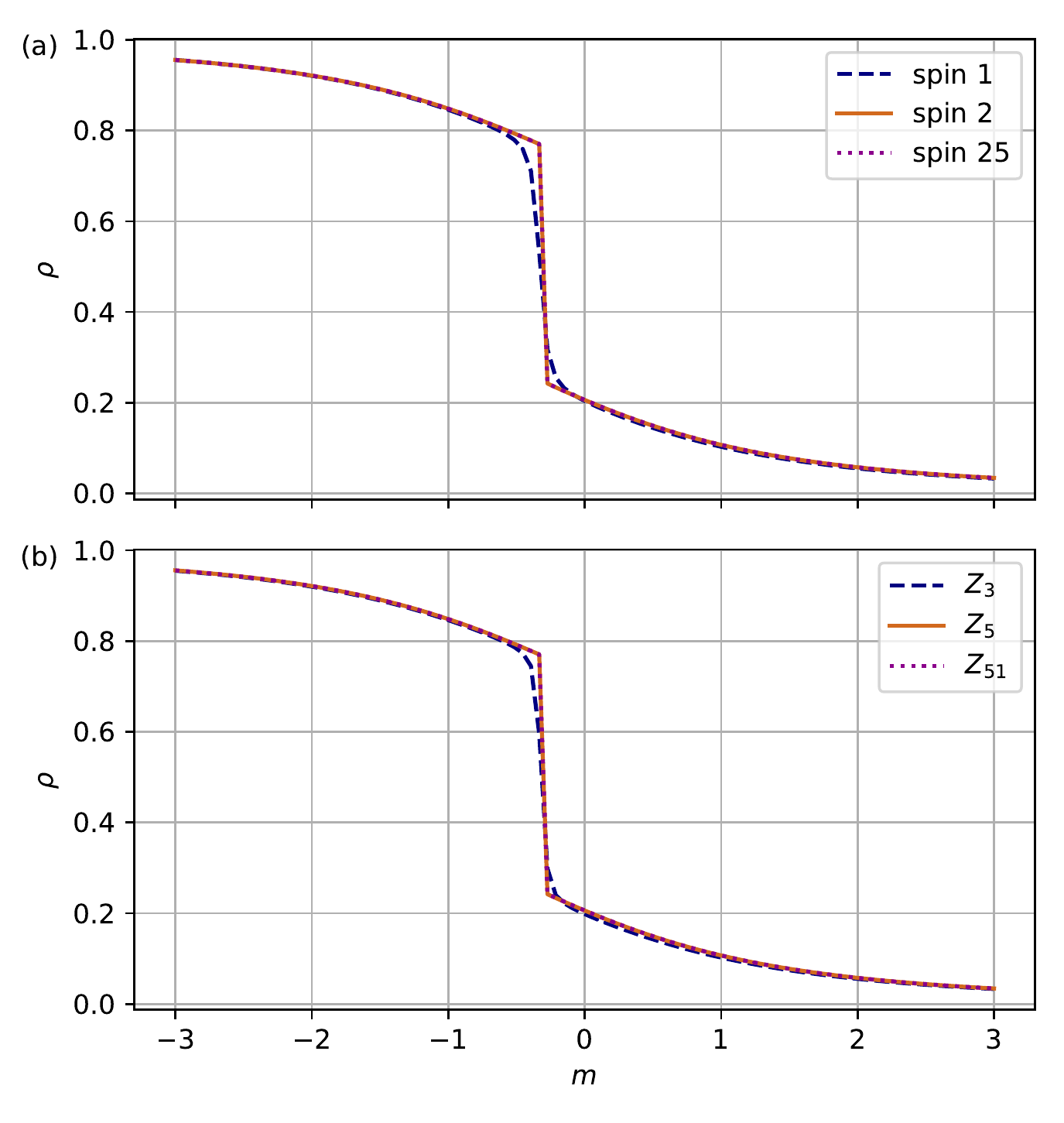}
    \caption{Particle density at zero density with $g_E^2=2$, $g_B^2=4$, $\kappa=1$ and different $m$ for $2\times 2$ using exact diagonalation for (a) quantum link model representation and (b) $\mathbb{Z}_N$ representation.
    }
    \label{fig:cv_charge_density}
\end{figure}

\begin{figure}
    \centering
    \includegraphics[width=\linewidth]{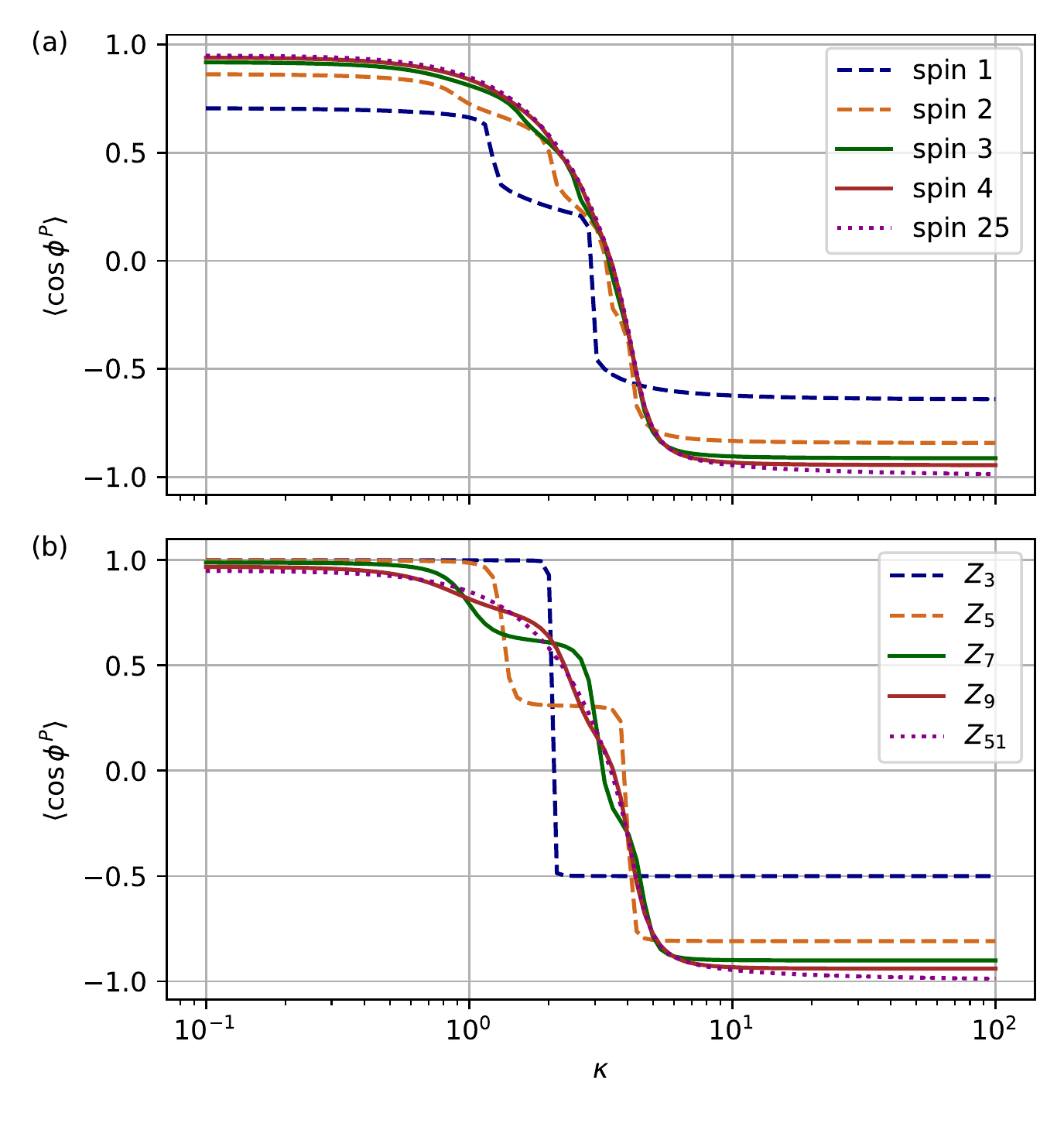}
    \caption{Average plaquette value at zero density with $g_E^2=0.01$, $g_B^2=1$ and $m=0.01$ for $2\times 2$ using exact diagonalization for (a) quantum link model representation and (b) $\mathbb{Z}_N$ representation.
    }
    \label{fig:cv_plaq}
\end{figure}

We investigate the effect of continuous variable on the order of phase transition by exact diagonalization on $2 \times 2$ lattice. We approach the $U(1)$ model by considering both the $\mathbb{Z}_N$ and the quantum link model(QLM). When $N \rightarrow \infty$ in $\mathbb{Z}_N$ and the spin cutoff in QLM goes to infinity, both models converge to the $U(1)$ model. Fig.~\ref{fig:cv_charge_density} shows that as one approaches the continuous variable the transition becomes sharper between charge crystal and vacuum, suggesting that it might go from second to order to first order. Fig.~\ref{fig:cv_plaq} shows that the magnetic transition becomes smoother in the continuous variable limit, suggesting that it might go from first order to second or higher order.

\section{Analytic Calculation in The Absence of Electric Term} \label{app:analytic}

\begin{figure}
    \centering
    \includegraphics{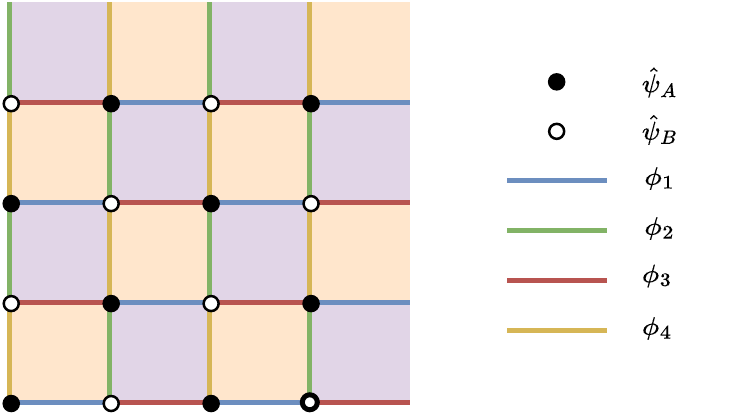}
    \caption{Illustration of the staggered configuration. A staggered flux pattern is formed by the staggered fermions.}
    \label{fig:staggere_illustra}
\end{figure}

\begin{figure}[h!]
    \centering
    \includegraphics[width=\linewidth]{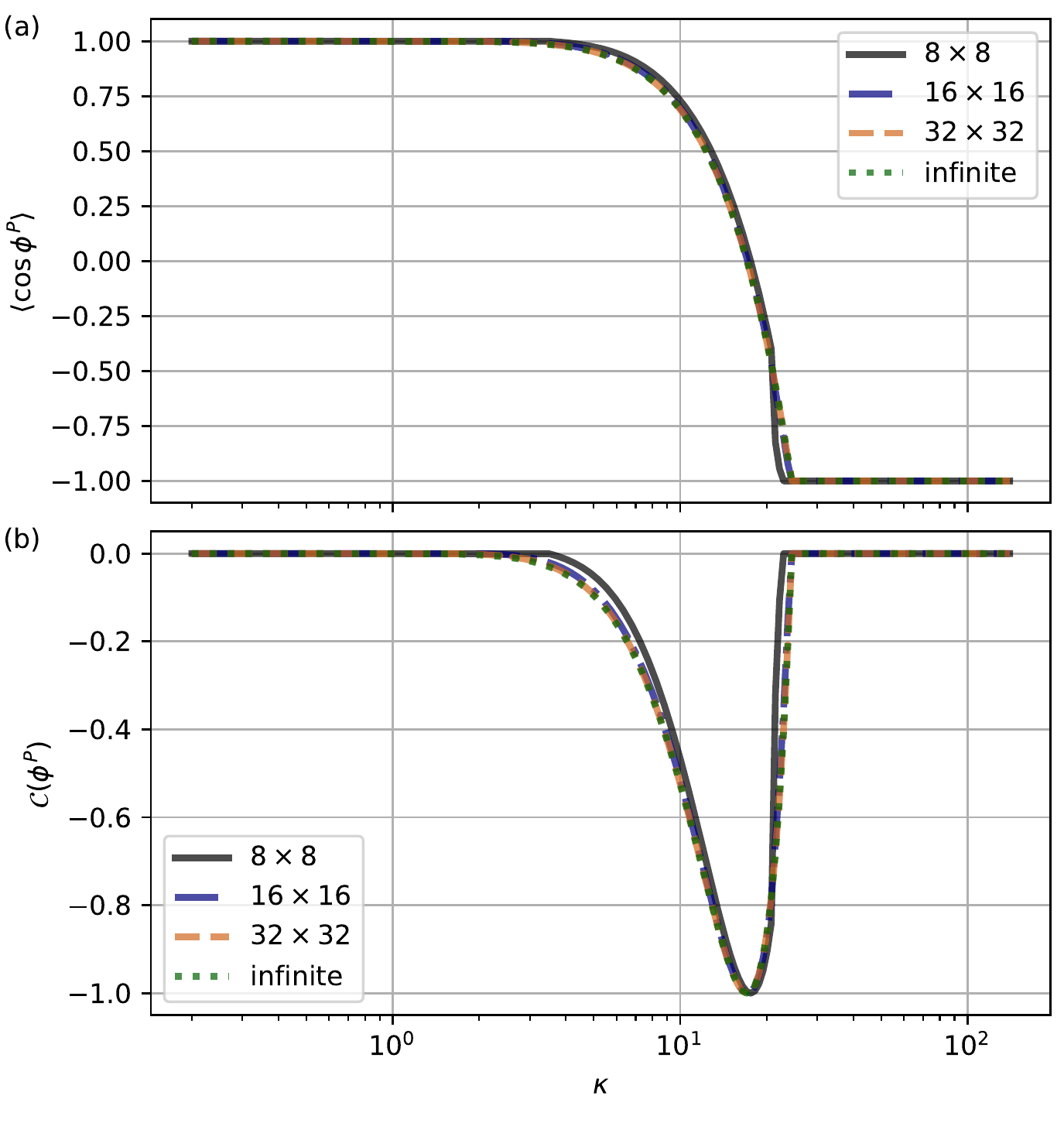}
    \caption{Magnetic phase transition at zero density with $g_E^2=0$, $g_B^2=1$ and $m=0.01$ for different system sizes with periodic boundary condition using the analytical method. The infinite system size result is calculated by numerically integrating over the Brillouin zone. (a) Average plaquette value as a function of $\kappa$. (b) Average nearest neighbor plaquette correlation as a function of $\kappa$.}
    \label{fig:topo_analytical}
\end{figure}

Here, we describe the analytical calculation of the Kogut-Susskind Hamiltonian in the absence of electric term ($g_E^2 = 0$). 

As described in \ref{sec:topo_phase}, the Hamitlonian of the system becomes 

\begin{equation}
    \hat H = \hat H_B + \hat H_M + \hat H_K,
\end{equation}
with the Gauss's law term.
Notice that all the terms commutes and they all commutes with the Gauss's law. This allows us to work with the eigenbasis of $\hat H_B$ (eigenbasis of $\hat U_{i, j}^{(w)}$ with eigenvalues $e^{i\phi_{i, j}^{(w)}}$) as
\begin{equation}
    \begin{aligned}
    \hat H_B &= -g_B^2\sum_{i,j} \cos{\phi^{P}_{i,j}}, \\
    \hat H_M &= m \sum_{i, j} (-1)^{i+j} \hat \psi_{i,j}^\dagger \hat \psi_{i,j}, \\
    \hat H_K &= -\kappa \sum_{i, j} \left[\hat \psi_{i,j}^\dagger e^{i \phi_{i,j}^{(x)}} \hat \psi_{i+1, j} + \hat \psi_{i,j}^\dagger e^{i \phi_{i,j}^{(y)}}   \hat \psi_{i, j+1} + \text{H.C.}\right],
    \end{aligned}
\end{equation}
where $\phi^{P}_{i,j} = \phi_{i,j}^{(x)}+\phi_{i+1,j}^{(y)}-\phi_{i,j+1}^{(x)}-\phi_{i,j}^{(y)}$ and $\phi_{i, j}^{(w)}\in (-\pi, \pi]$ is a $U(1)$ variable corresponds to the eigenvalues of $\hat U_{i,j}^{(w)}$. 
Notice that the definition of the Hamiltonian differs by some identity operator from the definition used in \ref{eq:ham} for a simpler analytical calculation.

In this basis, the Gauss's law term becomes 
\begin{equation}
    \hat A_{i,j}(\varepsilon)\ket{\psi} = \ket{\psi} \ \ \ \ \ \ \ \ \ \ \forall i, j, \varepsilon
\end{equation}
with 
\begin{equation}
\begin{aligned}
    \hat A_{i,j}(\varepsilon) =& \ket{\dots, \phi_{i,j}^{(x)}+\varepsilon, \phi_{i,j}^{(y)} + \varepsilon, \phi_{i-1,j}^{(x)}-\varepsilon, \phi_{i,j-1}^{(y)}-\varepsilon, \dots}\\
    &\bra{\dots, \phi_{i,j}^{(x)}, \phi_{i,j}^{(y)}, \phi_{i-1,j}^{(x)},\phi_{i,j-1}^{(y)}, \dots} e^{-i \varepsilon \hat q_{i,j}}
\end{aligned}
\end{equation}

We will try to solve the system without the Gauss's law first and then project it to the state that obeys the Gauss's law.

For periodic boundary condition, the total magnetic flux is zero. Due to the staggered fermion, the system permits a staggered flux with $\phi^{P}_{i, j} = - \phi^{P}_{i+1, j} = -\phi^{P}_{i, j+1}$. The most general configurations of the links are given by Fig.~\ref{fig:staggere_illustra}.

Then, the $\hat H_M$ term and the $\hat H_K$ terms forms a tight binding model with the hopping phase controlled by the variables on the links. 

Due to the staggered flux and the staggered mass, we need to choose a unit cell containing two adjacent sites. Here, we choose the two sites horizontally and label the sites A and B respectively. Then, the unit cells are spanned by the vectors $\hat h = [2, 0]^T$ and $\hat d = [1, 1]^T$.

Now, we rewrite the Hamiltonian as 
\begin{equation}
    \begin{aligned}
    \hat H_M &= m \sum_{\bm r} \left[\hat \psi_{\bm r, A}^\dagger \hat \psi_{\bm r, A} - \hat \psi_{\bm r, B}^\dagger \hat \psi_{\bm r, B}\right], \\
    \hat H_K &= -\kappa \sum_{\bm r} \Big[e^{i \phi_1}\hat \psi_{\bm r, B}^\dagger\hat \psi_{\bm r, A}  +  e^{-i \phi_3}\hat \psi_{\bm r-\hat h, B}^\dagger\hat \psi_{\bm r, A} \\
    & \ \ \ \ \ \ \ \ + e^{i \phi_4}\hat \psi_{\bm r - \hat h + \hat d, B}^\dagger\hat \psi_{\bm r, A}  + e^{-i \phi_2}\hat \psi_{\bm r - \hat d, B}^\dagger\hat \psi_{\bm r, A} \\
    & \ \ \ \ \ \ \ \ + \text{H.C.} \Big],
    \end{aligned}
\end{equation}
where $\bm r$ spans all the unit cells and $\phi = \phi_{i=1, j=1}^{P}$.

We can Fourier transform the Hamiltonian and obtain
\begin{equation}
    \begin{aligned}
    \hat H_M &= m \sum_{\bm k} \left[ \tilde \psi_{\bm k, A}^\dagger \tilde \psi_{\bm k, A} + \tilde \psi_{\bm k, B}^\dagger \tilde \psi_{\bm k, B}  \right]  \\
    \hat H_K &= -\kappa \sum_{\bm k} \tilde \psi_{\bm k, B}^\dagger \tilde \psi_{\bm k, A} \Big[e^{i \phi_1} +e^{i\phi_4 + i k_x - i k_y}\\
    & \ \ \ \ \ \ \ \ \ \ \ \ \ + e^{-i \phi_3 + 2 i k_x} + e^{-i\phi_2 + i k_x + i k_y} + \text{H.C.}\Big],
    \end{aligned}
\end{equation}
where $\bm k = (k_x, k_y)$ spans all the reciprocal space.
Then, we can solve for the energies 

\begin{align}
\begin{split}
    E^{\pm}&(k_x, k_y, \phi, \phi_x, \phi_y) = \pm2 \kappa \\
    &\Bigg[\cos^2 \left(k_x-\frac{\phi_x}{2}\right)+
    \cos^2 \left(k_y-\frac{\phi_y}{2}\right)\\
    &+2\cos \frac{\phi}{2}\cos \left(k_x-\frac{\phi_x}{2}\right) \cos \left(k_y-\frac{\phi_y}{2}\right)+\frac{m^2}{4\kappa^2}\Bigg]^{1/2},
\end{split}
\end{align}
where $\phi_x=\phi_1+\phi_3$, $\phi_y=\phi_2+\phi_4$, and $\phi = \phi_1 + \phi_2 - \phi_3 - \phi_4$ the plaquette flux.

The reciprocal lattice vectors are given by $\tilde h = [\pi, -\pi]^T$ and $\tilde d = [0, 2\pi]^T$.
Notice that the Brillouin zone, only spans half of the $k_x, k_y \in [0, 2\pi)$, and therefore we have

\begin{equation}
E_{\text{fermion}} = \sum_{i=0}^{L/2}\sum_{j=0}^L E^{-}\left(\frac{2 \pi i}{ L}, \frac{2 \pi j}{ L}, \phi, \phi_x, \phi_y\right).
\end{equation}
Notice that the $\phi_x$ and $\phi_y$ terms are just shifts of the $k_x$ and $k_y$ term, which has no effect and can be set to 0 in the thermodynamics limit when we need to integrate over the whole Brillouin zone.

The magnetic energy is just given by 
\begin{equation}
    E_{\text{magnetic}} = -g_B^2 L^2 \cos \phi.
\end{equation} 
Thus, for each $g_B^2$, $m$, and $\kappa$, we can solve for the $\phi^*$ that minimizes the energy. This analytical calculation has been confirmed to agree with a direct numerical calculation where each link is independently parameterized.

Now, we need to project our solution (in the form of $\ket{\phi, \psi_\phi^\text{fermion}}$ into the Gauss's law sector. This can be achieved by integrating over all possible Gauss's law transformation as
\begin{equation}
    \left[\prod_{i,j} \int \dd \varepsilon\hat A_{i,j}(\varepsilon) \right]\ket{\phi, \psi_\phi^\text{fermion}}.
\end{equation}
Since $G_{i, j}(\varepsilon)$ commutes with the Hamiltonian, the state after the integration is still an eigenstate of the Hamiltonian with the same energy, and it also preserves the plaquette variables. Thus, it is the ground state of the Hamiltonian that obeys the Gauss's law.

It turns out, when $m=0$, the fermion system is a conductor, while when $m\ne0$, it is a trivial insulator (with zero Chern number) independent of $\phi$. 

In Fig.~\ref{fig:topo_analytical}, we apply this method to compute (a) the average plaquette value 
\begin{equation}
\begin{aligned}
\langle\cos \phi^P\rangle &\equiv \frac{1}{L^2}\sum_{i, j}\langle \cos \phi_{i,j}^P\rangle \\
&= \frac{1}{2L^2}\sum_{i, j}\langle \hat P_{i,j}+\hat P_{i,j}^\dagger\rangle
\end{aligned}
\end{equation}
and (b) the average nearest neighbor plaquette correlation
\begin{equation}
\begin{aligned}
\mathcal{C}(\phi^P)&\equiv\frac{1}{2L^2}\sum_{\ev{i_1, j_1; i_2, j_2}}\langle \sin \phi_{i_1, j_1}^P \sin \phi_{i_2, j_2}^P \rangle  \\
&= -\frac{1}{8L^2}\sum_{\ev{i_1, j_1; i_2, j_2}}\langle (\hat P_{i_1, j_1}-\hat P_{i_1, j_1}^\dagger)(\hat P_{i_2, j_2}-\hat P_{i_2, j_2}^\dagger) \rangle,
\end{aligned}
\end{equation} where $\ev{i_1, j_1; i_2, j_2}$ refers to nearest-neighbor sites. 
From the figure, we can observe that the system has three phases
\begin{enumerate}
    \item The 0-flux phase, where $\langle\cos \phi^P\rangle=1$ and $\mathcal{C}(\phi^P)=0$. This means that all $\phi_P=0$.
    \item The stagger-flux phase, where $-1<\langle\cos \phi^P\rangle<1$ and $\mathcal{C}(\phi^P) < 0$. In this case we have $\phi_P \ne 0$ and adjacent $\phi_P$'s have opposite signs.
    \item The $\pi$-flux phase, where $\langle\cos \phi^P\rangle =-1$ and $\mathcal{C}(\phi^P) =0$. Here we have all $\phi_P=\pi$.
\end{enumerate}

Although, it is not completely clear from Fig.~\ref{fig:topo_analytical} that there is a phase transition from the $0$-flux region to the stagger-flux region, we can understand the existence of the phase transition by expanding both $E_\text{magnetic}$ and $E_\text{fermion}$ up to second order around $\phi=0$ (dropping constant term and) as
\begin{align}
    E_\text{magnetic} &\sim \frac{g_B^2 L^2}{2} \phi^2;\\
    E_\text{fermion} &\sim  \sum_{i=0}^{L/2}\sum_{j=0}^L \frac{\kappa\cos k_x \cos k_y}{4\sqrt{(\cos k_x + \cos k_y)^2+\frac{m^2}{4\kappa^2}}} \phi^2.
\end{align}
It can be shown that the $E_\text{fermion}$ term is negative and decreases with $\kappa$. Thus, when the two terms are added together, when $\kappa$ is small, the second derivative of the total energy is positive, meaning that $\phi=0$ is the ground state. As $\kappa$ increases, the second derivative changes from positive to negative; thus the $\phi=0$ is no longer the ground state. As the system changes from $\phi=0$ to $\phi\ne0$, a (second or higher order) phase transition must occur, as no analytical function can connect the two regions.

Similarly, around $\phi=\pi$, we can also expand the terms up to second order (dropping constant term and) as
\begin{align}
    E_\text{magnetic} &\sim -\frac{g_B^2 L^2}{2} (\phi-\pi)^2;\\
\begin{split}
    E_\text{fermion} &\sim  \sum_{i=0}^{L/2}\sum_{j=0}^L \frac{\kappa \cos^2 k_x \cos^2 k_y}{4\left(\cos^2 k_x + \cos^2 k_y + \frac{m^2}{4k^2}\right)^{\frac{3}{2}}}(\phi-\pi)^2.
\end{split}
\end{align}
In this case, the $E_\text{magnetic}$ term is negative while the $E_\text{fermion}$ term is positive. Thus, when $\kappa$ is large, the second derivative of the total energy is positive, meaning that $\phi=\pi$ is the ground state, while as $\kappa$ decreases, the second derivative changes sign, which again signatures a second (or higher order) phase transition.

Notice that the phase transition between the three phases is second (or higher) order, which we believe comes from the effect of continuous $U(1)$ variable. Consider the discrete $\mathbb{Z}_N$ model, where $\phi$ can only take discrete values from $[-\pi, \pi)$ with an interval of $2\pi/N$. In this case, as $\kappa$ increases, the system must jump discontinuously from different $\phi$ values, and therefore appears to be first order. However, as we allow $\phi$ to be a continuous variable it transitions smoothly from different values. In Appendix~\ref{app:cv}.

The result in this section is for $g_E^2=0$ only, but it is consistent with the neural network calculation in Fig.~\ref{fig:half_filling_topo}. When $g_E^2>0$, we believe that the system could still have three phases. In Appendix~\ref{app:magge}, we study the effect of $g_E^2$.

\section{Effect of Electric Term on Magnetic Phase Transition} \label{app:magge}

\begin{figure}[h!]
    \centering
    \includegraphics[width=\linewidth]{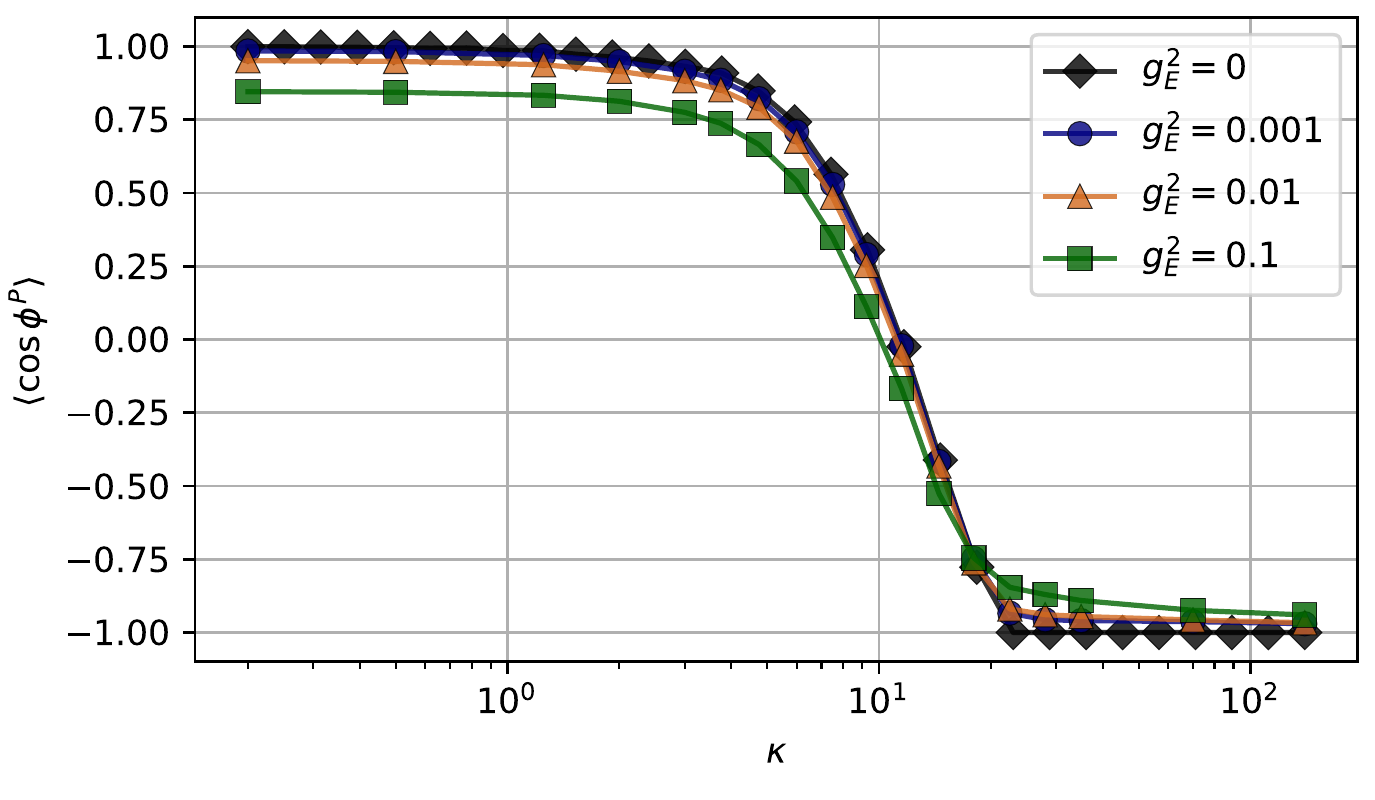}
    \caption{Magnetic effect with dynamical matter at zero density with $g_E^2=0.01$, $g_B^2=1$ and $m=0.01$ on a $4\times4$ system with open boundary condition. (a) Average plaquette observable as a function of $\kappa$ at zero density for different $g_E^2$. We observe that the neural network results approach the analytical result as we decrease $g_E^2$.
    }
    \label{fig:half_filling_ge0}
\end{figure}

We further study the effect of $g_E^2$ on the magnetic phase transition. 
Notice that when $g_E^2=0$, the wave function amplitude in the $\ket{E}$ basis does not decay to zero at infinity. Thus, we could not work with $g_E^2=0$ directly with the neural network.
However, we can still start with full Hamiltonian and gradually decrease $g_E^2$ to be 0.1, 0.01, 0.001, which allows us to see the transition as $g_E$ approaching to zero.
In addition, we calculate the $g_E^2=0$ result using tight binding model (see Appendix~\ref{app:analytic}) and parameterize each link independently. 
We observe that the neural network result for decreasing $g_E^2$ approaches the tight binding model result, suggesting that the neural network result is of high quality.

\end{document}